\documentclass[%
 reprint,
 amsmath,amssymb,
 aps,
prstab,
floatfix,
]{revtex4-2}

\usepackage{graphicx}
\usepackage{subfigure}
\usepackage{lipsum}
\usepackage{dcolumn}
\usepackage{bm}
\usepackage{textcomp}
\usepackage{siunitx}
\usepackage{stackengine}
\usepackage{pifont}
\newcommand{\mycomment}[1]{}

\usepackage{xcolor}
\definecolor{prlblue}{rgb}{0.176, 0.152, 0.57}
\usepackage[colorlinks=true, linkcolor=black, citecolor=prlblue, filecolor=black, urlcolor=prlblue]{hyperref}

\definecolor{light-gray}{gray}{0.8}
\usepackage[mathlines]{lineno}

\begin{document}

\preprint{APS/123-QED}

\title{Positron acceleration in plasma wakefields} 

\author{Gevy J. Cao}
 \altaffiliation{jiawei.cao@fys.uio.no}
\author{Carl A. Lindstr{\o}m}%
\author{Erik Adli}
\affiliation{%
 Department of Physics, University of Oslo, 0316 Oslo, Norway
}%

\author{S{\'e}bastien Corde}
\affiliation{
 LOA, ENSTA Paris, CNRS, Ecole Polytechnique, \\
 Institut Polytechnique de Paris, 91762 Palaiseau, France
}%
\author{Spencer Gessner}
\affiliation{%
SLAC National Accelerator Laboratory, Menlo Park, CA 94025, USA
}%

\date{\today}

\begin{abstract}
Plasma acceleration has emerged as a promising technology for future particle accelerators, particularly linear colliders. Significant progress has been made in recent decades toward high-efficiency and high-quality acceleration of electrons in plasmas. However, this progress does not generalize to acceleration of positrons, as plasmas are inherently charge asymmetric. Here, we present a comprehensive review of historical and current efforts to accelerate positrons using plasma wakefields. Proposed schemes that aim to increase the energy efficiency and beam quality are summarised and quantitatively compared. A dimensionless metric that scales with the luminosity-per-beam power is introduced, indicating that positron-acceleration schemes are currently below the ultimate requirement for colliders. The primary issue is \textit{electron motion}; the high mobility of plasma electrons compared to plasma ions, which leads to non-uniform accelerating and focusing fields that degrade the beam quality of the positron bunch, particularly for high efficiency acceleration. Finally, we discuss possible mitigation strategies and directions for future research.
\end{abstract}

\maketitle


\section{Introduction}
\label{sec:intro}

The high-energy-physics community is currently prioritizing the development of an electron--positron Higgs factory, as emphasized in recent reports from both the US Snowmass process \cite{snowmass_Efront} and the European Strategy for Particle Physics Update \cite{CERN-ESU-015,roadmap_2022_acc}. Linear electron--positron colliders provide clean collisions of elementary particles, suppress synchrotron radiation, and enable future upgrades to higher energies. However, if built using conventional technology---radio-frequency (rf) acceleration---these machines are typically very long and consequently very expensive. For this reason, advanced-accelerator technologies are being considered as a way to reduce the resources required to build such a collider.

Advanced accelerators aim to reduce the footprint by significantly increasing the accelerating gradient. Currently, two rf-based, mature linear-collider designs have been proposed: the International Linear Collider (ILC) \cite{ilc_overview,ilc_tdr1,ilc_tdr2,ilc_tdr3I,ilc_tdr3II,ilc_tdr4} and the Compact LInear Collider (CLIC) \cite{clic_cdr,clic_2016,clic_sum,clic_imp}. The ILC acceleration gradient is \SI{35}{MV/m}, with a total length is \SI{20}{km} including two \SI{6}{km} accelerator arms. This design allows collisions at $\sqrt{s}=\SI{250}{GeV}$, adequate for a Higgs factory. A 1-TeV collider using this technology would extend to at least \SI{40}{km}. CLIC aims to operate at an acceleration gradient of \SI{100}{MV/m} with a center-of-mass energy at $\sqrt{s}=\SI{380}{GeV}$. This collider would have a total length of \SI{11}{km}, with a potential upgrade to $\sqrt{s}=\SI{3}{TeV}$ and a total length of $\SI{50}{km}$. These designs are both pushing the limit of available resources. Beyond ILC and CLIC, other designs have been proposed, including the Cool Copper Collider (C$^3$) \cite{C3_2021}, which could reach up to \SI{120}{MeV/m}. Ultimately, the maximum achievable gradient in all the above machines is limited by electrical breakdown in the metallic rf cavity \cite{Grudiev2009}. Advanced accelerators, however, can surpass this limit by using structures that are more resistant to breakdown. 

Advanced-accelerator concepts include structure-based wakefield accelerators \cite{Jing:2022qbj, Lu:2022oin} as well as plasma-based accelerators. The latter makes use of the ``broken-down" nature of plasmas to overcome the gradient limit in rf accelerators. As a result, plasmas can sustain electric fields of order
\begin{equation}
\label{eq:gradient}
    E_0[\textmd{V/m}] \approx 96\sqrt{n_e [\textmd{cm}^{-3}]},
\end{equation}
which for typical plasma densities $n_e \approx 10^{14}$--\SI{e18}{cm^{-3}} range from 1 to \SI{100}{GV/m} \cite{Corde:2016natcom,Bohlen2022}. This field is up to a thousand times higher than in conventional accelerators.

Early ideas of accelerating particles in a plasma were proposed in 1956 \cite{Veksler:1956pxa,Fainberg:1956qxa}. However, the research field, in its modern form, started independently in 1979 with a seminal paper by Tajima and Dawson \cite{Tajima:1979bn} demonstrating that electrons could be accelerated in the plasma-density wave excited (or \textit{driven}) by an intense laser pulse. Five years later, Chen, Dawson \cite{Chen:1985ft} and Ruth \textit{et al.}~\cite{Ruth:1984pz} proposed to drive these waves using relativistic charged-particle beams. The electromagnetic fields in the plasma-density wave (or \textit{wake}) behind the laser or beam driver are known as \textit{plasma wakefields}.

\begin{figure*}[t]
  \centering
  \includegraphics[width=0.80\linewidth]{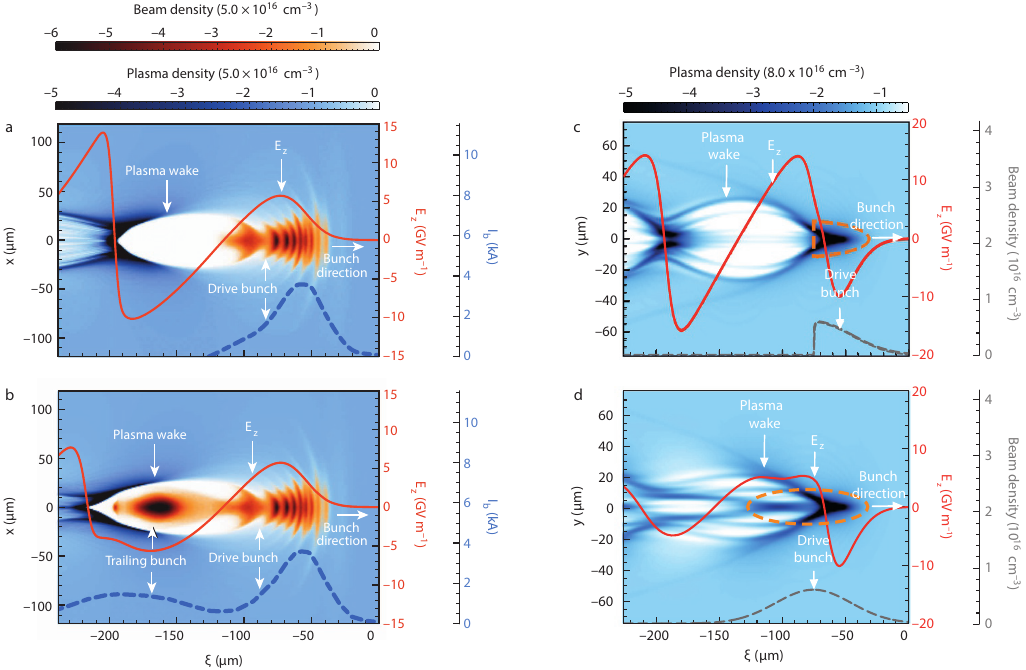}
  \caption{Particle-in-cell simulations of the plasma-density wave and on-axis longitudinal field $E_z$ excited by an electron or positron driver. (a) An electron driver excites a nonlinear plasma wake, or blowout, with strongly accelerating and focusing fields. (b) A trailing electron bunch is accelerated, extracting some of the energy in the wakefield; a process known as beam loading. (c) A positron drive bunch can also excite a nonlinear wake. Here, only the front half of a Gaussian is used, such that no positrons experience acceleration. (d) Using a full Gaussian bunch, the front half drives the wakefield and the rear half loads the wakefield and is accelerated. Adapted from Refs.~\cite{Litos:2014yqa} and \cite{Corde:2015zxa}.}
  \label{fig:blowout}
\end{figure*}

Initial concepts considered small perturbations of the plasma density, now known as the \textit{linear} regime \cite{Keinigs:1986sk}. Later, Rosenzweig \textit{et al.}~\cite{Rosenzweig:1991yx} realized that operating with stronger perturbations, in the so-called \textit{nonlinear} or \textit{blowout} regime, provided more favourable conditions for accelerating electrons with high efficiency and high beam quality. In this regime, plasma electrons are expelled radially outwards by an intense driver, creating a bubble-shaped sheath of plasma electrons surrounding a cavity containing only plasma ions [see Fig.~\ref{fig:blowout}(a)]. These ions, which are uniformly distributed and effectively immobile on the timescale of electron motion, attract the plasma electrons back toward the axis. The inward motion of the sheath electrons creates a longitudinal electric field that can accelerate electrons. Additionally, the exposed ion charge produces a transverse electric field that varies linearly with the transverse offset, thereby focusing electron bunches while preserving their area in transverse phase space (known as \textit{emittance} \cite{Floettmann:2003}). Acceleration extracts energy from the wakefield, which will therefore reduce in amplitude---a process known as \textit{beam loading} \cite{Katsouleas:1987yd}. This process can be used to shape the accelerating field [see Fig.~\ref{fig:blowout}(b)] such that all particles are accelerated uniformly \cite{tzoufras_2008}, allowing energy-efficient acceleration with low energy spread.

Experimental research into acceleration in plasma wakefields has progressed significantly over the past four decades. The first acceleration of electrons in a plasma was demonstrated at the Argonne National Lab in 1988~\cite{Rosenzweig_1988}. Later experiments demonstrated electron injection and acceleration in nonlinear plasma wakefields~\cite{Modena1995,Barov_2000}. Major milestones in beam-driven plasma-wakefield acceleration (PWFA) include: energy doubling of $\SI{42}{GeV}$ electrons~\cite{Blumenfeld:2007ph}; energy-efficient acceleration of an externally injected bunch~\cite{Litos:2014yqa}; and high-gradient, high-efficiency acceleration of electrons while preserving a low energy spread~\cite{Lindstrom:2021tkb}. Similarly, in laser-driven plasma-wakefield acceleration (LWFA), milestones include: generation of high-quality beams~\cite{Geddes:2004tb,Faure:2004tc,Mangles:2004ta}; $\SI{8}{GeV}$ energy gain~\cite{Gonsalves_2019}; and the demonstration of LWFA-based free-electron lasers~\cite{wang_2021_fel,labat_2022}. Several challenges still remain, such as reaching high-overall energy efficiency \cite{Pena_2023}, use of multiple stages \cite{Lindstrom:2021prab,Steinke:2016cyx}, ion motion \cite{Rosenzweig_2005,An_2017}, hosing and beam break-up (BBU) instabilities \cite{Whittum:1990cy,Lebedev:2017dcs,Lehe_2017,mehrling_2018,mehrling_2019}, spin polarization \cite{vieira_2011} and high repetition rate \cite{Cowley:2017skm,DArcy:2022zwq}. Briefly stated, ongoing experimental and theoretical research is rapidly maturing the technology, indicating that plasma acceleration of electrons may soon be compatible with a high-energy-physics application.

Nevertheless, plasma-based acceleration of electrons is not sufficient for a fully plasma-based electron--positron collider; acceleration of positrons is also required. Unfortunately, unlike in rf accelerators, the above-mentioned progress of electron acceleration in plasmas does not readily extend to positrons. Presently, the beam quality and energy efficiency achievable in plasma-based positron accelerator schemes, both in experiments and simulations, is insufficient to reach the requirements of a collider. 

Plasma is a unique accelerating medium in that it responds asymmetrically to particles of positive and negative charge. This is because plasmas are composed of lower-mass (more mobile) electrons and higher-mass (less mobile) ions---an aspect that is exploited in the blowout regime for electron acceleration. For positrons, however, the situation is not as fortunate. In nonlinear plasma wakefields driven by electrons (i.e., a blowout), the only region that both accelerates and focuses positrons is where the plasma electrons cross the axis [at $\xi=\SI{-200}{\mu m}$ in Fig.~\ref{fig:blowout}(a)]; a spatially very small region in which the plasma-electron density is highly non-uniform. This means that the accelerating and focusing fields are non-uniform and nonlinear, respectively, which induces large energy spread and emittance growth. All the favourable features of the blowout regime are therefore lost---previously referred to as the positron problem.

Considering instead nonlinear plasma wakefields driven by positrons, the situation is no better. In this ``suck-in" regime, plasma electrons are sucked into the positron bunch, after which electrons cross the axis and create a blowout-like structure [see Fig.~\ref{fig:blowout}(c)]. The resulting wakefield can be used to accelerate positrons and, if beam loaded, can also keep plasma electrons on axis such that a positron bunch can be focused [see Fig.~\ref{fig:blowout}(d)]. However, while this scheme can be energy efficient \cite{Corde:2015zxa}, the accelerating and focusing fields still vary transversely in a way that does not preserve low energy spread and low emittance. In short, neither the blowout nor the suck-in regime is ideal for positrons. The big question is: can we find a suitable regime that can accelerate positrons with high gradient, high efficiency and high beam quality?

In this review, we start by specifying the requirements of a collider in Sec.~\ref{sec:critical_requirements}. A history of experimental and theoretical progress on positron acceleration in plasma follows in Sec.~\ref{sec:history}. Several new schemes have recently been proposed to overcome the remaining challenges. These schemes are summarized in Sec.~\ref{sec:schemes}. To compare the performance of the schemes, a new dimensionless parameter proportional to the luminosity-per-power---characterizing both the positron bunch and the acceleration process---is employed. The resulting comparison is presented in Sec.~\ref{sec:comp}, which revealed a problem related to electron motion that currently limits the performance of plasma-accelerated positrons---a topic discussed in depth in Sec.~\ref{sec:positron-problem}. Finally, concluding remarks and an outlook is presented in Sec.~\ref{sec:conclusion}.

\section{Critical requirements for\\linear colliders}
\label{sec:critical_requirements}

The goal of plasma-based positron acceleration is to, in an affordable manner, deliver high-energy positrons for an electron--positron collider. Physics determines the required center-of-mass energy, typically in the range 0.25--\SI{15}{TeV}, as well as the required collision rate, or \textit{luminosity}, which is typically around \SI{e34}{\per\square\cm\per\s}. This luminosity can be calculated as
\begin{equation}
    \mathcal{L} \approx \frac{f N^2}{4\pi\sigma_x\sigma_y},
\end{equation}
where $f$ is the collision frequency, $N$ is the electron or positron bunch population (here assumed identical), and $\sigma_{x/y}$ is the root-mean-square (rms) beam size of the colliding bunches in the horizontal/vertical plane. Alternatively, it can be useful to express the luminosity as
\begin{equation}
    \label{eq:luminosity_alt}
    \mathcal{L} \approx \frac{1}{8\pi m_e c^2} \frac{P_\textmd{wall}}{\sqrt{\beta_x\epsilon_{nx}}}\frac{\eta N}{\sqrt{\beta_y\epsilon_{ny}}},
\end{equation}
where $P_\textmd{wall}$ is the wall-plug power required, $\eta$ is the wall-plug-to-beam energy-transfer efficiency, $\epsilon_{nx/ny}$ is the \textit{normalized} (i.e., energy-independent) emittance and $\beta_{x/y}$ is the beta function \cite{COURANT19581}, while $m_e$ and $c$ are the electron mass and speed of light in vacuum, respectively. Considering that the construction cost scales with the length of the linear collider, minimizing the construction cost requires high accelerating gradient; similarly, since the running cost scales with the wall-plug power, minimizing the running cost while maintaining luminosity [Eq.~\ref{eq:luminosity_alt}] requires high charge and low emittance (i.e., high beam quality) as well as high energy efficiency.

\subsection{Accelerating gradient}
\label{sec:accel_gradient}

The highest achievable gradient in an rf cavity is around \SI{200}{MV/m} \cite{dolgashev_2021}. To justify switching accelerator technology, a minimum accelerating field of \SI{1}{GV/m} is typically required for plasma accelerators. Equation~\ref{eq:gradient} indicates, therefore, that a plasma density of at least \SI{e14}{cm^{-3}} will be required. Likely, even higher in-plasma accelerating gradients ($>\SI{10}{GV/m}$) and therefore higher densities will be required ($>\SI{e16}{cm^{-3}}$), because the \textit{effective} gradient averaged longitudinally across multiple stages can be significantly reduced due to lengthy staging optics \cite{Lindstrom:2021prab,Steinke:2016cyx}. This minimum plasma density places restrictions on the length of the accelerating bunch $\sigma_z$: to be contained within the accelerating phase of the plasma wave, the bunch length must be less than approximately one plasma skin depth, $\sigma_z \lesssim k_p^{-1}$, where $k_p = \sqrt{n_e e^2/\epsilon_0 m_e c^2}$ is the plasma wavenumber, $\epsilon_0$ is the vacuum permittivity, and $m_e$ and $e$ are the electron mass and charge, respectively. This means that bunches must typically be shorter than \SI{50}{\micro\m} rms (assuming a plasma density of \SI{e16}{cm^{-3}}). The same argument applies to the drive bunch.

\subsection{Energy efficiency} 
\label{sec:energy_efficiency}

The combination of high particle energy, high charge and high collision frequency translates to high beam power. The wall-plug power needed to generate this beam power is defined by the energy-transfer efficiency. It is instructive to split this overall efficiency into three sub-efficiencies: 
\begin{equation}
    \label{eq:efficiencies}
    \eta = \eta_{\textmd{prod}} \times \eta_{\textmd{depl}} \times \eta_{\textmd{extr}},
\end{equation}
where $\eta_{\textmd{prod}}$ is the driver-production efficiency, or the fraction of the wall-plug power that ends up in the drive beam; $\eta_{\textmd{depl}}$ is the energy-depletion efficiency, or the fraction of the drive-beam energy transferred to the plasma wake; and $\eta_{\textmd{extr}}$ is the extraction efficiency, or the fraction of the wakefield energy extracted by the accelerating beam. The overall efficiency $\eta$ of conventional colliders is around 5-10\% \cite{ilc_overview,clic_cdr}.

The maximum achievable production efficiency depends on the type of driver; it is typically larger for electron drivers (as high as 50\%~\cite{clic_cdr}) compared to that for positrons, protons and laser pulses~\cite{Hooker:2014cza}. For electron-driven plasma accelerators, experiments have demonstrated a depletion efficiency above 50\% \cite{Pena_2023}, and simulations indicate that this can be extended beyond 90\%~\cite{Lotov_2005,Su_2023}. Such high depletion efficiencies require stable propagation of the driver, avoiding effects such as the \textit{hose instability}~\cite{Whittum:1990cy,mehrling_2017} and \textit{head erosion} (i.e., the divergence of the head of the driver~\cite{zhou_2007,blumenfeld_phd}). For laser drivers, the depletion efficiency scaling is somewhat more complex~\cite{Lu_2007_laserdep}, as laser pulses can evolve significantly if approaching energy depletion~\cite{shadwick2009}.

The extraction efficiency can be calculated using the ratio of the energy gained by the trailing bunch to the energy lost by the driver,
\begin{equation}
    \eta_{\textmd{extr}} = -\frac{Q_\textmd{trailing}\Delta \langle E_\textmd{trailing}\rangle}{Q_\textmd{driver}\Delta \langle  E_{\textmd{driver}}\rangle},
\end{equation}
where $Q$ is the charge and $\Delta \langle E \rangle$ is the change in centroid energy of the respective bunches. To compete with conventional machines, and assuming 50\% production and depletion efficiencies, a 20--40\% extraction efficiency is required.

\subsection{Beam quality}
\label{sec:beam_quality}

Beam quality directly affects the luminosity through two parameters: bunch charge and normalized emittance. Ultimately, there is no fixed requirement for charge and emittance---it is possible to have higher emittance as long as there is more charge (according to Eq.~\ref{eq:luminosity_alt}) and vice versa. However, conventional colliders typically use charges of order \SI{1}{nC} and normalized emittances of order 10 by \SI{0.01}{\milli\m\milli\radian} in the horizontal and vertical planes, respectively. These emittances are asymmetric, resulting in ``flat" beams at the collision point, to suppress disruptive beam--beam effects, or \textit{beamstrahlung} \cite{schulte_phd,Schulte:2016ijt}. Such requirements place tight constraints on preservation of charge and emittance throughout the accelerator, which can be particularly challenging in plasma accelerators \cite{Lindstrom_2022}.

It addition, the luminosity is indirectly affected by another beam quality: the energy spread. A small energy spread is desired to maintain a well-defined collision energy (i.e., a narrow \textit{luminosity spectrum}). A tighter restriction, however, comes from collider final-focusing systems, which can only provide sufficiently small beta functions ($\beta_y \lesssim \SI{1}{mm}$) if the energy spread is small \cite{Raimondi:2001cx}; typically less than 1\% rms. This problem, known as \textit{chromaticity}, also applies to transport between stages, where large energy spreads can lead to emittance growth \cite{migliorati_2013}. Note that these requirements apply to the \textit{uncorrelated} energy spread (i.e., within a longitudinal bunch slice); a \textit{correlated} energy spread, or \textit{chirp}, can potentially be removed by \textit{dechirping} prior to final focusing \cite{darcy_2019,shpakov_2019,Wu_2019}.

The last important beam quality is \textit{spin polarization}~\cite{Mane_2005}, which is required to study spin-dependent electroweak processes~\cite{Clendenin:1993pk}. Spin polarization can also be challenging to preserve in a plasma accelerator~\cite{vieira_2011}.

\subsection{Stability}
\label{sec:stability}

Two types of stability are required in a linear accelerator: avoidance of exponentially growing instabilities that arise from positive feedback loops (resonances), such as the beam-breakup instability \cite{Lau:1989,Lebedev:2016azx,Lebedev:2017dcs}; and operation within the error tolerance of all input parameters, which includes alignment and temporal synchronization. Instability manifests in the form of loss of beam quality, such as increased normalized emittance and eventually loss of charge. It can also affect luminosity beyond a direct effect on beam quality: for instance, significant transverse jitter at the interaction point may prevent collisions even if emittance and charge is preserved. Ultimately, the accelerator must maintain sufficient stability to ensure the effect on the luminosity is small.

\vspace{1em}

In summary, the above comprise a challenging list of ``top-down" requirements for any linear accelerator. The following section describes two decades of ``bottom-up" plasma-wakefield research towards delivering these requirements for positrons.

\section{A history of accelerating positrons in plasma wakefields}
\label{sec:history}

\begin{figure}[t]
  \centering
  \includegraphics[width=0.95\linewidth]{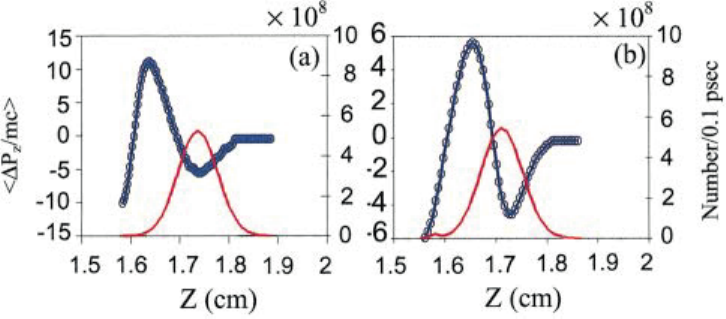}
  \caption{First PIC simulation comparing the energy change of electrons (a) and positrons (b) in a plasma wakefield, indicating an asymmetric response. From Ref.~\cite{Lee:2001bb}.}
  \label{fig:first-pic}
\end{figure}

Plasma-based positron-acceleration research started in the early 2000s. The first numerical study, performed by Lee \textit{et al.}~\cite{Lee:2001bb}, compared electron-driven and positron-driven nonlinear plasma wakes. The results showed that in a homogeneous plasma, a positron bunch drives comparatively lower-amplitude wakefields than those driven by an identical electron bunch (see Fig.~\ref{fig:first-pic}). However, they also found that in a hollow plasma channel \cite{Tajima:1983egt}, where no plasma exists on axis, the wakefield amplitudes can be more comparable between electrons and positrons. This section presents theoretical and experimental work focused on these two plasma profiles: homogeneous plasmas in Sec.~\ref{sec:homogeneous-plasma} and hollow plasma channels in Sec.~\ref{sec:hollow-plasma-channels}.

\begin{table*}[t]
    \centering
    \caption{Experimental milestones in plasma-based positron acceleration research. The year refers to year of publication.}
    \begin{ruledtabular}
        \begin{tabular}{p{0.05\textwidth}p{0.7\textwidth}p{0.2\textwidth}}
            \textbf{Year} & \textbf{Description} & \textbf{Reference} \\ \hline
             2001 & First plasma focusing of positrons & Ng \textit{et al.}~\cite{Ng:2001is} \\
             2003 & First guiding of positrons in a near-hollow plasma channel & Marsh \textit{et al.} \cite{Marsh:2003dnx} \\
             2003 & First broad-band deceleration and acceleration of positrons & Blue \textit{et al.}~\cite{Blue:2003nk} \\
             2003 & First meter-scale transport of positron bunches & Hogan \textit{et al.}~\cite{Hogan:2003bs} \\
             2008 & First observation of positron halo formation and emittance growth & Muggli \textit{et al.}~\cite{Muggli:2008zzb} \\
             2015 & First multi-GeV energy gain for positrons & Corde \textit{et al.}~\cite{Corde:2015zxa} \\
             2016 & First demonstration of a hollow-plasma-channel accelerator & Gessner \textit{et al.}~\cite{Gessner:2016bqz} \\ 
             2017 & First acceleration of a distinct positron bunch & Doche \textit{et al.}~\cite{Doche:2017jhd} \\
             2018 & First measurement of positron-driven transverse wakefields in a hollow channel & Lindstr{\o}m \textit{et al.}~\cite{Lindstrom:2018hhy} \\
             2023 & First efficient energy transfer between positron bunches in a hollow plasma channel & Gessner \textit{et al.}~\cite{Gessner:2023arxiv} \\
            \end{tabular}
        \end{ruledtabular}
    \label{tab:table_hist} 
\end{table*}

\begin{figure}[b]
    \includegraphics[width=\linewidth]{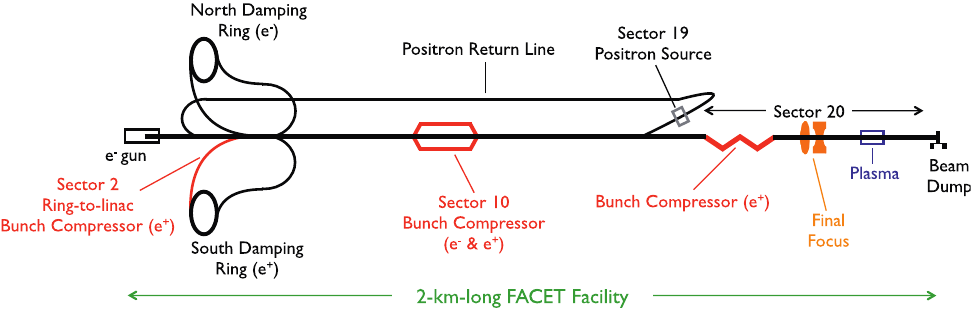}
    \caption{Schematic of the positron source at SLAC. Positron bunches are produced by sending electrons through a high-$Z$ target, subsequently transported through a return line to a damping ring. After damping, the positrons are accelerated and compressed by two bunch compressors before delivery to the experimental area, where plasma acceleration occurs. From Ref.~\cite{Corde:2015zxa}.}
    \label{fig:posi_ring}
\end{figure}

Only limited experimental research has been directed toward positron acceleration in plasma wakefields. This is due to a general lack of experimental facilities that can provide positron bunches with high charge and high energy. So far, all experiments have been performed at the SLAC National Accelerator Laboratory, which produced intense positron bunches for the Stanford Linear Collider (SLC) in the 1990s \cite{Seeman:1991wf}, as illustrated in Fig.~\ref{fig:posi_ring}. Selected experimental milestones are highlighted in Table~\ref{tab:table_hist}.

\subsection{Positron acceleration in homogeneous plasmas}
\label{sec:homogeneous-plasma}

In the 1990s, one of the greatest challenges for linear colliders was focusing beams to the sub-micron level in order to reach high luminosity. This prompted the launch of the Final Focus Test Beam (FFTB) facility \cite{Burke:1991xr,Berndt:1991ug} at SLAC, which delivered short electron and positron bunches at energies up to \SI{47}{GeV}. Several advanced focusing and acceleration techniques were also tested, initially including plasma lensing of electrons and positrons (the E-150 experiment \cite{Betz1991,Chen:1998ut}) and plasma-wakefield acceleration of electrons (the E-157 experiment \cite{ASSMANN1998396,Hogan:2000xq}). Later experiments continued the E-157 experiment by also investigating plasma-wakefield acceleration of positrons (the E-162 experiment \cite{Baird2000}).

\begin{figure}[t]
  \centering
  \includegraphics[width=0.85\linewidth]{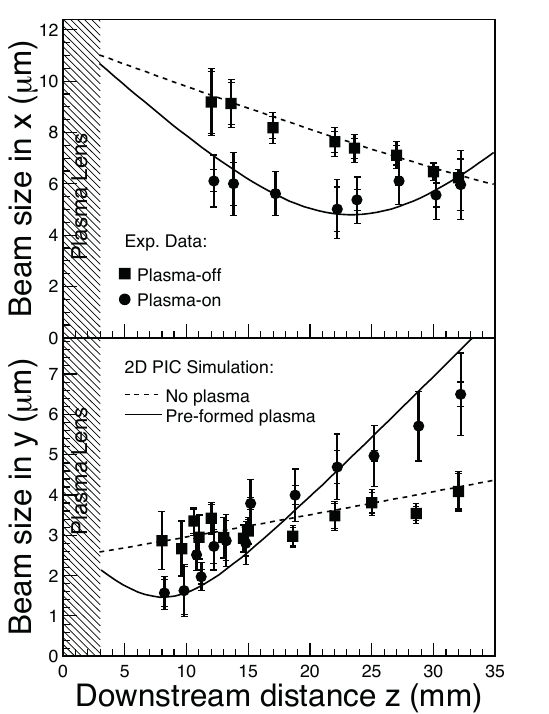}
  \caption{Plasma lensing of a positron bunch at FFTB. The beam size in the horizontal (top) and vertical plane (bottom) were measured using a wire scanner, showing a focused beam waist only 20 and 10 mm downstream, respectively. The measured beam envelopes match 2D PIC simulations (solid and dashed lines). The beam's cross-sectional area was reduced by a factor $2.0\pm0.3$ and focusing fields as high as $\SI{4}{MT/m}$ were achieved. From Ref.~\cite{Ng:2001is}.}
  \label{fig:plasma-lensing}
\end{figure}

Motivated by the promise of ultra-compact final focusing for linear colliders \cite{CHEN1998407}, the E-150 plasma-lens experiment demonstrated focusing of electrons and then of positrons in 2000. In the experiment, reported by Ng \textit{et al.}~\cite{Ng:2001is}, a \SI{28.5}{GeV} positron beam traversed a 3-mm-thick nitrogen gas jet ionized by the positron beam itself but assisted by an Nd:YAG laser pulse. The plasma densities were not measured at the time, but using a simulated value of $\SI{5e17}{cm^{-3}}$ yielded good agreement with experimental data. The beam density was around $\SI{2e16}{cm^{-3}}$, implying that the experiment was operated in the linear (or overdense) plasma-lens regime, in which plasma electrons neutralize the electric field of the positron bunch. The self-focusing effect was provided by the azimuthal magnetic field of the bunch. The main experimental results are shown in Fig.~\ref{fig:plasma-lensing}. This was the first experiment demonstrating positrons interacting with plasma wakefields.

\begin{figure}[t]
    \centering
    \includegraphics[width=\linewidth]{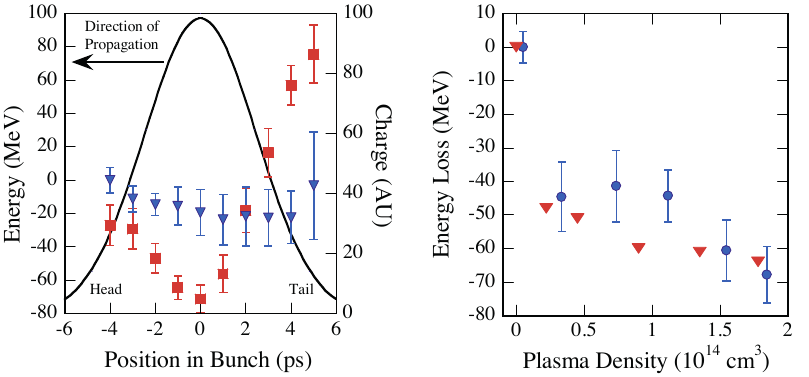}
    \caption{First observation of positron acceleration in plasma. The energy of various slices within the positron bunch was measured using a streak camera with a 1~ps temporal resolution (left plot). When the plasma is on (red), the bunch head loses energy while the tail gains energy; this is compared to when the plasma is off (blue). The peak accelerating field, averaged over \SI{1.4}{m}, is \SI{56}{MeV/m}. Moreover, a scan of plasma density (right plot) shows the experimental measurements (blue) as well as simulated predictions (red) of the change in centroid energy, indicating higher-amplitude wakefields at higher plasma densities. From Ref.~\cite{Blue:2003nk}.}
    \label{fig:history_uni_2003}
\end{figure}

\begin{figure}[h]
    \centering
    \includegraphics[width=\linewidth]{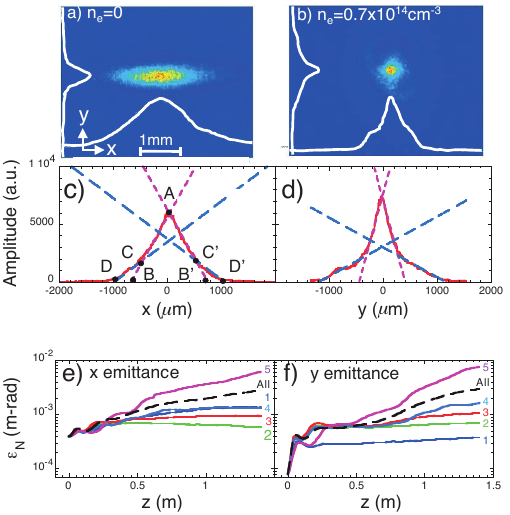}
    \caption{Halo formation and emittance growth for positrons in a plasma. Transverse profiles were measured without plasma (a, c) and with plasma (b, d). The fits to the profiles show that after propagation through a plasma, the fraction of charge contained in the surrounding halo (dashed blue lines) was significantly larger compared to the core (dashed purple lines). A matching PIC simulation shows a large emittance growth in the horizontal (e) and vertical plane (f), both for the full projected beam (dashed black line) and for various longitudinal slices (colored lines; numbered from head to tail). Adapted from Ref.~\cite{Muggli:2008zzb}.}
    \label{fig:history_uni_2008}
\end{figure}

Following successful plasma-wakefield experiments with electrons (E-157), the E-162 experiment demonstrated both meter-scale transport and acceleration of positron bunches. This experiment also made use of a $\SI{28.5}{GeV}$ beam, containing 1--$2\times10^{10}$ positrons compressed to a bunch length of \SI{700}{\hbox{\textmu}m} rms and focused to a beam size of \SI{40}{\hbox{\textmu}m} rms, but used a \SI{1.4}{m}-long lithium-vapor plasma source ionized by an ultraviolet laser to a density of up to $\SI{1.8e14}{cm^{-3}}$. In 2003, Blue \textit{et al.}~\cite{Blue:2003nk} demonstrated the first acceleration of positrons in a plasma, as shown in Fig.~\ref{fig:history_uni_2003}. Here, a streak camera and a Cherenkov radiator were used to measure energy loss and gain of different slices within the positron bunch. Using a similar setup, Hogan \textit{et al.}~\cite{Hogan:2003bs} showed that the focusing strength increased from the head to the tail of a positron bunch. The plasma density required to optimally focus the tail was found to be approximately 7~times lower than that needed for an identical electron bunch. This asymmetry occurs because the positron bunch attracts plasma electrons, resulting in an on-axis density spike with increasing electron density toward the tail of the bunch, as compared to the uniform ion density observed for electron bunches. 

Although the on-axis electron-density spike focuses positrons, it results in nonlinear focusing and rapid emittance growth. As a result, a halo of diverged positrons will form around the core of the bunch. In a subsequent FFTB experiment, Muggli \textit{et al.}~\cite{Muggli:2008zzb} investigated this effect by quantifying the fraction of positron charge contained in the halo, as illustrated in Figs.~\ref{fig:history_uni_2008}(a)--(d). This experiment showed that the halo contained as much as 40\% of the total charge after \SI{1.4}{m} of propagation. Supporting PIC simulations indicate that the normalized emittance increased by a factor 10--100, not only for the projected beam but also for individual longitudinal slices [see Figs.~\ref{fig:history_uni_2008}(e) and (f)].


After the shutdown of FFTB in 2006, a new facility was planned, reusing part of the accelerator but moving and upgrading the experimental area. The Facility for Advanced aCcelerator Experimental Tests (FACET) \cite{Hogan:2010zz,Clarke:2011za} started operation in 2012. 

In the intervening years, numerical studies investigated the transition between linear and nonlinear wakefields for positrons, as well as issues related to efficiency and beam quality. Lu \textit{et al.}~\cite{Lu_2005} found that the linear wakefield theory breaks down at lower beam densities for positrons compared to electrons. Another study by Zhou \textit{et al.}~\cite{Zhou_2006} found that an electron bunch drives a stronger wakefield than an equivalent positron bunch, in agreement with Lee \textit{et al.}~\cite{Lee:2001bb}. By investigating the dynamics of the plasma electrons, Zhou \textit{et al.} observed that when electrons flow into the positron bunch, the excess negative charge acts just like a (weaker) electron bunch with a positively charged head. In 2007, Lotov~\cite{Lotov_2007} demonstrated that high-efficiency (up to 60\%) positron acceleration is possible in an electron-driven plasma wakefield with optimized positron beam loading, though with energy spreads up to several percent. It was noted that the attainable efficiency is higher for weakly nonlinear wakes than for strongly nonlinear wakes. Complementary to this finding was a study by An \textit{et al.}~\cite{An:2010exa}, which demonstrated similarly high efficiencies and that the weakly nonlinear regime offers higher efficiency than the linear regime in exchange for a higher energy spread. Other studies showed that the positron emittance growth increases for higher plasma densities \cite{Li:2010moa}, and that the energy spread of a longitudinal slice can increase to the 10\% level after only \SI{50}{cm} of propagation in a plasma \cite{Gessner:2012zz}. 

\begin{figure}[t]
    \centering
    \includegraphics[width=0.98\linewidth]{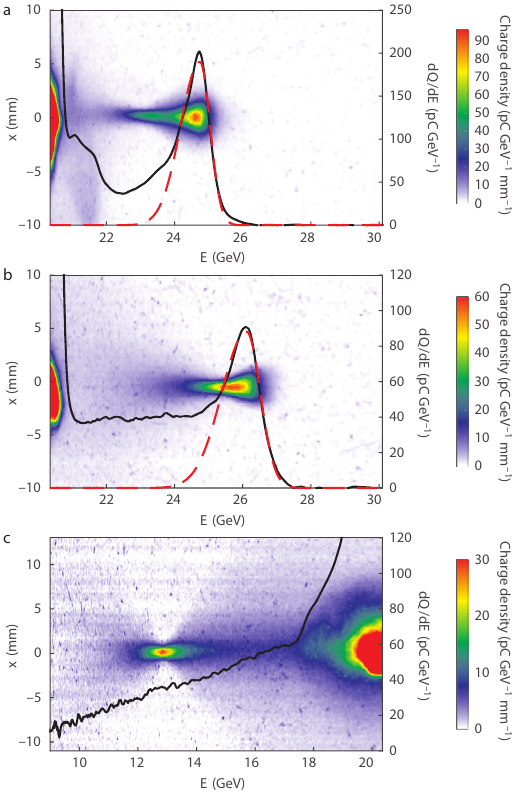}
    \caption{First multi-GeV acceleration of positrons in a plasma wakefield at FACET. (a) An imaging spectrometer focused at \SI{22.85}{GeV} shows accelerated positrons with a defined peak in the energy spectrum (black line) at \SI{24.75}{GeV} and an energy spread of 1.8\% rms based on a Gaussian fit (red dashed line). (b) The highest energy gain of the peak was approximately \SI{5.75}{GeV}; here the imaging energy was \SI{25.35}{GeV}. (c) To measure the energy-transfer efficiency, the continuous energy spectrum of the decelerated particles was also measured. From Ref.~\cite{Corde:2015zxa}.}
    \label{fig:history_uni_2015}
\end{figure}

In 2015, Corde \textit{et al.}~\cite{Corde:2015zxa} reached a major milestone at the FACET facility: experimental demonstration of high-gain (\SI{5.75}{GeV}), high-gradient (\SI{3.8}{GV/m}) and high-efficiency (30\%) acceleration of positrons in a plasma. Here, the incoming positron bunch had an energy of $\SI{20.35}{GeV}$, a charge of $\sim$\SI{2.2}{nC} and a bunch length of 30--\SI{50}{\hbox{\textmu}m}. Approximately 100--\SI{200}{pC} of charge was accelerated in a \SI{1.3}{m}-long lithium plasma at a density of $\SI{8e16}{cm^{-3}}$---a significantly higher density than in previous experiments, enabled by the use of shorter bunches. The results are illustrated in Fig.~\ref{fig:history_uni_2015}. In this experiment, the head of the positron bunch drove a strongly nonlinear wakefield while the tail loaded the wakefield, extracting a significant fraction of the energy deposited in the wake [Fig.~\ref{fig:blowout}(d)]. Without the presence of the bunch tail, the accelerating region is defocusing for positrons [see Fig.~\ref{fig:blowout}(c)] as the plasma electrons flow outward behind the bunch head, forming a blowout-like structure. However, some of the plasma electrons remain on axis due to the focusing field of the positron bunch tail [see Fig.~\ref{fig:blowout}(d) around $\xi=\SI{-100}{\hbox{\textmu}m}$], a process known as \textit{transverse beam loading}, resulting in an accelerating region which is focusing for the positrons. This scheme is referred to as \textit{self-loaded} plasma-wakefield acceleration. A final energy spread of 1.8\% rms was achieved, which for the 22\% relative energy gain corresponds to a uniformity of the accelerating field of around 8\% rms. 

\begin{figure*}[t]
    \centering
    \includegraphics[width=0.8\textwidth]{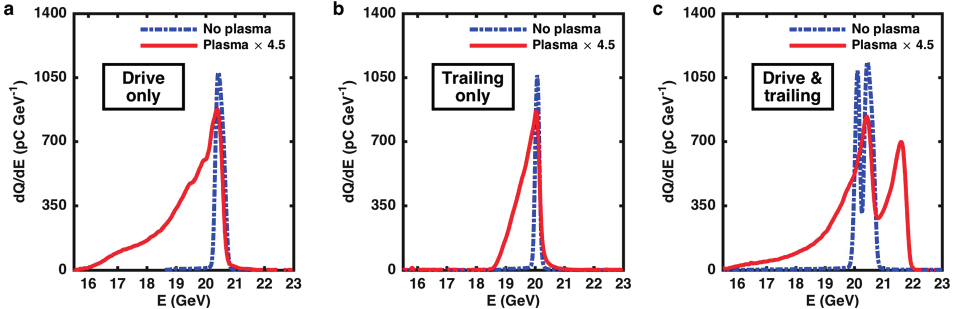}
    \caption{First acceleration in a plasma of a distinct positron bunch. The drive and trailing bunches were first sent individually through a plasma, shown in (a) and (b), respectively. Energy loss can be observed in both bunches with negligible energy gain, as the plasma wavelength is much longer than the bunch length. (c) However, when both bunches propagated through the plasma, approximately \SI{85}{pC} of the trailing-bunch charge was accelerated. The spectral peak at \SI{21.5}{GeV} was accelerated by \SI{1.45}{GeV} in \SI{1.3}{m}, corresponding to an accelerating gradient of \SI{1.12}{GV/m}. The drive-to-main efficiency was estimated to be 40\% with a final energy spread of 1.0\%. From Ref.~\cite{Doche:2017jhd} (CC BY 4.0).}
    \label{fig:history_uni_2017}
\end{figure*}

Up to this point, all positron experiments had been performed using single bunches---that is, bunches with a Gaussian-like current profile. However, to reduce the energy spread of the accelerated positrons, a driver--trailing bunch pair is required. With the FACET facility's ability to produce such bunch pairs, as used in PWFA experiments for electrons \cite{Litos:2014yqa}, a follow-up experiment was performed by Doche \textit{et al.}~\cite{Doche:2017jhd}. In this experiment, a chirped (i.e, time--energy correlated) positron bunch with a mean energy \SI{20.35}{GeV} traversed a ``W-shaped" magnetic chicane, within which the beam was energetically dispersed (i.e., there was a transverse--energy correlation). In the chicane, a notch collimator blocked the central part of the energy spectrum before the bunch was again undispersed, resulting in a double-bunch structure with a leading drive bunch (\SI{20.55}{GeV}) and a trailing bunch (\SI{20.05}{GeV}). The charges of the driver and trailing bunches were approximately \SI{480}{pC} and \SI{260}{pC}, respectively, with corresponding bunch lengths of \SI{30}{\hbox{\textmu}m} and \SI{40}{\hbox{\textmu}m}. The \SI{1.3}{m}-long lithium plasma was ionized to a density of $\SI{1e16}{cm^{-3}}$. Using this setup, the experiment demonstrated the first acceleration of a distinct positron bunch in a plasma wakefield driven by another positron bunch (see Fig.~\ref{fig:history_uni_2017}). Here, an energy-transfer efficiency of about 40\% and a final energy spread of 1.0\% rms were achieved; given the relative energy gain of 7\%, this energy spread corresponds to a field uniformity of approximately 14\%.

In the above experiment \cite{Doche:2017jhd}, the beam emittance was varied by inserting titanium foils of different thicknesses, as a way to investigate the transition between the nonlinear and quasi-linear regimes. This was motivated by the quality-preserving features of the quasi-linear regime, which could mitigate the emittance growth seen in strongly nonlinear wakefields. Measurements showed that higher emittances resulted in smaller energy gain; a sign of lower-amplitude plasma wakefields. Simulations using the experimental parameters indicate that the low-emittance beam ($100 \times \SI{10}{\milli\m\milli\radian}$) drove a nonlinear plasma wakefield while the high-emittance beam ($270 \times \SI{60}{\milli\m\milli\radian}$) drove a quasi-linear plasma wakefield. Additionally, a negative correlation was seen between the trailing-bunch charge and the amplitude of the wakefield, as well as between the charge and the energy spread---observations consistent with beam loading.

Finally, in a separate numerical study, Fujii \textit{et al.}~\cite{Fujii:2019qxb} examined the often-neglected, but important issue of positron-beam extraction at the end of acceleration; they found that a plasma-density down ramp can be detrimental and lead to increased divergence, in contrast to electrons, for which ramps lead to reduced divergence \cite{marsh_2005,Dornmair_2015,Xu:2016kia,ariniello_2019,Litos:2019tzi,Li_2019b,Zhao_2020}. An alternative method was therefore proposed: gradual reduction in wake amplitude via head erosion, as this keeps the positrons in phase with the focusing field. 

Summarizing the progress of positron acceleration in homogeneous plasmas, experiments have demonstrated acceleration with high gradient, high energy-transfer efficiency and reasonably low energy spread. However, acceleration of positron bunches with both low emittance and low energy-spread remains a major challenge.

\subsection{Positron acceleration in hollow plasma channels}
\label{sec:hollow-plasma-channels}

\begin{figure}[b]
    \centering
    \includegraphics[width=0.95\linewidth]{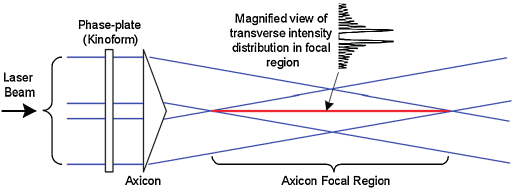}
    \caption{Experimental concept for creating a hollow plasma channel. A laser is used to ionize a gas (red line) by combining the use of an axicon, to create a longitudinally uniform plasma filament, and a kinoform, to create a higher-order Bessel profile in the transverse plane. From Ref.~\cite{Kimura:2011zz} (CC BY 3.0).}
    \label{fig:history_hollow_2011}
\end{figure}

\begin{figure*}[t]
    \centering
    \includegraphics[width=0.9\textwidth]{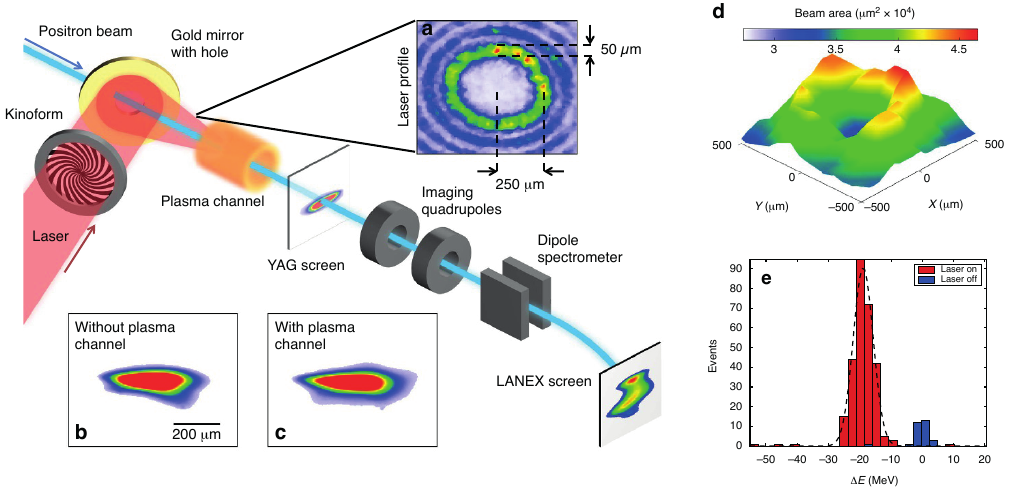}
    \caption{Experimental layout of the first demonstration of plasma wakefields in a hollow plasma channel. An ionizing laser passes through a kinoform, resulting in an annular transverse profile (a), and is coupled to the beam axis using a holed mirror. Shortly after, a positron bunch propagates through the channel. The transverse profile of the positron bunch is captured on a downstream yttrium-aluminium garnet (YAG) screen. Comparing images without (b) and with (c) a plasma channel, no difference in beam size was observed, indicating the absence of focusing fields in the channel. (d) The hollow-channel density profile was inferred by scanning the channel offset in both transverse planes with respect to the beam axis. When the beam interacts with the plasma of the channel wall, it experiences a focusing force, resulting in higher divergence and a larger beam size on the YAG screen. Aligning the beam to the channel centre, its energy spectrum was measured downstream using a dipole spectrometer and a LANEX screen. A histogram of the centroid energy loss for 315 shots is shown in (e). Adapted from Ref.~\cite{Gessner:2016bqz} (CC BY 4.0).}
    \label{fig:history_hollow_2016}
\end{figure*}

The alternative approach is to use a hollow plasma channel, which can be described as a tubular plasma surrounding an un-ionized (hollow) core. This concept was originally proposed by Tajima in 1983 \cite{Tajima:1983egt}, known at the time as the \textit{plasma-fiber accelerator}. The motivation was to increase the acceleration length, in two ways: by avoiding the reduced velocity of light in a plasma, thereby suppressing phase slippage between the accelerating electrons and the laser pulse (known as \textit{dephasing}); and to provide optical guiding of lasers, thereby suppressing the divergence of the wavefront of the laser (known as \textit{diffraction}). While initially based on \textit{overdense} plasmas ($\omega < \omega_p$, where $\omega$ and $\omega_p$ are the laser and plasma frequencies, respectively), the concept was later extended to using \textit{underdense} plasmas ($\omega > \omega_p$), which provide favourable laser-guiding characteristics~\cite{Barnes:1987,Katsouleas:1992yn,Chiou_1995}.

Later work by Chiou and Katsouleas \cite{chiou_1998} highlighted that hollow plasma channels provide several advantages: transversely uniform accelerating fields, which enable low energy spread; zero focusing fields within the channel, which enable preservation of emittance; as well as high energy efficiency through beam loading. However, around the same time, Schroeder \textit{et al.}~\cite{Schroeder:1999cb} calculated that bunches travelling off-axis in the channel would excite a transverse wakefield that acts to deflect the bunch, and could therefore lead to a beam-breakup instability.

The first proposal to use hollow plasma channels for positron acceleration came from Lee \textit{et al.}~in 2001 \cite{Lee:2001bb}, motivated by the higher-amplitude wakefields achievable compared to those in a homogeneous plasma. This led to hollow channels being proposed for the positron arm of plasma-based electron--positron colliders, including in the \textit{plasma afterburner} concept proposed in 2002 \cite{Lee:2002bh,Raubenheimer_2004}.

Motivated by these collider concepts, a first attempt at realizing the hollow channel was performed as part of the E-162 \cite{Baird2000} positron experiment at FFTB. In 2003, Marsh \textit{et al.}~\cite{Marsh:2003dnx} reported the propagation of positron bunches through a meter-scale near-hollow plasma channel, comparing it to propagation through a homogeneous plasma. In this experiment, the channel was produced by blocking the central portion of the UV laser that ionized the plasma \SI{200}{ns} before the arrival of the positron bunch---an approach that produced an on-axis density depression rather than a truly hollow channel. To properly exploit the advantages of a hollow channel, complete absence of plasma in the channel would be required. 

Several ideas were proposed regarding how to realize the hollow channel experimentally. Kirby \textit{et al.}~\cite{Kirby:2009zza} proposed inserting an circular obstruction into a gas jet. However, another approach by Fan \textit{et al.}~\cite{Fan_2000} was found to be more promising: ionizing the gas using a tubular laser pulse created using a combination of an axicon and a spiral phase plate (known as a \textit{kinoform}~\cite{Andreev_1996}). This axicon--kinoform setup, which produces a plasma that is longitudinally uniform and a high-order Bessel function transversely, is illustrated in Fig.~\ref{fig:history_hollow_2011}. 

In 2011, Kimura \textit{et al.}~\cite{Kimura:2011zz} performed the first self-consistent simulations of positron acceleration in hollow plasma channels produced by a kinoform optic, finding that \SI{1}{m}-long plasma channels with a density of order $\SI{e16}{cm^{-3}}$ could be produced and would sustain accelerating fields as high as \SI{3}{GV/m}. Another conclusion was that the positron beam could in principle ionize the gas within the channel, but this can be avoided with a sufficiently low beam density (e.g., larger than \SI{20}{\hbox{\textmu}m} rms beam size and \SI{20}{\hbox{\textmu}m} rms bunch length for a \SI{2.9}{nC} bunch, assuming hydrogen gas). These simulations laid the groundwork for hollow-channel experiments at FACET.

Using the kinoform method to generate a hollow channel, in a 2014 experiment performed at FACET, Gessner \textit{et al.}~\cite{Gessner:2016bqz} demonstrated the first plasma wakefields driven by positrons in a truly hollow plasma channel. The experimental setup and results are shown in Fig.~\ref{fig:history_hollow_2016}. An \SI{8}{cm}-long hollow channel with a radius of \SI{250}{\hbox{\textmu}m} and a thickness of \SI{50}{\hbox{\textmu}m} was formed in lithium vapor at a density of $\SI{8e16}{cm^{-3}}$. This channel was ionized by a Ti:sapphire laser with a $J_7$ Bessel profile (using a 7-step kinoform), which arrived \SI{3}{ps} prior to the arrival of the positron bunch. The bunch had an energy of \SI{20.35}{GeV}, a charge \SI{0.53}{nC}, and a length of \SI{35}{\hbox{\textmu}m} rms. Measurements of the beam size were used to show that the channel center was fully devoid of ionized gas [see Fig.~\ref{fig:history_hollow_2016}(d)]. When the beam propagated through the channel, changes in the energy spectrum of approximately \SI{20}{MeV} were observed [see Fig.~\ref{fig:history_hollow_2016}(e)], implying longitudinal wakefields of approximately \SI{230}{MV/m}---in agreement with PIC simulations.

\begin{figure}[t]
    \centering
    \includegraphics[width=0.93\linewidth]{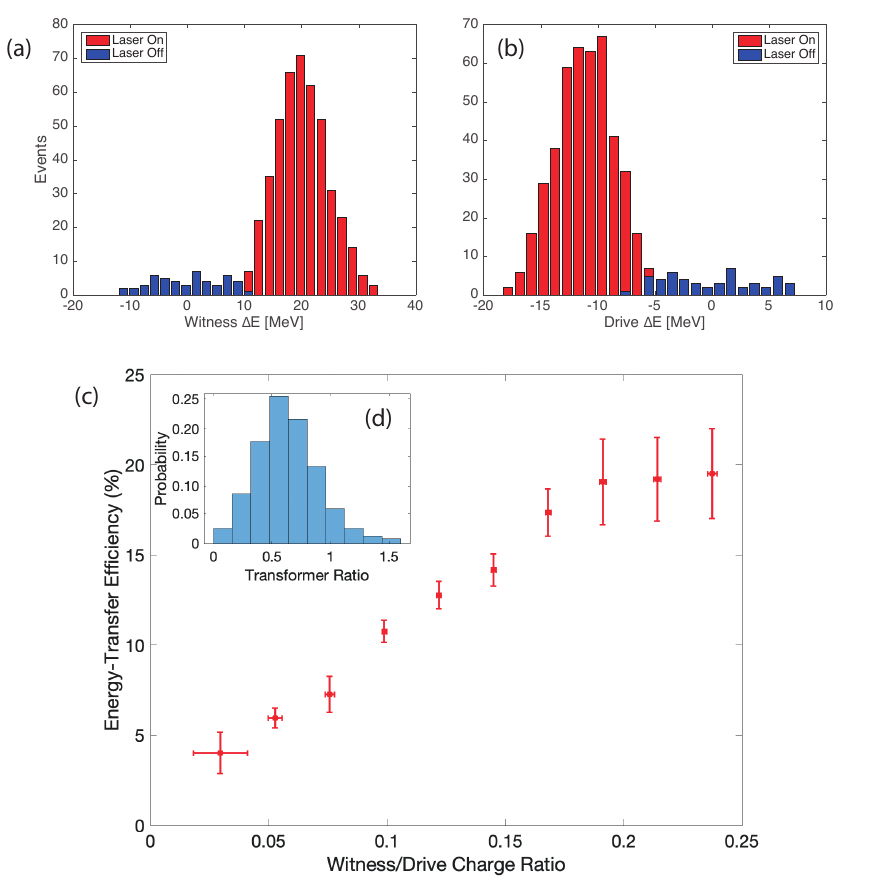}
    \caption{First positron acceleration in a hollow plasma channel. Energy gain of the trailing bunch (a) and energy loss of the drive bunch (b) were measured for 500 shots (red bars), and compared to spectra measured without plasma (blue bars). At a bunch separation of \SI{330}{\hbox{\textmu}m}, the energy-transfer efficiency (c) and transformer ratio (d) were calculated using the centroid energy change of the bunches. Adapted from Refs.~\cite{Gessner:2016bqz} (CC BY 3.0) and \cite{Gessner:2023arxiv} (CC BY 4.0). }
    \label{fig:HC_2016}
\end{figure}

Following these results, the first acceleration of a positron bunch in a hollow plasma channel was demonstrated in 2016 \cite{Gessner:2016uof,gessner_phd}. The experiment was performed using FACET's two-bunch configuration, where a separate positron trailing bunch was accelerated in the wake of a positron drive bunch. A drive bunch of energy \SI{20.5}{GeV} and a trailing bunch of energy \SI{20.1}{GeV}, with a combined charge of \SI{560}{pC}, were sent into the hollow plasma channel. As in the single-bunch experiments, a kinoform was used to create the channel with a similar radius but now \SI{25}{cm} long and with a reduced plasma density of $\SI{3e15}{cm^{-3}}$. A linear wakefield was excited in the channel, accelerating the trailing bunch by approximately \SI{20}{MeV}, corresponding to a peak gradient of around \SI{70}{MeV/m}. The energy-transfer efficiency reached a maximum of 18\% when the drive-to-trailing-bunch charge ratio was approximately 5:1, with a median transformer ratio of 0.55 \cite{Gessner:2023arxiv}. The experimental results are depicted in Fig.~\ref{fig:HC_2016}.

While at this point, hollow-channel positron acceleration appeared to be a working solution, one major problem remained: the transverse instability, as noted by Schroeder \textit{et al.}~\cite{Schroeder:1999cb} in 1999. Misaligned beams propagating off-axis induce a strong transverse wakefield that quickly deflects the beam---a problem that is aggravated by the lack of on-axis focusing fields. The transverse wakefield $W_x$ (per offset $\Delta x$ of the driving particle) at a short distance $z$ behind a particle is fundamentally linked to the longitudinal wakefield $W_z$ through
\begin{equation}
    \label{eq:short-range-wake-theorem}
    \frac{W_x}{\Delta x} = -\frac{\kappa(a,b)}{a^2}\int_0^z W_z dz',
\end{equation}
also known as the \textit{short-range wake theorem} \cite{Lebedev:2017dcs}, where $\kappa(a,b)$ is a numerical coefficient dependent on the inner and outer channel radii $a$ and $b$, respectively \cite{gessner_phd,Lindstrom:2018hhy}. Equation~\ref{eq:short-range-wake-theorem} implies that the transverse wakefield increases with larger offsets $\Delta x$, which leads to a resonance and therefore an instability \cite{Penn:2017uyh}. It also implies that if the hollow-channel radius is reduced to increase the longitudinal wakefield (scaling approximately as $W_z \sim 1/a$), the transverse wakefield increases even more rapidly ($W_x \sim 1/a^3$ assuming the above scaling for $W_z$). The resulting instability inevitably leads to catastrophic beam loss.

\begin{figure}[t]
    \centering
    \includegraphics[width=0.87\linewidth]{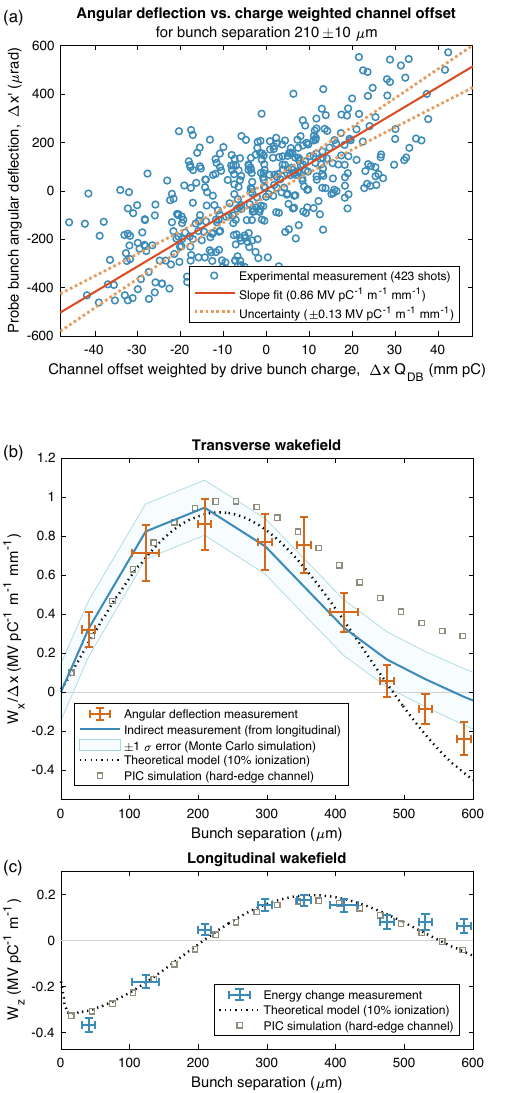}
    \caption{First measurement of the transverse wakefields in a hollow plasma channel. (a) The slope (red line) of the correlation between angular deflection and the channel offset multiplied by the drive-bunch charge (blue points) can be used to calculate the amplitude of the transverse wakefield. (b) This measurement (orange error bars) is compared to an analytical model (black dotted line) and PIC simulations (gray squares). (c) The longitudinal wakefield was also measured and used for an indirect estimate of the transverse wakefield [blue area in (b)]. The bunch separation was measured using electro-optical sampling. From Ref.~\cite{Lindstrom:2018hhy} (CC BY 4.0).}
    \label{fig:HC_2018}
\end{figure}

The effect of a misaligned beam was measured experimentally in the 2016 two-bunch experiments at FACET. By observing the transverse deflection of the trailing bunch when propagating in a misaligned channel, Lindstr{\o}m \textit{et al.}~\cite{Lindstrom:2018hhy} measured the transverse wakefield in a hollow plasma channel. Using the same beam and plasma parameters as reported in Refs.~\cite{Gessner:2016uof,gessner_phd,Gessner:2023arxiv}, the deflection angle of the trailing bunch was measured with the downstream spectrometer screen and correlated with the channel offset [see Fig.~\ref{fig:HC_2018}(a)]. This offset was measured by imaging the transverse profile of the ionizing laser using downstream cameras. The main source of the channel offset was a 30--\SI{40}{\hbox{\textmu}m} rms transverse laser jitter (the positron beam jittered by less than \SI{5}{\hbox{\textmu}m} rms). The wakefield measurement was repeated at various drive-to-trailing bunch separations (50--\SI{600}{\hbox{\textmu}m}), peaking at a separation of \SI{200}{\hbox{\textmu}m}, in good agreement with both analytical models and PIC simulations [see Fig.~\ref{fig:HC_2018}(b)]. A second, indirect estimation of the transverse wakefield was performed by utilizing Eq.~\ref{eq:short-range-wake-theorem} and a measurement of the longitudinal wakefield [see Fig.~\ref{fig:HC_2018}(c)]. Ultimately, these measurements confirmed the presence of the strong transverse wakefields and the tendency for hollow channels to be unstable.

In parallel, several aspects and variations of the hollow-channel concept were studied with numerical simulations. Yi \textit{et al.}~\cite{Yi:2013upa, Yi:2014gta} proposed a proton-driven hollow channel for positron (and proton) acceleration, in order to accelerate by up to \SI{1}{TeV} in a single accelerator stage, using the in-flowing electrons to keep the bunch focused. This was later extended to using trains of proton bunches by Li \textit{et al.}~\cite{Li_2019}. Amorim \textit{et al.}~\cite{Amorim:2016prc} proposed a scheme where an intense, tightly-focused positron drive bunch causes strong ion motion, creating a (near-)hollow channel with wakefields that can both accelerate and focus positrons. Moreover, Golovanov \textit{et al.}~\cite{Golovanov_2017} extended the analytical description of hollow channels from linear \cite{Schroeder:1999cb,gessner_phd} to nonlinear wakefields. Finally, Wu \textit{et al.}~proposed using hollow channels to both dechirp \cite{Wu_2019_hc} and linearize \cite{Wu_2023} electron/positron bunches in longitudinal phase space.

\vspace{1em}

Since the end of the FACET program in 2016, no other plasma-based positron-acceleration experiments have been performed. The possibility for FACET-II~\cite{facet_ii}---the next-generation facility that started operation in 2021---to deliver positrons is presently unclear. In LWFA, positron-acceleration experiments have not yet been performed, partly due to the challenges of sourcing and injecting positron beams. That said, experiments have demonstrated generation of intense ultra-relativistic positron beams using high-energy electrons from LWFAs \cite{lwfa_gahn,lwfa_Chen,lwfa_Chen2,lwfa_Sarri}, which may be used for future positron experiments.

In summary, plasma-based positron acceleration was successfully demonstrated in experiments, showing high gradients ($>$GV/m), high gains ($>$GeV), and high efficiency ($\sim$40\%). However, while reasonable charge was accelerated (100--\SI{200}{pC}), the energy spread per energy gain (i.e., the field uniformity) was too high ($\sim$10\%), as was the final emittance---where measured, the emittance growth was large or the propagation unstable. Nevertheless, while sufficient beam quality has not yet been demonstrated, the experiments and theoretical investigations have inspired a wave of new schemes (Sec.~\ref{sec:schemes}).

\section{Proposed schemes}
\label{sec:schemes}

The minimum objective for any plasma-based positron accelerator is to create a region that is both accelerating and focusing. Firstly, to generate a longitudinal field that accelerates positrons, there must be a net radial current of plasma electrons outwards within each longitudinal slice of the bunch. Secondly, to generate a transverse field that focuses positrons, there must be a net negative charge density (i.e., surplus plasma electrons) locally within the bunch.

The proposed schemes, summarized below, all fulfill the minimum objective. These either optimize or modify existing schemes tested in experiments (i.e., homogeneous plasmas and hollow plasma channels; see Sec.~\ref{sec:history}); modifying the driver, the plasma or both. Here we discuss the principles behind each scheme, their advantages and limitations, as well as example parameters from PIC simulations. An important difference between previous experiments and the proposed schemes is that the former were all positron-driven, whereas the latter are all electron-driven. 

Note that only schemes with externally injected, relativistic positron bunches are reviewed, which excludes several proposed schemes for generating and injecting positron bunches \cite{wang_2006,wang_2008,wang_2009,Firouzjaei:2017kfh,Sahai:2018ksu,Xu:2019zov,Fujii:2019qxb,Martinez:2022idh,Liu_2022b,Amorim:2023bof,Sugimoto_2023}. Positron sources are not discussed in this section, but recent research shows potential for a novel positron source that offers small emittance~\cite{hessami2023}.



\subsection{Homogeneous plasmas with Gaussian beams}
\label{sec:hom_gaussian}

In homogeneous plasmas, two regimes are considered: (1) the quasi-linear regime, where fields are strong but several advantages of the linear regime are retained; and (2) the nonlinear regime, which can support even stronger fields and higher bunch charges. While these regimes have all been studied extensively both in theory and in experiments, recent numerical work has focused on finding optimal parameters. 

\subsubsection{Quasi-linear regime}
\label{scheme:qlin}

\begin{figure}[b]
    \includegraphics[width=0.95\linewidth]{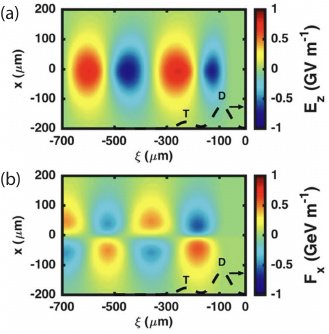}
    \caption{(a) Accelerating field and (b) transverse focusing force in a quasi-linear plasma wakefield excited by a \SI{20.55}{GeV} positron driver (labelled ``D") with emittance $270 \times \SI{60}{\milli\m\milli\radian}$ and beam density $n_b/n_0=0.6$. Here, the plasma density is $n_0 = \SI{e16}{cm^{-3}}$. A \SI{20.05}{GeV} trailing positron bunch (labelled ``T") with the same emittance and a charge of $\sim$\SI{130}{pC} reached a peak beam density of $n_b/n_0=0.15$. From Ref.~\cite{Doche:2017jhd} (CC BY 4.0).}
    \label{fig:linear}
\end{figure}

The linear regime characterizes plasma-density perturbations $\delta n$ that are small compared to the ambient plasma density $n_0$. In this case, the first-order or the linear perturbation term, $\delta n/n_0$, dominates over higher-order terms, which means the plasma-electron motion can be described by the wave equation
\begin{equation}
    \label{eq:linear_theory}
    \left(\frac{\partial^2}{\partial t^2} + \omega_p^2\right)\frac{\delta n}{n_0} = S_{\textmd{driver}},
\end{equation}
where $\omega_p$ is the plasma frequency and $S_{\textmd{driver}}$ is a driver source term, either from a particle beam ($S_\textmd{beam}=-\omega_p^2n_b/n_0$ where $n_b$ is the peak beam density) or a laser beam ($S_\textmd{laser}=\frac{1}{2} c^2 \nabla^2 a^2$ where $a^2$ is the normalized laser intensity) \cite{Esarey:1996}. The solution to Eq.~\ref{eq:linear_theory} is a density perturbation that is zero in front of the driver and sinusoidal behind it. The resulting electric fields, as described by Keinigs \textit{et al.}~\cite{Keinigs:1986sk}, are also sinusoidal in the longitudinal direction, but evanescent in the transverse (see Fig.~\ref{fig:linear}). In the linear regime, the response of the plasma electrons to positrons and electrons is completely symmetric, with a phase difference of \SI{180}{\degree}. This symmetry breaks when the density perturbation approaches the ambient density ($\delta n/n_0 \approx 1$)---often known as the quasi-linear regime.

High efficiency can be achieved with beam loading, which in the linear regime is simply a destructive interference between the fields of the driver and trailing bunch. Efficient beam loading therefore requires the charge of the trailing bunches to be approximately that of the driver. For wide drivers (with beam size $\sigma_r > k_p^{-1}$) the field extends to the size of the beam, which means that the trailing bunch must have matching beam size to extract the wakefield energy \cite{hue_2021}. On the other hand, if the driver is narrow ($\sigma_r < k_p^{-1}$), the fields extend transversely over a characteristic range $k_p^{-1}$ regardless of beam size---it is therefore possible to beam load with potentially very narrow trailing bunches ($\sigma_r \ll k_p^{-1}$). Low correlated energy spread is possible in the linear regime through beam loading with a tailored current profile \cite{Katsouleas:1987yd}, whereas low uncorrelated energy spread can be achieved by using narrow bunches, for which all particles experience only the near-uniform fields close to the axis. For laser drivers, the transverse extent of the wakefields is directly determined by the laser intensity profile. This means the exact shape of the wakefields can be controlled using higher-order laser modes \cite{cormier2011}, potentially enabling acceleration of higher positron charge with both low correlated and uncorrelated energy spread through beam loading. 

Low emittance also implies narrow bunches. As an example, assume collider-level emittances --- \SI{0.5}{\milli\m\milli\radian} averaged over both transverse planes, and beams with energies of order \SI{10}{GeV} in a quasi-linear plasma wave at a density of \SI{1e16}{cm^{-3}}: the transverse beam size will be approximately $\sigma_r \approx 0.01 k_p^{-1}$~\cite{hue_2021}. In principle, emittance will be preserved in the quasi-linear regime, as the focusing fields are linear close to the axis. However, when combined with the charge required for high-efficiency beam loading, the beam density quickly exceeds the plasma density. This means the trailing bunches do not operate in the quasi-linear, but instead in the nonlinear regime. High beam density results in an on-axis electron spike which can make it challenging to preserve beam emittance. A quantitative description of this beam-density limit is given in Sec.~\ref{sec:positron-problem}. Additionally, it is worth noting that a quasi-linear wake may evolve due to head erosion of the driver \cite{huephd_2020}.

Positron acceleration experiments in the quasi-linear regime by Doche \textit{et al.}~\cite{Doche:2017jhd} (described in Sec.~\ref{sec:homogeneous-plasma}) demonstrated gradients of order \SI{1}{GV/m}, charge of around \SI{100}{pC}, energy efficiency of around 40\%, energy spread (per gain; i.e., field uniformity) of order 10\%, for emittances of around $270 \times \SI{60}{\milli\m\milli\radian}$. Here, the emittance was not shown to be preserved. In simulations by Hue \textit{et al.}~\cite{hue_2021}, optimized for emittance and uncorrelated energy-spread in the quasi-linear regime, an emittance as low as $\SI{0.64}{\milli\m\milli\radian}$ and a charge of $\SI{4}{pC}$ can be accelerated with a high gradient of \SI{1.3}{GV/m}, a high efficiency of 30\% and good quality (emittance preserved and less than 1\% uncorrelated energy spread). In this study, the correlated energy spread was not fully preserved, but assumed removed by dechirping prior to collisions.

In summary, the quasi-linear regime can almost deliver collider relevant positron beams with low emittance and low energy spread at high gradient and efficiency, falling short only with respect to accelerated charge---this is a few orders of magnitude lower than required for colliders. The charge can in principle be increased, but to maintain the same beam density (to stay in the quasi-linear regime), the emittance must increase proportionally.

\subsubsection{Nonlinear regime}
\label{scheme:nonlin}

The nonlinear regime is defined by density perturbations of order the plasma density or higher. These waves are driven by either particle beams with densities exceeding the plasma density ($n_b/n_0 \geq 1$) or lasers with normalized intensity exceeding unity ($a^2 \geq 1$). The induced transverse plasma-electron motion is significant in this regime, which results in bubble formation and a more compact region both transversely and longitudinally, in which positrons can be both accelerated and focused---here, the accelerating gradient is higher and the focusing forces larger compared to the linear regime. 

The total volume of the favourable region for positrons depends on the strength of the nonlinearity, which is often quantified by the \textit{normalized driver charge} \cite{Rosenzweig_2004,Barov_2004}: 
\begin{equation}
    \label{eq:normalized-charge}
    \tilde{Q}=\frac{k_p^3 N}{n_0} = 4 \pi r_e k_p N,
\end{equation}
where $N$ is the bunch population and $r_e$ is the classical electron radius. In the \textit{strongly nonlinear} regime \cite{Rosenzweig:1991yx} ($\tilde{Q} \gg 1$) the volume is negligible, whereas in the \textit{weakly nonlinear} regime \cite{Lu_2005} ($\tilde{Q} \approx 1$; also known as \textit{moderately nonlinear} \cite{hue_2021,Cao:2022zkb} or \textit{quasi-nonlinear} regime \cite{Muggli_2012,LONDRILLO2014236,Marocchino:2016tfe}) the volume is small but non-negligible. Lotov \cite{Lotov_2007} and An \textit{et al.}~\cite{An:2010exa} found that $\tilde{Q} \approx 1$--10 (or equivalent for lasers \cite{Liu_2022a}) provides the optimum conditions for positron acceleration.

\begin{figure}[b]
    \includegraphics[width=\linewidth]{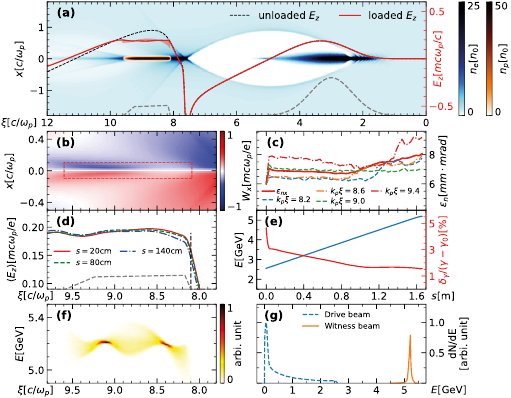}
    \caption{Positron acceleration in nonlinear wakefields, driven by an electron bunch of charge \SI{534}{pC} ($\tilde{Q}=2$, $n_b/n_0=27$), length $\sigma_z=\SI{40}{\micro\m}$ and beam size $\sigma_r=\SI{5}{\micro\m}$. The plasma density is \SI{7.8e15}{cm^{-3}}. (a) Density map of the plasma, electrons and positrons (beam current shown as dashed gray lines), and the on-axis longitudinal field (red line). (b) Transverse force seen by the positrons (dashed red box). (c) Emittance evolution of the positrons, both slice (dashed) and projected (solid red). (d) The accelerating field, shown at several propagation distances, is largely stable. (e) The energy (blue line) and relative energy spread (red line) of the positron bunch throughout the propagation. (f) Final longitudinal phase space of the positron bunch, indicating a low correlated energy spread. (g) Final energy spectra of the two bunches, indicating high driver energy depletion. From Ref.~\cite{zhou2022positron}.}
    \label{fig:nonlin_flattop} 
\end{figure}

In the presence of a high-density positron bunch, beam loading occurs [see Fig.~\ref{fig:nonlin_flattop}(a)]: the energy is extracted from the accelerating field (longitudinal beam loading), and moreover, the length of the focusing region is extended (transverse beam loading) because the positrons keep the electron close to the axis---an effect often known as self-loading \cite{Corde:2015zxa} (also seen in Fig.~\ref{fig:blowout}). Ultimately, this effect allows high-efficiency acceleration of higher-charge bunches (similar to the drive-bunch charge). However, as shown in Fig.~\ref{fig:nonlin_flattop}(b), this self-loading leads to nonlinear transverse fields and consequently higher emittances for Gaussian bunches. Therefore, while its focusing fields are different from those in the quasi-linear regime, the nonlinear regime shares the same trade-off: higher charge and efficiency implies higher emittance and energy spread because of nonlinear fields.

As an example, an optimized simulation, shown in Fig.~\ref{fig:nonlin_flattop}, indicate that an emittance of approximately $\SI{8}{\milli\m\milli\radian}$ can be preserved when accelerating $\SI{102}{pC}$ positron bunches at an energy efficiency of 26\% and an accelerating gradient of \SI{1.6}{GV/m} \cite{zhou2022positron}. Here, emittance preservation in longitudinally non-uniform focusing fields was achieved using slice-by-slice matching. The energy spread was then dominated by the uncorrelated energy spread, at a level of 2.39\% [see Fig.~\ref{fig:nonlin_flattop}(f)]. Another example used a bunch of charge $\SI{23}{pC}$ and an emittance of approximately \SI{1}{\milli\m\milli\radian}, resulting in an efficiency of 40\% in an accelerating gradient of \SI{4.8}{GV/m} \cite{hue_2021}. In this example, the uncorrelated energy spread was 1\% rms, but the projected energy spread was larger (at the level of 10\% rms). 

A limitation of this regime is the degree to which the field structure can depend on the exact driver-beam density---in particular in the transition between the weakly and strongly nonlinear regime. This implies strict tolerances on driver parameters as well as the need to avoid significant evolution of the driver during propagation, which may have implications on the energy-depletion efficiency of the driver [see Fig.~\ref{fig:nonlin_flattop}(g)]. 

A variation on this scheme utilizes the nonlinear wakefield for simultaneous acceleration of electrons. This can increase the overall efficiency---simulations indicate at least 10\% higher efficiency---by sharing the energy extraction between the trailing electron and positron bunches~\cite{zhou2022positron}. Alternatively, Wang \textit{et al.}~\cite{wang_2021} argue that highly efficient electron beam loading in the strongly nonlinear regime can also create an elongated on-axis plasma-electron filament, which can be used for positron acceleration, and that the width of this filament can be controlled through the plasma temperature.

This regime can potentially be tested experimentally also without a conventional positron source, by using an injection scheme proposed by Wang \textit{et al.}~\cite{wang_2006,wang_2008,wang_2009}. Here, the electron driver passes through a foil prior to the plasma, in which electron--positron pairs are produced. Further studies have shown that while the trapping conditions are excellent, the trapping efficiency decreases as the gap between the foil and plasma wakefield increases. Such a gap may be necessary for practical experiments~\cite{Fujii:2019qxb}.

In summary, the nonlinear regime can offer either higher acceleration gradient or higher charge compared to the quasi-linear regime, but also remains limited by the trade-off between charge (and thereby efficiency) and beam quality (i.e., emittance and energy spread).

\subsection{Modified drivers: Donut-shaped laser or electron beams}
\label{scheme:donut}

\begin{figure}[b]
    \includegraphics[width=0.99\linewidth]{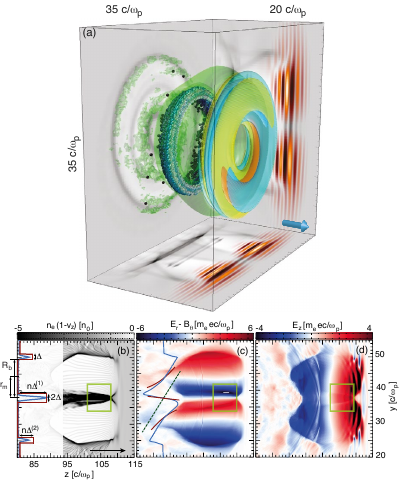}
    \caption{(a) Simulated donut wakefields driven by a Laguerre-Gaussian laser propagating in the direction of the arrow. (b) The plasma-electron density is perturbed by the laser. Here, $r_m$ is the radius of peak laser intensity, $R_b$ the diameter of the blowout, and $\Delta$ is the approximate width of the electron sheath and radius of the electron filament. (c) The transverse fields are focusing for positrons on axis. A lineout at $z=105c/\omega_p$ (blue line) is compared to a theoretical model (red line). (d) The corresponding longitudinal field is accelerating for positrons in the front half of the blowout structure. The green box in (b--d) represents this accelerating and focusing region. Adapted from Ref.~\cite{vieira_2014}.}
    \label{fig:dd} 
\end{figure}

It is possible to create an accelerating and focusing region for positrons even in the strongly nonlinear regime, by modifying the transverse shape of the driver. Vieira and Mendon\c{c}a~\cite{vieira_2014} proposed to use donut-shaped (or Laguerre-Gaussian \cite{mendonca_2014,Wang_2020}) laser pulses for this purpose, as illustrated in Fig.~\ref{fig:dd}. Similar wakefields can also be excited by a donut-shaped electron bunch \cite{jain_2015,vieira_2016} or overlapping but non-neutral electron and positron bunches \cite{Silva_2023}. Such drivers create an electron filament that can focus positrons by guiding plasma electrons through the hollow core of the driver and onto the axis. Nonlinear accelerating wakefields are created by expelling plasma electrons outwards, much like in a regular blowout.

The main advantage of this scheme is that the on-axis electron filament exists also when the fields are strongly nonlinear, independently of whether the fields are beam loaded. This allows the accelerating field to be significantly higher, as well as more charge to be accelerated, compared to that achieved in the quasi-linear and weakly nonlinear regime with Gaussian drivers (see Sec.~\ref{sec:hom_gaussian}). Another advantage is the additional degrees of freedom available for tailoring the electron filament---as an example, Yu \textit{et al.}~\cite{Yu_2014} showed that the strength and shape of the on-axis focusing can be controlled by varying the relative intensity of higher-order laser modes.


In principle, the donut wakefield allows acceleration of very low emittance and low energy-spread bunches, due to the linear focusing fields and uniform accelerating fields close to the axis [see Fig.~\ref{fig:dd}(c)]. Jain \textit{et al.}~\cite{jain_2015} found that for a donut-shaped electron driver, emittances as low as $\approx\SI{0.04}{\milli\m\milli\radian}$ and energy spreads less than 0.4\% rms can be preserved. Here, the accelerating gradient was \SI{8.9}{GV/m}. However, the accelerated bunch had a charge of only \SI{14}{pC}, whereas the driver had \SI{5.2}{nC}, leading to very low energy efficiency (approx.~0.17\%). If the wakefield is beam loaded more strongly with a higher-charge positron bunch, the beam density is consequently higher, which alters the shape of the transverse focusing fields---exactly the same problem encountered in the weakly nonlinear regime (see Sec.~\ref{scheme:nonlin}). The resulting trade-off between efficiency and beam quality is discussed in Sec.~\ref{sec:positron-problem}.

Another potential issue is that any evolution of the driver may impact the shape of the wakefields. While donut-shaped lasers can maintain their approximate shape, and thereby a region that focuses and accelerates positrons, until energy is nearly depleted (several hundred plasma wavelengths)~\cite{vieira_2014}, the exact shape of the wakefield will evolve---not optimal for beam-quality preservation. However, Pathak \textit{et al.}~\cite{Pathak_2016} note that a filamentation instability can occur, leading to transverse breakup of the laser pulse; this instability can be suppressed by using a parabolic plasma-density profile. 

Unlike Laguerre-Gaussian laser beams, which have non-zero angular momentum, donut-shaped electron bunches typically do not. As a result, these bunches can collapse onto the axis, forming a regular blowout wake. Jain \cite{Jain_2019} found that this collapse happens for electron drivers with a donut radius smaller than approximately $1.8 k_p^{-1}$, but these drivers can propagate stably for larger donut radii. Moreover, use of non-zero-divergence electron drivers can lead to head erosion~\cite{zhou_2007,blumenfeld_phd}. Finally, similar to the filamentation instability observed for laser pulses, the azimuthal Weibel instability may also be a problem, causing filamentation of the electron driver~\cite{Su_1987}.

A semi-optimized parameter set, balancing beam quality and efficiency, was obtained by Hue \textit{et al.}~\cite{hue_2021}: a positron bunch of charge \SI{189}{pC} and equilibrium emittance of $\SI{1.5}{\milli\m\milli\radian}$ can be accelerated with 3.5\% efficiency at a gradient of approximately \SI{40}{GV/m}. A conclusion from these simulations was that the donut width should be optimized (here, $\sigma_r\approx0.4k_p^{-1}$), as this leads to more transversely uniform accelerating fields. Note that these (single-step) simulations were performed with electron drivers with a donut radius of $1k_p^{-1}$, which may therefore have suffered an on-axis collapse.


No positron experiments have yet operated with this scheme. However, donut-shaped laser pulses can be generated \cite{Allen_1992,Wang_2020} and have been interacted with plasmas: an experiment by Nakanii \textit{et al.}~\cite{Nakanii_2016} showed no acceleration of electrons, but a strong decomposition of laser modes (i.e., the filamentation instability). Donut-shaped electron bunches can also be produced~\cite{bubley_2002,Bubley:2006ia}, but have not yet been injected into plasma.


In summary, donut-shaped drivers can provide higher fields and higher charge than Gaussian drivers, but are ultimately limited by the same efficiency--quality trade-off. Additionally, propagation of these drivers can be very unstable.

\subsection{Modified plasmas: Inhomogeneous channels}
\label{cat3}

As an alternative to modifying the profile of the driver, a region for positron acceleration and focusing can also be created in the nonlinear regime by modifying the profile of the plasma. Three such schemes have been proposed, all electron-driven and in the nonlinear regime: (1) a finite-radius plasma channel; (2) a two-column plasma channel with additional ionization from a co-propagating laser; and (3) a thin hollow channel filled with hot plasma electrons, created through ion motion.

\subsubsection{Finite-radius plasma channel}
\label{scheme:fr}

\begin{figure}[t]
    \includegraphics[width=0.9\linewidth]{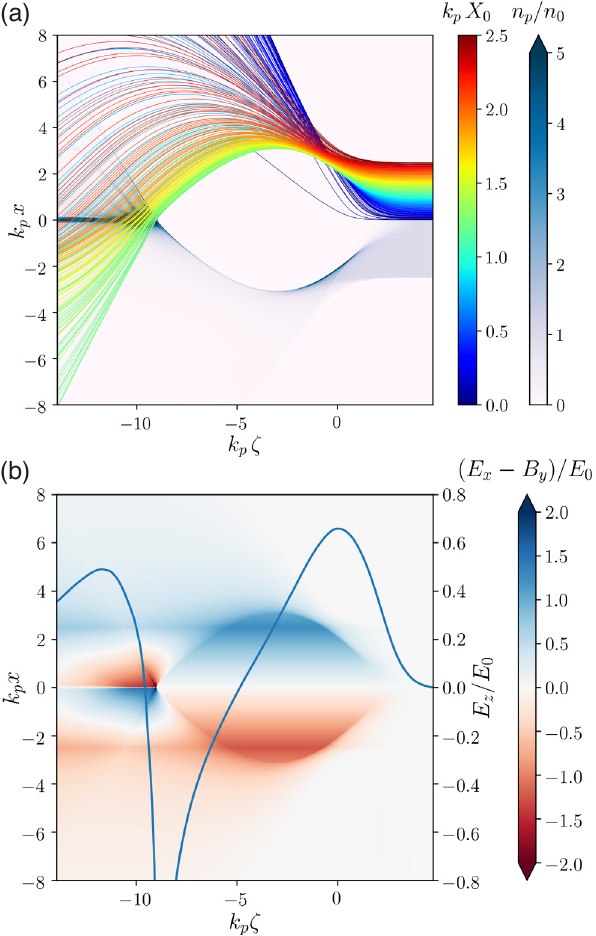}
    \caption{Simulation of a finite-radius plasma channel, showing (a) plasma-electron trajectories (colored based on initial radius $X_0$) and density (blue color map) of a plasma channel of radius $k_p R_p = 2.5$, driven by a high-charge ($\tilde{Q}\approx44)$ electron bunch with beam size $k_p\sigma_r=0.3$ and bunch length $k_p\sigma_{\xi}=\sqrt{2}$; (b) the corresponding transverse (red--blue color map) and on-axis longitudinal (blue line) wakefields. From Ref.~\cite{Diederichs:2019wnl} (CC BY 4.0).}
    \label{fig:finite-radius-setup} 
\end{figure}

The singularity-like electron spike seen in the strongly nonlinear regime, which creates a volume too small to accelerate and focus positrons, is a consequence of the highly coherent motion of plasma electrons. One way to distribute longitudinally where the electrons cross the axis is to use a finite-radius channel, proposed by Diederichs \textit{et al.}~\cite{Diederichs:2019wnl}. Here, the plasma-column radius must be smaller than the maximum blowout radius ($R_p \lesssim R_b$). This results in electrons outside the channel experiencing a nonlinear focusing force from the plasma ions, which leads to decoherence of the electron motion, as illustrated in Fig.~\ref{fig:finite-radius-setup}(a). Collectively, these electrons form an elongated filament on axis: an extended region of accelerating and focusing fields [as shown in Fig.~\ref{fig:finite-radius-setup}(b)].

\begin{figure}[t]
    \includegraphics[width=\linewidth]{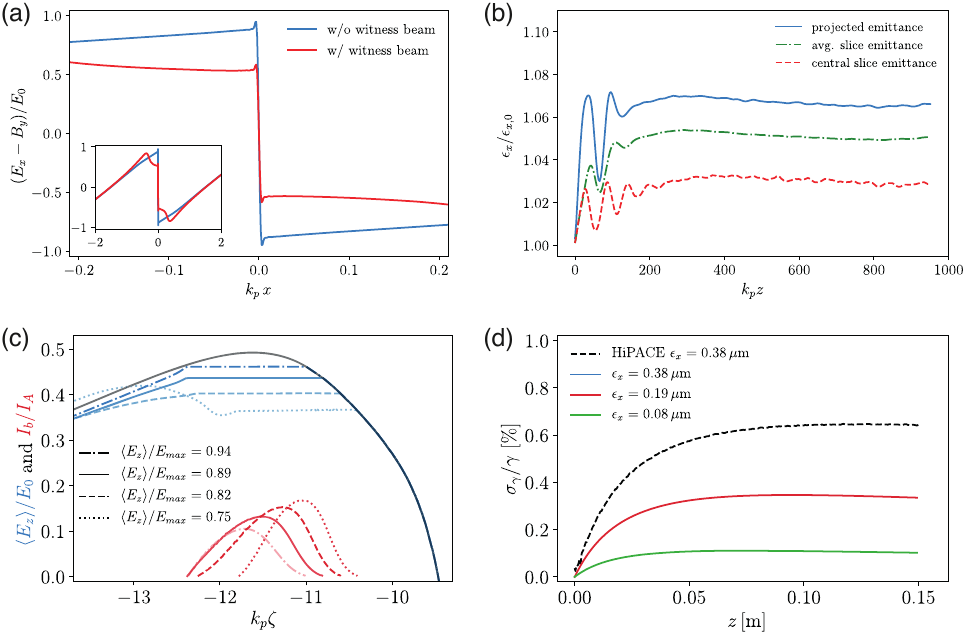}
    \caption{Beam quality in the finite-radius scheme, showing: (a) the step-like transverse focusing fields at $k_p\xi=-11.6$ (zoomed-out in the inset); (b) emittance evolution for a beam with an initial normalized emittance of $k_p\epsilon_x=0.1$; (c) beam-loading optimisation and the resulting bunch-current profiles; and (d) the uncorrelated energy-spread evolution for beams with different emittances. Adapted from Refs.~\cite{Diederichs:2019wnl} and \cite{Diederichs:2020hri} (CC BY 4.0)}.
    \label{fig:finite-radius-quality} 
\end{figure}

Although the plasma electrons are spread out longitudinally, they are still highly localized transversely. A thin on-axis filament forms, resulting in what appears as a step-like focusing field close to the axis: $(E_r - c B_\phi)/E_0 \approx - \alpha\:\textmd{sgn}(r)$~\cite{Diederichs:2019wnl}, where $\alpha$ is the normalized amplitude of the transverse field close to the axis, $\textmd{sgn}(r)$ is the sign function and $r$ is the radial position---this field is shown in Fig.~\ref{fig:finite-radius-quality}(a). Since this focusing field is nonlinear, the emittance will not be fully preserved for a Gaussian transverse profile. However, by quasi-matching (i.e., approximate matching, leading to emittance growth of a few \%) to the non-Gaussian equilibrium profile \cite{lotov_2017}, which has a beam size given by
\begin{equation}
    \sigma_r^3 \approx 1.72\frac{\epsilon_r^2}{k_p\alpha\gamma},
\end{equation}
where $\gamma$ is the Lorentz factor of the beam, the emittance stays approximately preserved to within a few percent, as shown in Fig.~\ref{fig:finite-radius-quality}(b). This quasi-matching condition assumes a constant $\alpha$ (i.e., a radius-independent focusing field), which sets an upper limit on the beam emittance. A unique feature of this scheme is that even when beam loaded, the transverse fields maintain their step-like shape (with a decreased amplitude $\alpha$), which implies that emittance can still be preserved.

Energy spread can be minimized by using an optimized current profile \cite{Diederichs:2020hri}, as shown in Fig.~\ref{fig:finite-radius-quality}(c), producing near-zero correlated energy spreads. However, the uncorrelated energy spread will depend on the emittance [see Fig.~\ref{fig:finite-radius-quality}(d)], as the beam samples the transversely non-uniform accelerating field. Keeping the uncorrelated energy spread below 1\%, an optimized parameter set provides a positron bunch with charge \SI{52}{pC}, an emittance of $\SI{0.38}{\milli\m\milli\radian}$ and projected energy spread of 0.7\% rms (corresponding to 0.86\% per gain), accelerated in a field of gradient \SI{30}{GV/m} with an energy efficiency of 3\% at a plasma density of \SI{5e17}{cm^{-3}}. Another parameter set \cite{Diederichs:2019wnl}, not optimized for energy spread, has an emittance of $\SI{0.75}{\milli\m\milli\radian}$, a charge of \SI{84}{pC} and energy efficiency of 4.8\% (but with a few-percent-level energy spread). This parameter set was recently combined with a warm plasma (\SI{50}{eV})~\cite{diederichs2023_temp2}, which linearizes the focusing field, allowing preservation of emittances as low as $0.002 k_p^{-1}$ (i.e., $\SI{0.015}{\milli\m\milli\radian}$ at \SI{5e17}{cm^{-3}}).

It is unclear if high efficiency is attainable in this scheme. Electrons in a bunch repel inward-moving plasma electrons, slowing down their inward motion and thereby extracting energy from them. Conversely, positrons in a bunch attract outward-moving plasma electrons, extracting energy by slowing down their outward motion. However, in the accelerating and focusing region for positrons in a finite-radius plasma channel, there are overlapping populations (or annuli) of inward- and outward-moving plasma electrons at each longitudinal position. This incoherent motion reduces the field energy available to the positrons, and therefore the energy efficiency. 

Both the driver and the trailing bunch propagate stably in the channel. Transverse instability of the driver can be suppressed by ion motion and energy spread \cite{Diederichs:2022pjj}, whereas longitudinally non-uniform and transversely nonlinear focusing fields suppress instabilities of the trailing bunch \cite{Diederichs:2022yfd}.

Finite-radius channels can be experimentally realized using for instance laser ionization with axicons \cite{Green_2014} or by beam ionization \cite{OConnell_2006}. Plans exist to demonstrate this scheme at FACET-II \cite{facet_ii} (the E-333 experiment~\cite{slac_2022}), assuming positrons and electrons can be delivered simultaneously. 

In summary, the finite-radius scheme can support very high quality and acceleration gradients, but likely not high efficiency.

\subsubsection{Laser-augmented blowout in a two-column plasma}
\label{scheme:lab}

\begin{figure}[t]
    \includegraphics[width=0.95\linewidth]{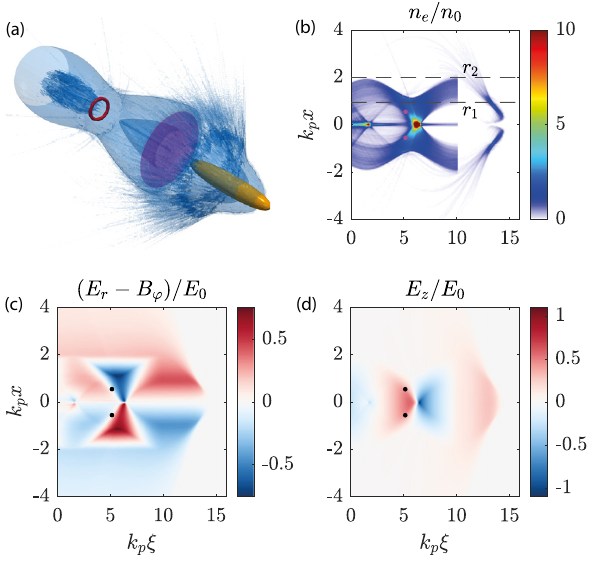}
    \caption{(a) In the laser-augmented blowout scheme, two plasma columns are made: a Gaussian electron bunch (yellow) beam ionizes a thin column, and a trailing laser pulse (purple) ionizes a wider column. The trailing positron bunch (red) is donut or ring-shaped such that the entire bunch is inside the blowout sheath at the beginning of the second bubble, also shown in (b). Both the (c) focusing force and (d) accelerating field are shown, indicating the location of the positron bunch (black dots). Adapted from Ref.~\cite{reichwein_2022} (CC BY 4.0).}
    \label{fig:lab} 
\end{figure}

The laser-augmented scheme, proposed by Reichwein \textit{et al.}~\cite{reichwein_2022}, uses an alternative geometry to the on-axis plasma-electron filament normally used for positron focusing. Using a combination of beam ionization and a trailing laser pulse for additional ionization of a wider channel, the singularity-like electron-density spike behind a blowout is widened transversely (as opposed to longitudinally as in the finite-radius scheme). This enables focusing and acceleration of a \textit{ring-shaped} positron bunch, as illustrated in Fig.~\ref{fig:lab}. 

This scheme is unique in its use of the blowout sheath for focusing, but this also increases the transverse size, and thereby the emittance, of the positron bunch. Simulations show that for a bunch of charge \SI{15}{pC}, the emittance saturates at about $\SI{31}{\milli\m\milli\radian}$ and the energy spread at 1.7\% (3.4\% per gain), while accelerating in a field of gradient \SI{20}{GV/m} with an efficiency of approximately 5.5\%. The charge and efficiency can potentially be increased by using a cone-shaped beam, matching the shape of the electron sheath at the head of the second bubble, although this may be challenging to realize in experiments.

\subsubsection{Thin, warm, hollow plasma channel}
\label{scheme:twhc}

\begin{figure}[t]
    \includegraphics[width=0.97\linewidth]{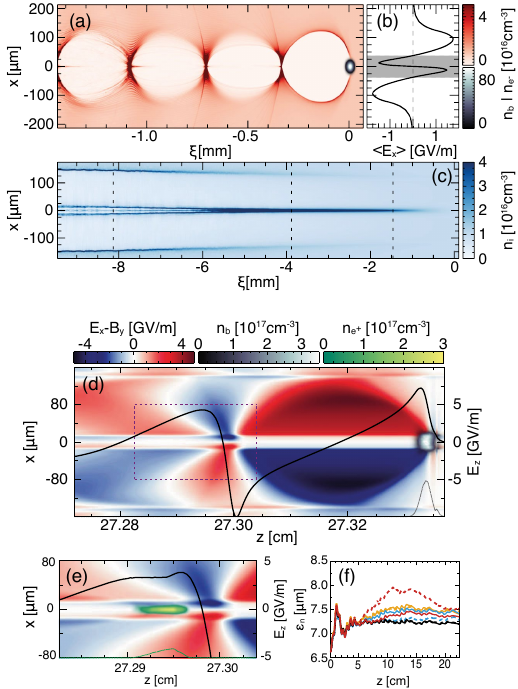}
    \caption{Simulated thin, warm hollow channel development, showing: (a) early-stage plasma density perturbation, no hollow channel has been formed at this time; (b) the average transverse fields, focusing for the ions (inside the gray box); and (c) the start of on-axis ion accumulation and hollow channel formation. A second electron beam creates (d) the transverse wakefields and the on-axis longitudinal field (black line). The dashed box corresponds to the region around where the positron bunch can be accelerated. (e) If beam loaded with an optimized current profile, the field is partly flattened. (f) The emittance evolution of the positron bunch shows only marginal emittance growth. Adapted from Ref.~\cite{silva_2021}.}
    \label{fig:twhc} 
\end{figure}

Hollow-channel acceleration has two main challenges: beam-breakup instability from transverse wakefields \cite{Schroeder:1999cb,Lindstrom:2018hhy}, and unwanted ionization of gas on axis \cite{Kimura:2011zz}. Ion motion \cite{Rosenzweig_2005} can be used to create a truly hollow plasma channel from a homogeneous plasma. As discussed in Sec.~\ref{sec:hollow-plasma-channels}, Amorim \textit{et al.}~\cite{Amorim:2016prc} proposed to use an intense positron bunch to create such a channel and use it for positron acceleration with nonlinear fields. Making this concept more experimentally viable, Silva \textit{et al.}~\cite{silva_2021} proposed using two intense electron bunches---one to generate the hollow channel and the other to create accelerating and focusing fields for positrons.

As illustrated in Fig.~\ref{fig:twhc}(a--c), the formation of the channel starts with the first drive beam creating a nonlinear blowout, where the longitudinally averaged transverse fields focus the ions. After initially accumulating on axis, the ions diverge and create a thin, hollow structure around the axis. At this point, the wakefield has decayed, leaving plasma electrons with high temperature (around 2--\SI{9}{keV} in the example). The second drive beam then creates the wakefields used in positron acceleration. The combination of high plasma-electron temperature, which spreads out the inward-moving sheath electrons, and the thin hollow channel, which traps them on axis, results in an extended region with positron focusing and acceleration, as shown in Fig.~\ref{fig:twhc}(d). The transverse nonlinearity and longitudinal non-uniformity of this focusing field suppresses the beam-breakup instability.

Beam loading of the accelerating field with a Gaussian bunch produces a non-negligible energy spread ($\sim$10\% rms in this example). Similarly, the emittance cannot be fully preserved for a Gaussian transverse profile without some loss of charge. However, by tailoring the positron beam-current profile and using a flat-top transverse profile, the energy spread saturates around 4\% rms (6\% per gain) and the emittance is approximately preserved at $\sim$\SI{7.4}{\milli\m\milli\radian} (10\% growth before saturation) [see Fig~\ref{fig:twhc}(e--f)]. For a charge of $\SI{100}{pC}$ accelerating in a $\SI{3.5}{GV/m}$ field, the energy-transfer efficiency from the second driver to the a positron bunch is approximately 4.7\%.

A clear advantage of this scheme is its experimentally realizable method of creating a hollow channel with stable positron acceleration. While the dynamics may be somewhat complex, the implementation is not: it only requires two electron drivers and a homogeneous plasma. Plans exist to demonstrate this scheme at FACET-II---the E-337 experiment \cite{slac_2022}. On the other hand, a disadvantage is that the overall efficiency of the scheme is limited by the need for two drivers, since the energy in the first driver (generating the hollow channel) cannot be extracted by the positron bunch. 

In summary, the thin, warm, hollow plasma channel scheme offers a simple experimental setup, but may suffer from low energy efficiency.

\subsection{Modified drivers and plasmas: Hollow channels with asymmetric drivers}
\label{cat3}

\begin{figure*}[t]
    \includegraphics[width=0.75\textwidth]{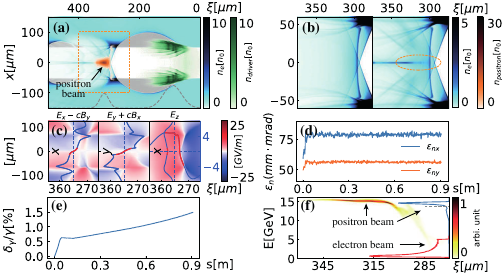}
    \caption{Simulated hollow channel with an asymmetric driver, showing: (a) plasma-density perturbation (blue color map) by a \SI{2}{nC} electron drive bunch (green color map) that has reached two-beamlet equilibrium profile, loaded by a \SI{640}{pC} positron bunch (orange color map); (b) plasma-electron density without and with beam loading in the region around the positron bunch [dashed orange box in (a)]; (c) the loaded wakefields and lineouts (blue lines), indicating $\pm2\sigma$ of the positron bunch (red lines); (d) emittance evolution in both $x$ and $y$ planes; (e) energy spread evolution for positrons at $\xi > \SI{305}{\micro\m}$; (f) spectra of the driver (red line) and the positron bunch (blue line) as well as the longitudinal phase space of the positron bunch (colorbar). From Ref.~\cite{zhou_2021}.}
    \label{fig:hcad} 
\end{figure*}

All the schemes above modify either the driver shape or the plasma channel to create a region of excess plasma electrons for positron focusing. This region is not present in a hollow channel, which leads to instabilities for both the driver and the trailing positron bunch. However, by combining modified drivers and modified plasmas, it is possible to exploit the benefits of hollow plasma channels while stabilizing the driver and producing a stable focusing region for the positron bunch. One scheme, proposed by Zhou \textit{et al.}~\cite{zhou_2021}, achieves this by driving nonlinear wakefields in a hollow channel using a transversely asymmetric electron driver.

This scheme utilizes the self-induced quadrupole wakefields generated in a hollow channel during the propagation of an asymmetric driver, for which $\sigma_x > \sigma_y$ (i.e., the beam size is larger in the horizontal than the vertical plane). As the driver propagates, the quadrupole field defocuses the driver in $x$ and focuses it in $y$ until it splits into two beamlets that reach the channel boundary in $x$, as illustrated in Fig.~\ref{fig:hcad}(a). At this point, two half-blowout structures are created---one on each side of the channel---where plasma electrons are expelled much like in the nonlinear regime. A stable, equilibrium driver shape is reached when the focusing field from the exposed plasma ions balances the defocusing quadrupole field.

Focusing of positrons is achieved by driving a nonlinear plasma wakefield, such that some plasma electrons become relativistic and enter the hollow channel, resulting in a near-uniform electron density. A similar region was also identified by Yi \textit{et al.}~\cite{Yi:2013upa,Yi:2014gta}. Here, the longitudinal wakefield is also accelerating for positrons. When this field is beam loaded, in order to reach high energy-transfer efficiency, an on-axis electron density spike appears, as shown in Fig.~\ref{fig:hcad}(b). Figure~\ref{fig:hcad}(c) shows the resulting focusing and accelerating fields.

A simulation with stable acceleration until driver depletion shows that \SI{490}{pC} of charge can be accelerated with a gradient of \SI{4.9}{GV/m} at an efficiency of 33\%. After an initial growth of 20--30\%, the emittance stabilizes at $79 \times \SI{56}{\milli\m\milli\radian}$, shown in Fig.~\ref{fig:hcad}(d). The initial emittance growth may be suppressed if the driver is injected with its equilibrium profile rather than evolving to it. Figures~\ref{fig:hcad}(e--f) show a final energy spread of 1.6\% rms after 4.4 GeV of gain from 10.2 GeV, giving an energy spread per gain (i.e., field uniformity) of 5.3\% rms; this can potentially be improved with more optimal current profiles than a Gaussian. In principle, the charge and efficiency can be further increased by overlapping a positron bunch with a similarly shaped electron bunch \cite{zhou_2022_hcob}, although this may not be compatible with the above focusing method, as the overlapping electrons would quickly be defocused.

The main advantages of this scheme are its stability during propagation and strong accelerating gradients that are near-uniform transversely---the latter being a common feature of hollow channels \cite{chiou_1998}. However, it is unclear whether this field will remain uniform if loaded with lower-emittance, higher-density positron bunches than the given parameters. The scheme is generally well suited for experimental demonstration, as the complex equilibrium driver shape self-generates from easily available Gaussian beams. That said, experiments may suffer from unwanted beam ionization of on-axis gas, since the scheme requires nonlinear and consequently strong wakefields---this will restrict the choice of gas species as well as the beam and plasma densities.

In summary, this scheme offers stable propagation of the driver and stable acceleration of the positron bunch with high charge and high efficiency. However, the trade-off between beam quality and efficiency remains similar to other schemes---a topic explored in more detail in Secs.~\ref{sec:comp} and \ref{sec:positron-problem} below.

\section{Comparison of schemes}
\label{sec:comp}

We have, up to this point, discussed the various proposed schemes and their key parameters separately (Sec.~\ref{sec:schemes}). Table~\ref{tab:table2} summarizes these values for the positron-acceleration schemes, as well as electron-acceleration schemes and relevant experiments. However, without a common metric, it is non-trivial to compare them to each other or to the requirements for a collider. As argued in Sec.~\ref{sec:critical_requirements}, the two key metrics are: accelerating gradient, which affects the collider footprint; and luminosity-per-power, which affects the running costs.

\begin{table*}[t]
\caption{\label{tab:table2} Key parameters of the plasma accelerator and accelerated beam in each of the proposed positron-acceleration schemes (see Sec.~\ref{sec:schemes}). Electron-acceleration schemes and conventional technology are listed for comparison. The parameter $\Delta\phi_e$ represents the phase advance, or degree of plasma-electron motion, inside the positron bunch (see Sec.~\ref{sec:positron-problem}).}
\begin{ruledtabular}
\begin{tabular}{lcccccccccc}
    & Density & Gradient & Charge & Energy & Emittance & En.~spread & Uncorr.  & Fin.~energy &  \\
    \textit{Scheme} & (\SI{}{cm^{-3}}) & (\SI{}{GV/m}) & (pC) & efficiency & (\SI{}{\milli\m\milli\radian}) & per gain & en.~spread & (GeV) & $\Delta\phi_e$\footnote{For electrons, this represents the phase advance if positrons were the focusing species, instead of ions.} & Ref. \\ \hline
    
    Quasi-linear regime (sim.)& \SI{5e16}{} & 1.3 & 4.3 & 30\% & 0.64 & $\sim$10\%\footnote{The correlated energy spread was not optimized and not given in the reference.} & 0.7\% & 1 & 0.77 & \cite{hue_2021} \\
    Quasi-linear regime (exp.)& \SI{1e16}{} & 1 & 85 & 40\% & 127\footnote{The final emittance was not measured, but is here assumed to be preserved.}& $\sim$14\% & n/a & 21 & 0.51 & \cite{Doche:2017jhd} \\
    Nonlinear regime & \SI{7.8e15}{} & 1.6 & 102 & 26\% & 8 & 2.4\% & n/a & 5.2 & 7.6 & \cite{zhou2022positron} \\
    Donut driver (\#1) & \SI{5e16}{} & 8.9 & 13.6 & 0.17\% & 0.036 & 0.3\% & n/a & 35.4 & 0.50 & \cite{jain_2015}\\
    Donut driver (\#2) & \SI{5e16}{} & 40 & 189 & 3.5\% & 1.5\footnote{A non-evolving driver was used in the simulation.} & 6\% & 1\% & 1 & 7.1 & \cite{hue_2021}\\
    Finite-radius channel (cold) & \SI{5e17}{} & 30 & 52 & 3\% & 0.38 & 0.86\% & 0.73\% & 5.5 & 34 & \cite{Diederichs:2020hri}\\
    Finite-radius channel (warm) & \SI{5e17}{} & 30 & 84 & 4.8\% & 0.015 & n/a{\footnote{The energy spread per gain is not given, likely on the few-percent level, but it can be optimized as shown in Ref.~\cite{Diederichs:2020hri}.}} & $\sim$0.01\% & 1.1 & 269 & \cite{diederichs2023_temp2}\\
    Laser-augmented blowout & \SI{2e17}{} & 20 & 15 & 5.5\% & 31 & 3.4\% & n/a & $\sim$10 & 0.67 & \cite{reichwein_2022}\\
    Thin, warm, hollow channel & \SI{1e16}{} & 3.5 & 100 & 4.7\%\footnote{Accounting for the energy loss of the electron driver that creates the hollow channel (not done here) would reduce the overall efficiency.} & 7.4 & 6\% & n/a & 1.45 & 2.0 & \cite{silva_2021} \\
    Asymmetric hollow channel & \SI{3.1e16}{} & 4.9 & 490 & 33\% & 67 & 5.3\% & n/a& 14.6 & 6.5 & \cite{zhou_2021}\\ \hline
    $e^-$ nonlinear regime (sim.)& \SI{2e16}{} & $-10$ & 800 & 37.5\% & 0.133\footnote{Emittance is assumed to be preserved. Ion motion was not simulated, but can likely be avoided with a sufficiently heavy gas species.} & 1.1\% & $\lesssim$1\% & 1500 & 292 & \cite{Chen_2020}\\ 
    $e^-$ nonlinear regime (exp.) & \SI{1.2e16}{} & $-1.4$ & 40 & 22\% & 2.8 & 1.6\% & n/a & 1.1 & 3.0 & \cite{lindstrom_2023} \\ \hline
    Conv.~technology (CLIC) & n/a & 0.1 & 596 & 28.5\%\footnote{Here using the rf-to-beam transfer efficiency, as this is most comparable to the wake-to-beam extraction efficiency.} & 0.11 & 0.35\% & n/a & 1500 & n/a & \cite{clic_cdr} \\ 
\end{tabular}
\end{ruledtabular}
\end{table*}

\begin{figure*}[t]
    \centering
    \includegraphics[width=0.9\textwidth]{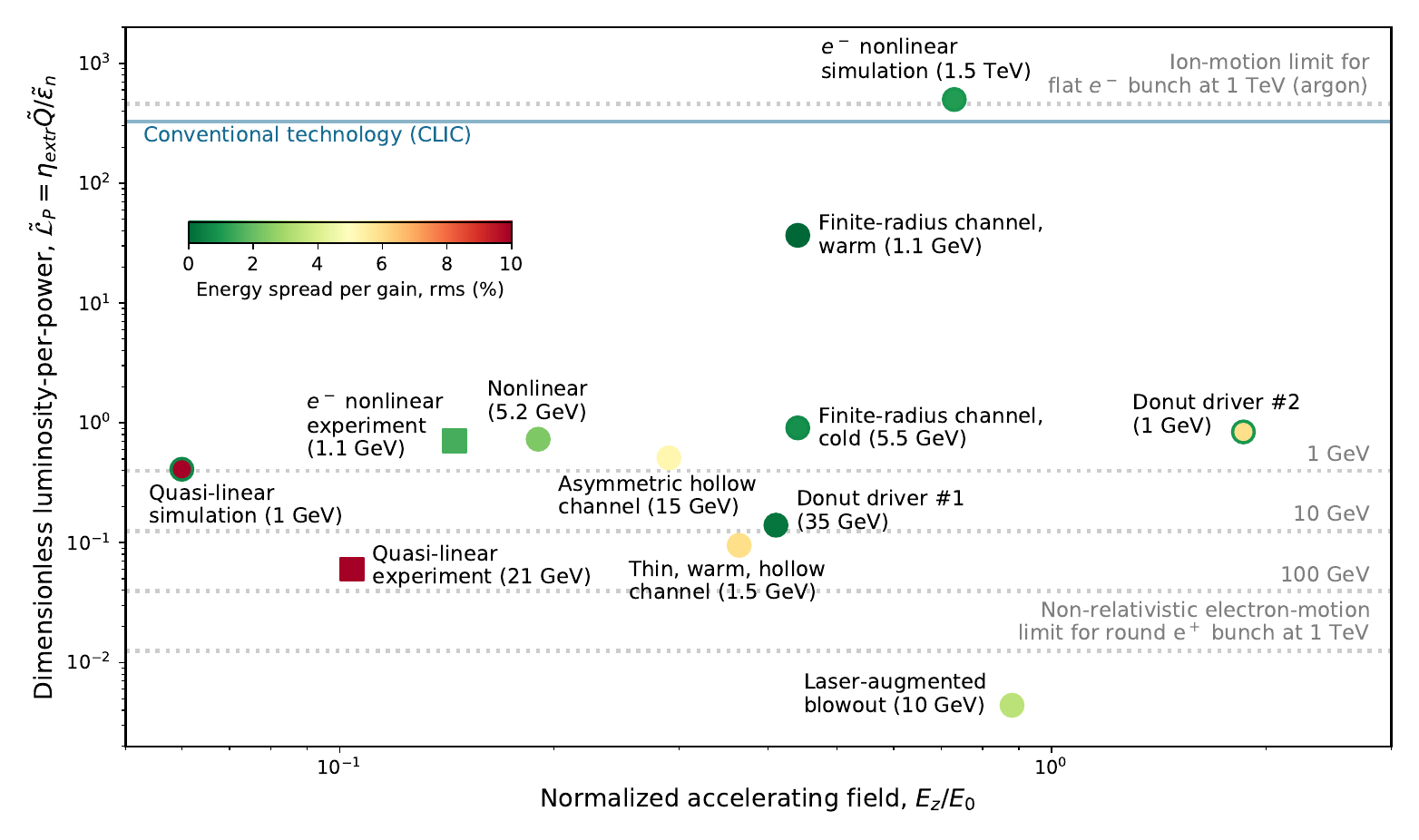}
    \caption{Comparison of the dimensionless luminosity-per-power versus the normalized accelerating field for all proposed positron-acceleration schemes, as well as the nonlinear blowout electron-acceleration scheme and relevant experimental results (see Table~\ref{tab:table2}). The energy spread per gain (red-yellow-green color map; the inner and outer circles represent the projected and uncorrelated energy spreads, respectively) and final energy (parenthesis) of each simulation/experiment are indicated. Conventional technology is represented by CLIC parameters (blue line). Estimated limits on the luminosity-per-power based on the motion of plasma electrons and ions, which depend on beam energy and ion mass, are indicated (gray dotted lines).}
    \label{fig:comp}
\end{figure*}

Combining Eqs.~\ref{eq:luminosity_alt} and \ref{eq:efficiencies}, and assuming that the colliding electron and positron bunches are identical, we express the luminosity-per-power as
\begin{equation}
    \label{eq:luminosity_per_power}
    \frac{\mathcal{L}}{P_\textmd{wall}} \approx \frac{1}{8\pi m_e c^2} \frac{\eta_{\textmd{prod}}\eta_{\textmd{depl}}}{\sqrt{\beta_x\beta_y}}\frac{\eta_{\textmd{extr}}N}{\sqrt{\epsilon_{nx}\epsilon_{ny}}}.
\end{equation}
Here, the production efficiency $\eta_{prod}$ can be assumed to be identical across all proposed schemes, as all are electron-driven. The driver-depletion efficiency $\eta_{depl}$, which may vary somewhat between schemes, is assumed also to be similar. Lastly, the interaction-point beta functions $\beta_x$ and $\beta_y$ are determined mainly by the beam-delivery system, which sets constraints on the energy spread (i.e., $<1$\% rms), but can otherwise be assumed to be independent of scheme. That leaves the wake-to-beam extraction efficiency $\eta_{extr}$, the bunch population $N$ and the normalized emittances $\epsilon_{nx}$ and $\epsilon_{ny}$ for comparison. To produce a metric for meaningful comparison, we define the \textit{dimensionless luminosity-per-power}
\begin{equation}
    \label{eq:normalized_luminosity_per_power}
    \tilde{\mathcal{L}}_P \equiv 4\pi r_e\frac{\eta_\mathrm{extr}N}{\sqrt{\epsilon_{nx}\epsilon_{ny}}} = \frac{\eta_\mathrm{extr}\tilde{Q}}{\tilde{\epsilon}_n},
\end{equation}
where $\tilde{\epsilon}_n = k_p \sqrt{\epsilon_{nx}\epsilon_{ny}}$ is the dimensionless normalized emittance and $\tilde{Q} = 4 \pi r_e k_p N $ is the normalized charge (Eq.~\ref{eq:normalized-charge}). This metric scales as the luminosity-per-power with a factor $H_D$ difference (typically between 1.5--2); a parameter that captures beam--beam effects at the interaction point \cite{schulte_phd}.

An important feature of this dimensionless luminosity-per-power is its independence of the plasma density. Plasma-wakefield simulations can in general be scaled to different plasma densities, resulting in higher accelerating gradients for higher densities ($E_z \sim k_p$, where $k_p \sim \sqrt{n_0}$), simultaneously giving lower charges ($N \sim k_p^{-1}$) and lower emittances ($\epsilon_n \sim k_p^{-1}$). However, the efficiency, the normalized charge ($\tilde{Q}$) and the dimensionless normalized emittance ($\tilde{\epsilon}_n$), which together define $\tilde{\mathcal{L}}_P$, are all independent of plasma density. Ultimately, this means that simulations at different densities can be directly compared and that there is no gain in operating at either higher or lower plasma density, at least in terms of luminosity-per-power. Nevertheless, since the accelerating gradient does scale with plasma density, it is meaningful instead to compare this gradient normalized by the wave-breaking field $E_0=m_ec\omega_p/e$ (Eq.~\ref{eq:gradient})---equivalent to scaling all simulations to the same density.

Figure~\ref{fig:comp} compares all the proposed schemes, showing the dimensionless luminosity-per-power versus the normalized accelerating field, based on the values in Table~\ref{tab:table2}. 
Note that while these represent the best current values, further optimization may still be possible, as discussed in Sec.~\ref{sec:positron-problem}. 

We observe that several schemes perform similarly with respect to luminosity-per-power: the cold finite-radius channel, donut driver, nonlinear regime, asymmetric hollow channel and quasi-linear regime all reach $\tilde{\mathcal{L}}_P \approx 0.4$--0.9. On the other hand, the normalized accelerating field varies significantly between these schemes in the range $E_z/E_0 \in [0.06,1.9]$, where the donut driver, laser-augmented blowout, and finite-radius channel schemes provide the highest gradients for a given plasma density. In terms of energy spread (not optimized for all schemes), the schemes perform at varying levels, with the donut driver, finite-radius channel and nonlinear regime currently providing the most collider-relevant energy spreads. An exception is the warm finite-radius channel scheme, which reaches $\tilde{\mathcal{L}}_P \approx 37$ while maintaining a high accelerating field and a low (slice) energy spread.

Comparing to conventional technology, here represented by CLIC ($\tilde{\mathcal{L}}_P \approx 300$), almost all the proposed positron-acceleration schemes perform worse in luminosity-per-power by at least 2.5 orders of magnitude. The warm finite-radius channel scheme, however, is only 1 order of magnitude lower in luminosity-per-power. Plasma-accelerated electrons are at the level of conventional technology ($\tilde{\mathcal{L}}_P \approx 500$), at least in simulations without ion motion.

Why do we in general observe such a large difference between plasma-acceleration of positrons and electrons? Is it possible to surpass the currently highest achieved luminosity-per-power, and if so, how? This topic is discussed in detail in Sec.~\ref{sec:positron-problem} below.


\section{The positron problem: Plasma-electron motion and transverse beam loading}
\label{sec:positron-problem}

The discrepancy in performance between electron and positron acceleration can in large part be explained by the ratio in mass between plasma ions and electrons for many of the schemes considered in this review. Lighter plasma particles have lower inertia, leading to comparatively more motion within the accelerated positron bunch. The motion of plasma electrons within the positron bunch leads to variation in the plasma-electron density, which in turn disrupts the quality of the accelerated bunch. This effect is a potential limitation on the density of the loaded positron bunch, and therefore a limitation on the achievable luminosity of electron-positron colliders. In the end of this section, we consider schemes and conditions that exceed this limitation, but nevertheless appear to preserve the quality of the accelerated positron bunch.

\subsection{The ideal case}
\label{sec:ideal-case}

The ideal plasma-based positron accelerator is similar to the standard nonlinear blowout for electron acceleration: the focusing fields must vary linearly in the transverse directions to preserve the emittance, and the accelerating fields must be uniform in both the transverse and longitudinal directions to preserve the uncorrelated and correlated energy spread, respectively. For emittance preservation, we specifically require \cite{Lu_2006_pop,Lu_2006}
\begin{equation}
    \nabla_\perp(E_r - v_zB_\phi) = \frac{1}{\epsilon_0} (\rho - J_z/c) = \textmd{const},
\end{equation}
where $\rho$ is the charge density (providing \textit{passive} plasma lensing \cite{Chen:1987}) and $J_z$ is the axial current density (providing \textit{active} plasma lensing \cite{vanTilborg_2015}). This means that either both $\rho$ and $J_z$ need to be transversely uniform, or, more generally, that any variation in $\rho$ must be matched by a corresponding variation in $J_z$. Longitudinally uniform focusing fields [$\partial_z (E_r-v_z B_\phi) = 0$] are not strictly necessary, as the beam emittance can still be preserved with slice-by-slice matching \cite{Benedetti_2017}, assuming the fields are linear within each slice. However, the Panofsky-Wenzel theorem \cite{panofsky_wenzel}
\begin{equation}
    \label{eq:panof_wenz}
    \partial_z(E_r-v_z B_\phi) = \nabla_\perp E_z,
\end{equation}
states that in order to preserve energy spread transversely ($\nabla_\perp E_z$=0), the focusing fields must be uniform longitudinally [$\partial_z(E_r-v_z B_\phi)=0$]. This generalizes the restriction on $(\rho - J_z/c)$ from being constant transversely to being constant everywhere within the accelerated bunch. Lastly, longitudinally uniform accelerating fields ($\partial_z E_z = 0$) can be obtained through precise shaping of the current profile---optimal beam loading \cite{tzoufras_2008}.

\subsection{Ion and electron motion}
All the above criteria are normally satisfied in the blowout regime for electrons, where the ion-charge density is constant everywhere and there is no axial current density anywhere within the accelerating bunch. There is, however, an important exception: if the charge density of the electron bunch is sufficiently high to induce ion motion, the ion-charge density will no longer be constant within the bunch. If the beam density $n_b$ is sufficiently high to move the ions toward the axis within the timescale of the full bunch length $\Delta\zeta \approx \sqrt{2\pi}\sigma_z$, emittance may no longer be preserved. Rosenzweig \textit{et al.}~\cite{Rosenzweig_2005} calculated the \textit{phase advance} of the ion motion (for round electron bunches) to be
\begin{equation}
    \label{eq:ion-motion-limit}
    \Delta\phi_i \simeq k_i \Delta\zeta = \sqrt{\frac{\mu_0 e^2}{2} \frac{Z \sigma_z N}{m_i} \sqrt{\frac{r_e \gamma n_0}{\epsilon_{nx}\epsilon_{ny}}}},
\end{equation}
where $k_i$ is the plasma-ion wavenumber in the focusing field of the electron bunch, $Z$ is the ion charge state, $m_i$ is the mass of the ion, $\gamma$ is the relativistic factor of the beam particles, and $\mu_0$ is the permeability of free space. An on-axis density spike will form when ions are focused onto the axis, which should be avoided: $\Delta\phi_i \lesssim \pi/2$, often referred to as the \textit{ion-motion limit}.

\begin{figure}[t]
    \centering
    \includegraphics[width=\linewidth]{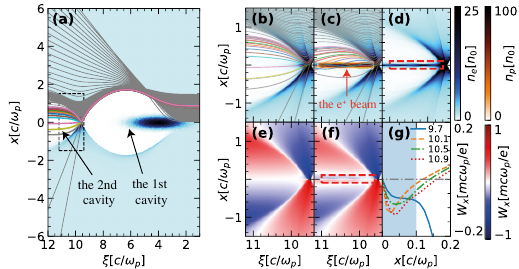}
    \caption{Simulations demonstrating plasma-electron motion in a nonlinear wakefield, (a) showing plasma-electron densities (blue color map) and trajectories (gray and colored lines) driven by an intense electron bunch. The corresponding plasma-electron density and trajectories are shown for the positron-loading region [dashed box in (a)], both unloaded (b) and loaded (c--d) by an intense positron bunch (orange color map). This extracts energy from the wake (i.e., longitudinal beam loading) but also significantly changes the trajectory and distribution of some plasma electrons inside the positron bunch (red dashed box), leading to electron oscillations with a phase advance of approximately $2\pi$. The corresponding focusing field changes (i.e, transverse beam loading) from being defocusing when unloaded (e) to being focusing when loaded (f). Transverse lineouts (g) show that in the presence of the positron bunch these fields are nonlinear away from the axis and nonuniform longitudinally. From Ref.~\cite{zhou2022positron}.}
    \label{fig:electron-motion}
\end{figure}

Exactly the same dynamics occur for plasma electrons in the presence of a high-density positron bunch, as illustrated in Fig.~\ref{fig:electron-motion}. In this case, we substitute the ion mass $m_i$ for the electron mass, specifically $\gamma_{pe} m_e$, where $\gamma_{pe}$ is the Lorentz factor of the plasma electrons, set $Z=1$ as plasma electrons are always singly charged, and change the focusing background-ion density $n_0$ to the local background density (i.e., the difference between the electron and ion densities $\Delta n =|n_e - n_i|$):
\begin{equation}
    \label{eq:electron-motion-limit}
    \Delta\phi_e \simeq k_e \Delta\zeta = \sqrt{\frac{\mu_0 e^2}{2} \frac{\sigma_z N}{\gamma_{pe} m_e} \sqrt{\frac{r_e \gamma \Delta n}{\epsilon_{nx}\epsilon_{ny}}}},
\end{equation}
where $k_e$ is the plasma-electron wavenumber in the focusing field of the positron bunch. 
The corresponding \textit{electron-motion limit}, $\Delta\phi_e \lesssim \pi/2$, is approximately equivalent to the limit $k_e \sigma_z \approx 1$ (up to a factor $\sqrt{\pi/8} \approx 0.63$), as discussed in Ref.~\cite{hue_2021}. 

The electron phase advance $\Delta\phi_e$ is calculated for each positron-acceleration scheme and the nonlinear blowout scheme for electrons and displayed in Table~\ref{tab:table2}.
We note that some of the schemes discussed in this review preserve positron beam quality even though the plasma-electron phase advance exceeds $\pi/2$. This suggests alternative strategies for exceeding the electron-motion limit.  
Before discussing these strategies, one important question remains: how does electron motion affect the luminosity-per-power of colliders based on positron acceleration concepts? 

\subsection{An electron-motion ``limit" to the dimensionless luminosity-per-power}
We observe that the phase advance of plasma electrons (Eq.~\ref{eq:electron-motion-limit}) depends on the same ratio of charge to emittance, (i.e., $N/\sqrt{\epsilon_{nx}\epsilon_{ny}}$), exactly like the dimensionless luminosity-per-power (Eq.~\ref{eq:normalized_luminosity_per_power}). Crucially, this means that the luminosity-per-power can be expressed as the electron-motion phase advance:
\begin{equation}
    \label{eq:positrons-lumi-per-power}
    \tilde{\mathcal{L}}_P^{e^{+}} \simeq \sqrt{\frac{16\pi}{\gamma}} (\Delta \phi_e)^2 \left(\frac{\eta_{\mathrm{extr}}}{k_p \sigma_z}\right)\gamma_{pe} \sqrt{\frac{n_0}{\Delta n}}.
\end{equation}
The ratio of extraction efficiency to normalized bunch length, in the case of optimally beam-loaded bunch, is typically of order unity, i.e.~$\eta_{\mathrm{extr}}/k_p \sigma_z = \mathcal{O}(1)$. As examples, in Ref.~\cite{zhou2022positron} (nonlinear regime for positrons) the ratio is $\sim$0.6 and in Ref.~\cite{Lindstrom:2021tkb} (nonlinear regime for electrons) the ratio is $\sim$0.9. Note in particular the unfavourable energy dependence in Eq.~\ref{eq:electron-motion-limit} ($\sim1/\sqrt{\gamma}$), which results from smaller matched beam sizes ($\sim1/\gamma^{1/4}$; leading to higher beam densities) at higher energy. The charge density ratio varies with scheme, but is typically no more than one order of magnitude away from unity: $\Delta n / n_0 = \mathcal{O}(0.1-10)$.

Combining these ratios ($\eta_{\mathrm{extr}}/k_p \sigma_z \approx 1$ and $\Delta n / n_0 \approx 1$) with the conventional phase-advance limit ($\Delta \phi_e \approx \pi/2$) and assuming non-relativistic plasma electrons ($\gamma_{pe} \approx 1$), we get an estimated upper bound on the dimensionless luminosity-per-power for plasma-based positron accelerators of $\tilde{\mathcal{L}}_P^{e^{+}} \approx 17.5/\sqrt{\gamma}$, or about 0.4 for \SI{1}{GeV} and 0.013 for \SI{1}{TeV}. This range is indeed consistent with the dimensionless luminosity-per-power found across most the proposed schemes, as shown in Fig.~\ref{fig:comp}. 
Note that some schemes exceed this limit, indicating that they operate outside the above assumptions. These will be discussed further in Sec.~\ref{sec:outlook}. 

For electrons, the dimensionless luminosity-per-power differs from that for positrons by the mass ratio of the plasma ions and electrons and the ion-charge state. By comparing Eqs.~\ref{eq:ion-motion-limit} and \ref{eq:electron-motion-limit} we therefore find that the ion-motion limit on the dimensionless luminosity-per-power for electrons is
\begin{equation}
    \label{eq:electrons-lumi-per-power-ratio}
    \tilde{\mathcal{L}}_P^{e^{-}} = \frac{m_i}{Z \gamma_{pe} m_e}\sqrt{\frac{\Delta n}{n_0}} \tilde{\mathcal{L}}_P^{e^{+}},
\end{equation}
which is a factor of 73350 larger for singly ionized argon ions compared to plasma electrons where $\Delta n = n_0$.
Taking also into account that flat beams ($\epsilon_{nx} \gg \epsilon_{ny}$) have a larger phase advance by a factor $\sqrt{2}$, as argued by Rosenzweig \textit{et al.}~\cite{Rosenzweig_2005}, the limit for flat beams is consequently a factor 2 lower than for round beams:
\begin{equation}
    \label{eq:flat-vs-round-beams}
    \tilde{\mathcal{L}}_P^\mathrm{flat} \approx \frac{1}{2}\tilde{\mathcal{L}}_P^\mathrm{round}.
\end{equation}
The resulting ion-motion limit for flat beams is indicated in Fig.~\ref{fig:comp}, which corresponds well to that reached in simulations as well as that of conventional technology. 

In short, the root of the positron problem is the comparatively low mass of plasma electrons, leading to complex motion and therefore nonlinear focusing fields (transverse beam loading) inside high-density positron bunches. This causes degradation of the beam quality, ultimately making simultaneous high-efficiency and high-quality acceleration challenging.

\subsection{Outlook: raising the electron-motion limit} \label{sec:outlook}

So how can we increase the luminosity-per-power beyond the values detailed in Fig.~\ref{fig:comp}?
Most of the positron-acceleration schemes are designed starting from the ideal case (as discussed in Sec.~\ref{sec:ideal-case}); creating quality-preserving wakefields and then increasing the beam density until electron motion occurs. This may explain why many schemes reach similar luminosity-per-power. However, since electron motion is inevitable for the positron beam densities required for linear colliders, all future schemes must include transverse beam loading as an integral part of their design.

Equation~\ref{eq:positrons-lumi-per-power} straightaway motivates four main strategies of improvement: (1) tolerating larger electron phase advance, increasing $\Delta\phi_e$; (2) reaching high efficiencies with shorter bunch lengths, increasing $\eta_{\mathrm{extr}}/k_p \sigma_z$; (3) using relativistic electrons for focusing the positron beam, increasing $\gamma_{pe}$; and (4) using a low on-axis excess electron density, decreasing $\Delta n/n_0$.

At first glance, increasing the phase advance significantly beyond $\pi/2$ implies tolerating multiple electron oscillations within the positron bunch. However, simulations of the finite-radius channel do not show this feature, despite the fact $\Delta \phi_e \approx 34$, or approximately 5 plasma electron oscillations. The plasma electrons do not oscillate within the volume of the accelerated positron bunch because they already have large transverse momentum as they return towards the beam axis. The dense positron bunch further accelerates plasma electrons toward the axis such that their momenta carry them well beyond the positron beam volume and their subsequent return to axis occurs behind the positron bunch. The $\Delta \phi_e$ limit still exists for a sufficiently long positron bunch in this scheme, but the first return of plasma electrons within the bunch may occur at $\Delta \phi_e \gg \pi/2$.

Using warm plasmas, the effective electron phase advance $\Delta \phi_e$ can be increased even further~\cite{diederichs_temperature_2023,diederichs2023_temp2}. As demonstrated in Fig.~\ref{fig:comp}, this warm finite-radius channel scheme is currently the most promising strategy for going beyond the electron-motion limit.

In the homogeneous-plasma nonlinear regime, plasma electrons can also oscillate significantly within the positron bunch, as shown in Fig.~\ref{fig:electron-motion}(d). Note that the electron oscillations in and of themselves are not necessarily problematic: the resulting, often non-uniform charge distribution, is. Therefore, surviving multiple electron oscillations will likely require finding an equilibrium positron-bunch profile that results in uniform electron density inside the bunch.

Achieving high efficiency with significantly shorter bunches, all while maintaining low energy spread (through optimal beam loading), is another interesting strategy. As noted by Tzoufras \textit{et al.}~\cite{tzoufras_2008}, in a highly nonlinear wake for electrons, it is possible to achieve a high energy efficiency with higher charge at lower accelerating field or lower charge at higher accelerating field. In the latter case, the bunch length can be significantly reduced. The concept of accelerating short positron bunches at high efficiencies has already been exploited in the donut-driver scheme \cite{hue_2021}. Nevertheless, beam loading with short, high-current positron bunches is challenging because of the trade-off between efficiency and beam quality. The use of shorter bunches also comes with practical challenges related to their production and coherent-synchrotron-radiation effects in chicanes. Shorter bunches provide an advantage for beam--beam interactions because they reduce the deleterious effects of beamstrahlung on the luminosity spectrum~\cite{Barklow2023}. Energy-recovery techniques \cite{Schroeder:2016phw,zhou2022positron} that use additional laser or electron beam pulses to extract energy from the wake can also be used to increase efficiency.



The use of highly relativistic plasma electrons is a way to effectively symmetrize the mass of plasma electron and ions. One way to achieve this would be using an external source of high-energy, counter-propagating electrons~\cite{Lindstrom_phd}---a scheme with similarities to \textit{electron lenses}~\cite{Shiltsev_1999} used in proton colliders---although the power required to maintain such a stream of electrons may be problematic.

Lastly, by reducing the excess charge density of the electron filament that focuses the positron bunch, the matched beta function can be increased. This reduces the beam density for a given charge and emittance, mitigating the issues related to electron motion. An extreme case of this approach is the hollow channel (discussed in Sec.~\ref{sec:hollow-plasma-channels}), where $\Delta n = 0$, which can in principle provide beams that reach high luminosity-per-power, but suffers from a catastrophic transverse instability~\cite{Schroeder:1999cb,Lindstrom:2018hhy}.

As an alternative, we could in principle use an \textit{anti-plasma} \cite{gessner_phd}, which would effectively swap the electron-motion limit (Eq.~\ref{eq:positrons-lumi-per-power}) for the ion-motion limit (Eq.~\ref{eq:electrons-lumi-per-power-ratio}). Unfortunately, this particular solution is presently neither technologically nor economically feasible. 

More broadly, whether the above strategies can be used, individually or in combination, to develop a scheme that provides competitive luminosity-per-power is currently an open question, and a potential topic of future research.

\section{Conclusion}
\label{sec:conclusion}

The overarching goal of accelerating positrons in plasma wakefields is to reduce the footprint and cost of future electron--positron colliders. This imposes a number of strict requirements on the positron accelerator: high accelerating gradient ($>$1--\SI{10}{GV/m}); high energy efficiency (5--10\% from wall plug to beam); high beam quality including high charge (nC-scale), low emittance ($<$\SI{1}{\milli\m\milli\radian}) and low energy spread ($<$1\%); as well as sufficient stability. An important combined metric is the luminosity-per-beam power. Plasma-based electron acceleration appears able to meet these requirements (including the luminosity-per-power), but this is currently not the case for positrons.

Major progress on plasma acceleration for positrons has been made over the previous two decades, since the first theoretical investigations around 2000. The first experiments were performed at SLAC's FFTB facility, which showed that positrons can indeed be both focused and accelerated in a plasma. Subsequent work branched into two main directions: acceleration in homogeneous plasmas, and acceleration in hollow plasma channels---the latter promised better beam quality. After numerous theoretical advancements and several years of commissioning the FACET test facility at SLAC, major experimental milestones were reached. Simultaneous high-efficiency and high energy gain (multi-GeV) positron acceleration was demonstrated in a homogeneous plasma. Moreover, acceleration of positrons in a laser-ionized hollow plasma channel was demonstrated, albeit with significantly less energy gain. However, strong nonlinear focusing field in the homogeneous plasma scheme caused large emittance growth, whereas in the hollow channel scheme, strong transverse wakefields caused an instability which rapidly deflected the accelerating positron bunch. As a result, while experiments partly met several collider requirements, the accelerated positron bunches were generally not suitable for a collider.

To remedy this shortfall, several new positron-acceleration schemes have been proposed. These schemes create favourable conditions for positron acceleration either by further optimizing the homogeneous plasma scheme, or by modifying the shape of the driver, the plasma, or both. A new common metric---dimensionless luminosity-per-power $\tilde{\mathcal{L}_P} = \eta_{\mathrm{extr}} \tilde{Q}/{\tilde{\epsilon}_n}$---is introduced here to compare the seven proposed schemes. The resulting comparison indicates that most of the proposed schemes perform similarly within 2 orders of magnitude in luminosity-per-power. However, they are at least 2.5 orders of magnitude below that of collider proposals using conventional technology (e.g., CLIC) and the nonlinear blowout scheme for plasma-accelerated electrons. An exception is the finite-radius channel with warm plasma, which is currently the most promising scheme, reaching within 1 order of magnitude of the required dimensionless luminosity-per-power.

The key limitation that affects the majority of positron-acceleration schemes is that of complex electron motion within the positron bunch, which arises from high beam densities---effectively acting as a strong lens for the plasma electrons. The resulting nonlinear focusing fields lead to degradation of the positron beam quality, exactly equivalent to the effects of ion motion on electrons. However, since the mass of plasma electrons is significantly lower than that of plasma ions, this disruptive motion occurs for correspondingly lower beam densities, which can explain the observed discrepancy between positron and electron acceleration in plasma.

While alternative plasma-based collider concepts have been proposed that circumvent the positron problem altogether, including asymmetric plasma--RF hybrid colliders~\cite{foster_2023} and $\gamma$--$\gamma$ colliders~\cite{Telnov:1998vs,rosenzweig:1996,adli_2019}, several strategies do exist for overcoming the electron-motion challenge. These may include: increasing the temperature of the plasma; imparting large-transverse momenta to converging plasma electrons; maintaining a uniform distribution of plasma electrons within a high-density positron bunch; using relativistic (and effectively heavier) plasma electrons; achieving uniform and high efficiency acceleration also with very short bunches; and sustaining a decreased excess electron charge density to increase the matched beta function---or perhaps something more exotic. Regardless of strategy, any future scheme for positron acceleration will inevitably face electron motion and should therefore be designed from the start to tolerate it or even exploit it.


\begin{acknowledgements}
This work was supported by the Research Council of Norway (NFR Grant No.~313770). Computations were performed on resources provided by Sigma2; the National Infrastructure for High Performance Computing and Data Storage in Norway.
\end{acknowledgements}

\nocite{*}
\clearpage
\bibliography{references}

\providecommand{\noopsort}[1]{}\providecommand{\singleletter}[1]{#1}%
\begin{thebibliography}{207}%
\makeatletter
\providecommand \@ifxundefined [1]{%
 \@ifx{#1\undefined}
}%
\providecommand \@ifnum [1]{%
 \ifnum #1\expandafter \@firstoftwo
 \else \expandafter \@secondoftwo
 \fi
}%
\providecommand \@ifx [1]{%
 \ifx #1\expandafter \@firstoftwo
 \else \expandafter \@secondoftwo
 \fi
}%
\providecommand \natexlab [1]{#1}%
\providecommand \enquote  [1]{``#1''}%
\providecommand \bibnamefont  [1]{#1}%
\providecommand \bibfnamefont [1]{#1}%
\providecommand \citenamefont [1]{#1}%
\providecommand \href@noop [0]{\@secondoftwo}%
\providecommand \href [0]{\begingroup \@sanitize@url \@href}%
\providecommand \@href[1]{\@@startlink{#1}\@@href}%
\providecommand \@@href[1]{\endgroup#1\@@endlink}%
\providecommand \@sanitize@url [0]{\catcode `\\12\catcode `\$12\catcode `\&12\catcode `\#12\catcode `\^12\catcode `\_12\catcode `\%12\relax}%
\providecommand \@@startlink[1]{}%
\providecommand \@@endlink[0]{}%
\providecommand \url  [0]{\begingroup\@sanitize@url \@url }%
\providecommand \@url [1]{\endgroup\@href {#1}{\urlprefix }}%
\providecommand \urlprefix  [0]{URL }%
\providecommand \Eprint [0]{\href }%
\providecommand \doibase [0]{https://doi.org/}%
\providecommand \selectlanguage [0]{\@gobble}%
\providecommand \bibinfo  [0]{\@secondoftwo}%
\providecommand \bibfield  [0]{\@secondoftwo}%
\providecommand \translation [1]{[#1]}%
\providecommand \BibitemOpen [0]{}%
\providecommand \bibitemStop [0]{}%
\providecommand \bibitemNoStop [0]{.\EOS\space}%
\providecommand \EOS [0]{\spacefactor3000\relax}%
\providecommand \BibitemShut  [1]{\csname bibitem#1\endcsname}%
\let\auto@bib@innerbib\@empty
\bibitem [{\citenamefont {Narain}\ \emph {et~al.}(2022)\citenamefont {Narain} \emph {et~al.}}]{snowmass_Efront}%
  \BibitemOpen
  \bibfield  {author} {\bibinfo {author} {\bibfnamefont {M.}~\bibnamefont {Narain}} \emph {et~al.},\ }\bibfield  {title} {\bibinfo {title} {{The Future of US Particle Physics - The Snowmass 2021 Energy Frontier Report}},\ }\href@noop {} {\bibfield  {journal} {\bibinfo  {journal} {arXiv e-prints}\ } (\bibinfo {year} {2022})},\ \Eprint {https://arxiv.org/abs/2211.11084} {arXiv:2211.11084} \BibitemShut {NoStop}%
\bibitem [{CER(2020)}]{CERN-ESU-015}%
  \BibitemOpen
  \href {https://doi.org/10.17181/ESU2020} {\emph {\bibinfo {title} {{2020 Update of the European Strategy for Particle Physics}}}}\ (\bibinfo  {publisher} {CERN Council},\ \bibinfo {address} {Geneva},\ \bibinfo {year} {2020})\BibitemShut {NoStop}%
\bibitem [{\citenamefont {{Adolphsen}}\ \emph {et~al.}(2022)\citenamefont {{Adolphsen}} \emph {et~al.}}]{roadmap_2022_acc}%
  \BibitemOpen
  \bibfield  {author} {\bibinfo {author} {\bibfnamefont {C.}~\bibnamefont {{Adolphsen}}} \emph {et~al.},\ }\bibfield  {title} {\bibinfo {title} {{European Strategy for Particle Physics -- Accelerator R\&D Roadmap}},\ }\href@noop {} {\bibfield  {journal} {\bibinfo  {journal} {arXiv e-prints}\ } (\bibinfo {year} {2022})},\ \Eprint {https://arxiv.org/abs/2201.07895} {arXiv:2201.07895} \BibitemShut {NoStop}%
\bibitem [{\citenamefont {Barish}\ and\ \citenamefont {Brau}(2013)}]{ilc_overview}%
  \BibitemOpen
  \bibfield  {author} {\bibinfo {author} {\bibfnamefont {B.}~\bibnamefont {Barish}}\ and\ \bibinfo {author} {\bibfnamefont {J.~E.}\ \bibnamefont {Brau}},\ }\bibfield  {title} {\bibinfo {title} {{The International Linear Collider}},\ }\href {https://doi.org/10.1142/S0217751X13300391} {\bibfield  {journal} {\bibinfo  {journal} {Int. J. Mod. Phys. A}\ }\textbf {\bibinfo {volume} {28}},\ \bibinfo {pages} {1330039} (\bibinfo {year} {2013})}\BibitemShut {NoStop}%
\bibitem [{ilc(2013{\natexlab{a}})}]{ilc_tdr1}%
  \BibitemOpen
  \href@noop {} {\emph {\bibinfo {title} {{The International Linear Collider Technical Design Report - Volume 1: Executive Summary}}}},\ \bibinfo {type} {Tech. Rep.}\ (\bibinfo {year} {2013})\ \Eprint {https://arxiv.org/abs/1306.6327} {arXiv:1306.6327} \BibitemShut {NoStop}%
\bibitem [{ilc(2013{\natexlab{b}})}]{ilc_tdr2}%
  \BibitemOpen
  \href@noop {} {\emph {\bibinfo {title} {{The International Linear Collider Technical Design Report - Volume 2: Physics}}}},\ \bibinfo {type} {Tech. Rep.}\ (\bibinfo {year} {2013})\ \Eprint {https://arxiv.org/abs/1306.6352} {arXiv:1306.6352} \BibitemShut {NoStop}%
\bibitem [{ilc(2013{\natexlab{c}})}]{ilc_tdr3I}%
  \BibitemOpen
  \href@noop {} {\emph {\bibinfo {title} {{The International Linear Collider Technical Design Report - Volume 3.I: Accelerator R\&D in the Technical Design Phase}}}},\ \bibinfo {type} {Tech. Rep.}\ (\bibinfo {year} {2013})\ \Eprint {https://arxiv.org/abs/1306.6353} {arXiv:1306.6353} \BibitemShut {NoStop}%
\bibitem [{ilc(2013{\natexlab{d}})}]{ilc_tdr3II}%
  \BibitemOpen
  \href@noop {} {\emph {\bibinfo {title} {{The International Linear Collider Technical Design Report - Volume 3.II: Accelerator Baseline Design}}}},\ \bibinfo {type} {Tech. Rep.}\ (\bibinfo {year} {2013})\ \Eprint {https://arxiv.org/abs/1306.6328} {arXiv:1306.6328} \BibitemShut {NoStop}%
\bibitem [{\citenamefont {Abramowicz}\ \emph {et~al.}(2013)\citenamefont {Abramowicz} \emph {et~al.}}]{ilc_tdr4}%
  \BibitemOpen
  \bibfield  {author} {\bibinfo {author} {\bibfnamefont {H.}~\bibnamefont {Abramowicz}} \emph {et~al.},\ }\href@noop {} {\emph {\bibinfo {title} {{The International Linear Collider Technical Design Report - Volume 4: Detectors}}}},\ \bibinfo {type} {Tech. Rep.}\ (\bibinfo {year} {2013})\ \Eprint {https://arxiv.org/abs/1306.6329} {arXiv:1306.6329} \BibitemShut {NoStop}%
\bibitem [{cli(2012)}]{clic_cdr}%
  \BibitemOpen
  \href {https://doi.org/10.5170/CERN-2012-007} {\emph {\bibinfo {title} {{A Multi-TeV Linear Collider Based on CLIC Technology}: {CLIC Conceptual Design Report}}}},\ \bibinfo {type} {Tech. Rep.}\ (\bibinfo {year} {2012})\BibitemShut {NoStop}%
\bibitem [{\citenamefont {{CLIC}}\ and\ \citenamefont {{CLICdp collaborations}}(2016)}]{clic_2016}%
  \BibitemOpen
  \bibfield  {author} {\bibinfo {author} {\bibnamefont {{CLIC}}}\ and\ \bibinfo {author} {\bibnamefont {{CLICdp collaborations}}},\ }\bibfield  {title} {\bibinfo {title} {{Updated baseline for a staged Compact Linear Collider}},\ }\href@noop {} {\bibfield  {journal} {\bibinfo  {journal} {arXiv e-prints}\ } (\bibinfo {year} {2016})},\ \Eprint {https://arxiv.org/abs/1608.07537} {arXiv:1608.07537} \BibitemShut {NoStop}%
\bibitem [{\citenamefont {{CLIC}}\ and\ \citenamefont {{CLICdp collaborations}}(2018)}]{clic_sum}%
  \BibitemOpen
  \bibfield  {author} {\bibinfo {author} {\bibnamefont {{CLIC}}}\ and\ \bibinfo {author} {\bibnamefont {{CLICdp collaborations}}},\ }\bibfield  {title} {\bibinfo {title} {{The Compact Linear Collider (CLIC) - 2018 Summary Report}},\ }\href@noop {} {\bibfield  {journal} {\bibinfo  {journal} {arXiv e-prints}\ } (\bibinfo {year} {2018})},\ \Eprint {https://arxiv.org/abs/1812.06018} {arXiv:1812.06018} \BibitemShut {NoStop}%
\bibitem [{cli(2018)}]{clic_imp}%
  \BibitemOpen
  \href {https://doi.org/10.23731/CYRM-2018-004} {\emph {\bibinfo {title} {{The Compact Linear Collider (CLIC) - Project Implementation Plan}}}},\ \bibinfo {type} {Tech. Rep.}\ (\bibinfo {year} {2018})\ \Eprint {https://arxiv.org/abs/1903.08655} {arXiv:1903.08655} \BibitemShut {NoStop}%
\bibitem [{\citenamefont {{Bai}}\ \emph {et~al.}(2021)\citenamefont {{Bai}}, \citenamefont {{Barklow}}, \citenamefont {{Bartoldus}}, \citenamefont {{Breidenbach}}, \citenamefont {{Grenier}}, \citenamefont {{Huang}} \emph {et~al.}}]{C3_2021}%
  \BibitemOpen
  \bibfield  {author} {\bibinfo {author} {\bibfnamefont {M.}~\bibnamefont {{Bai}}}, \bibinfo {author} {\bibfnamefont {T.}~\bibnamefont {{Barklow}}}, \bibinfo {author} {\bibfnamefont {R.}~\bibnamefont {{Bartoldus}}}, \bibinfo {author} {\bibfnamefont {M.}~\bibnamefont {{Breidenbach}}}, \bibinfo {author} {\bibfnamefont {P.}~\bibnamefont {{Grenier}}}, \bibinfo {author} {\bibfnamefont {Z.}~\bibnamefont {{Huang}}}, \emph {et~al.},\ }\bibfield  {title} {\bibinfo {title} {{C$^3$: A ``Cool'' Route to the Higgs Boson and Beyond}},\ }\href@noop {} {\bibfield  {journal} {\bibinfo  {journal} {arXiv e-prints}\ } (\bibinfo {year} {2021})},\ \Eprint {https://arxiv.org/abs/2110.15800} {arXiv:2110.15800} \BibitemShut {NoStop}%
\bibitem [{\citenamefont {Grudiev}\ \emph {et~al.}(2009)\citenamefont {Grudiev}, \citenamefont {Calatroni},\ and\ \citenamefont {Wuensch}}]{Grudiev2009}%
  \BibitemOpen
  \bibfield  {author} {\bibinfo {author} {\bibfnamefont {A.}~\bibnamefont {Grudiev}}, \bibinfo {author} {\bibfnamefont {S.}~\bibnamefont {Calatroni}},\ and\ \bibinfo {author} {\bibfnamefont {W.}~\bibnamefont {Wuensch}},\ }\bibfield  {title} {\bibinfo {title} {New local field quantity describing the high gradient limit of accelerating structures},\ }\href {https://doi.org/10.1103/PhysRevSTAB.12.102001} {\bibfield  {journal} {\bibinfo  {journal} {Phys. Rev. ST Accel. Beams}\ }\textbf {\bibinfo {volume} {12}},\ \bibinfo {pages} {102001} (\bibinfo {year} {2009})}\BibitemShut {NoStop}%
\bibitem [{\citenamefont {Jing}\ and\ \citenamefont {Ha}(2022)}]{Jing:2022qbj}%
  \BibitemOpen
  \bibfield  {author} {\bibinfo {author} {\bibfnamefont {C.}~\bibnamefont {Jing}}\ and\ \bibinfo {author} {\bibfnamefont {G.}~\bibnamefont {Ha}},\ }\bibfield  {title} {\bibinfo {title} {{Roadmap for Structure-based Wakefield Accelerator (SWFA) R\&D and its challenges in beam dynamics}},\ }\href {https://doi.org/10.1088/1748-0221/17/05/T05007} {\bibfield  {journal} {\bibinfo  {journal} {J. Instrum.}\ }\textbf {\bibinfo {volume} {17}}\bibinfo  {number} { (05)},\ \bibinfo {pages} {T05007}}\BibitemShut {NoStop}%
\bibitem [{\citenamefont {Lu}\ \emph {et~al.}(2022)\citenamefont {Lu} \emph {et~al.}}]{Lu:2022oin}%
  \BibitemOpen
\bibfield  {number} {  }\bibfield  {author} {\bibinfo {author} {\bibfnamefont {X.}~\bibnamefont {Lu}} \emph {et~al.},\ }\bibfield  {title} {\bibinfo {title} {{Advanced RF Structures for Wakefield Acceleration and High-Gradient Research}},\ }in\ \href@noop {} {\emph {\bibinfo {booktitle} {{2022 Snowmass Summer Study}}}}\ (\bibinfo {year} {2022})\ \Eprint {https://arxiv.org/abs/2203.08374} {arXiv:2203.08374} \BibitemShut {NoStop}%
\bibitem [{\citenamefont {Corde}\ \emph {et~al.}(2016)\citenamefont {Corde}, \citenamefont {Adli}, \citenamefont {Allen}, \citenamefont {An}, \citenamefont {Clarke}, \citenamefont {Clausse} \emph {et~al.}}]{Corde:2016natcom}%
  \BibitemOpen
  \bibfield  {author} {\bibinfo {author} {\bibfnamefont {S.}~\bibnamefont {Corde}}, \bibinfo {author} {\bibfnamefont {E.}~\bibnamefont {Adli}}, \bibinfo {author} {\bibfnamefont {J.}~\bibnamefont {Allen}}, \bibinfo {author} {\bibfnamefont {W.}~\bibnamefont {An}}, \bibinfo {author} {\bibfnamefont {C.~I.}\ \bibnamefont {Clarke}}, \bibinfo {author} {\bibfnamefont {B.}~\bibnamefont {Clausse}}, \emph {et~al.},\ }\bibfield  {title} {\bibinfo {title} {{High-field plasma acceleration in a high-ionization-potential gas}},\ }\href {https://doi.org/10.1038/ncomms11898} {\bibfield  {journal} {\bibinfo  {journal} {Nat. Commun.}\ }\textbf {\bibinfo {volume} {7}},\ \bibinfo {pages} {11898} (\bibinfo {year} {2016})}\BibitemShut {NoStop}%
\bibitem [{\citenamefont {Bohlen}\ \emph {et~al.}(2022)\citenamefont {Bohlen}, \citenamefont {Br\"ummer}, \citenamefont {Gr\"uner}, \citenamefont {Lindstr\o{}m}, \citenamefont {Meisel}, \citenamefont {Staufer} \emph {et~al.}}]{Bohlen2022}%
  \BibitemOpen
  \bibfield  {author} {\bibinfo {author} {\bibfnamefont {S.}~\bibnamefont {Bohlen}}, \bibinfo {author} {\bibfnamefont {T.}~\bibnamefont {Br\"ummer}}, \bibinfo {author} {\bibfnamefont {F.}~\bibnamefont {Gr\"uner}}, \bibinfo {author} {\bibfnamefont {C.~A.}\ \bibnamefont {Lindstr\o{}m}}, \bibinfo {author} {\bibfnamefont {M.}~\bibnamefont {Meisel}}, \bibinfo {author} {\bibfnamefont {T.}~\bibnamefont {Staufer}}, \emph {et~al.},\ }\bibfield  {title} {\bibinfo {title} {In situ measurement of electron energy evolution in a laser-plasma accelerator},\ }\href {https://doi.org/10.1103/PhysRevLett.129.244801} {\bibfield  {journal} {\bibinfo  {journal} {Phys. Rev. Lett.}\ }\textbf {\bibinfo {volume} {129}},\ \bibinfo {pages} {244801} (\bibinfo {year} {2022})}\BibitemShut {NoStop}%
\bibitem [{\citenamefont {Veksler}(1956)}]{Veksler:1956pxa}%
  \BibitemOpen
  \bibfield  {author} {\bibinfo {author} {\bibfnamefont {V.~I.}\ \bibnamefont {Veksler}},\ }\bibfield  {title} {\bibinfo {title} {{Coherent principle of acceleration of charged particles}},\ }in\ \href {https://s3.cern.ch/inspire-prod-files-b/b20cafeae29613c9c573b4cb95f74131} {\emph {\bibinfo {booktitle} {{CERN Symposium on High-Energy Accelerators and Pion Physics}}}}\ (\bibinfo {year} {1956})\ p.~\bibinfo {pages} {80}\BibitemShut {NoStop}%
\bibitem [{\citenamefont {Fainberg}(1956)}]{Fainberg:1956qxa}%
  \BibitemOpen
  \bibfield  {author} {\bibinfo {author} {\bibfnamefont {I.~B.}\ \bibnamefont {Fainberg}},\ }\bibfield  {title} {\bibinfo {title} {{The Use Of Plasma Waveguides As Accelerating Structures In Linear Accelerators}},\ }in\ \href {https://s3.cern.ch/inspire-prod-files-6/6e0714d7648114855b3740d8d984d0c4} {\emph {\bibinfo {booktitle} {{CERN Symposium on High-Energy Accelerators and Pion Physics}}}}\ (\bibinfo {year} {1956})\ p.~\bibinfo {pages} {84}\BibitemShut {NoStop}%
\bibitem [{\citenamefont {Tajima}\ and\ \citenamefont {Dawson}(1979)}]{Tajima:1979bn}%
  \BibitemOpen
  \bibfield  {author} {\bibinfo {author} {\bibfnamefont {T.}~\bibnamefont {Tajima}}\ and\ \bibinfo {author} {\bibfnamefont {J.~M.}\ \bibnamefont {Dawson}},\ }\bibfield  {title} {\bibinfo {title} {{Laser electron accelerator}},\ }\href {https://doi.org/10.1103/PhysRevLett.43.267} {\bibfield  {journal} {\bibinfo  {journal} {Phys. Rev. Lett.}\ }\textbf {\bibinfo {volume} {43}},\ \bibinfo {pages} {267} (\bibinfo {year} {1979})}\BibitemShut {NoStop}%
\bibitem [{\citenamefont {Chen}\ and\ \citenamefont {Dawson}(1985)}]{Chen:1985ft}%
  \BibitemOpen
  \bibfield  {author} {\bibinfo {author} {\bibfnamefont {P.}~\bibnamefont {Chen}}\ and\ \bibinfo {author} {\bibfnamefont {J.~M.}\ \bibnamefont {Dawson}},\ }\bibfield  {title} {\bibinfo {title} {{The plasma wake field accelerator}},\ }\href {https://doi.org/10.1063/1.35301} {\bibfield  {journal} {\bibinfo  {journal} {AIP Conf. Proc.}\ }\textbf {\bibinfo {volume} {130}},\ \bibinfo {pages} {201} (\bibinfo {year} {1985})}\BibitemShut {NoStop}%
\bibitem [{\citenamefont {Ruth}\ \emph {et~al.}(1985)\citenamefont {Ruth}, \citenamefont {Chao}, \citenamefont {Morton},\ and\ \citenamefont {Wilson}}]{Ruth:1984pz}%
  \BibitemOpen
  \bibfield  {author} {\bibinfo {author} {\bibfnamefont {R.~D.}\ \bibnamefont {Ruth}}, \bibinfo {author} {\bibfnamefont {A.~W.}\ \bibnamefont {Chao}}, \bibinfo {author} {\bibfnamefont {P.~L.}\ \bibnamefont {Morton}},\ and\ \bibinfo {author} {\bibfnamefont {P.~B.}\ \bibnamefont {Wilson}},\ }\bibfield  {title} {\bibinfo {title} {{A plasma wake field accelerator}},\ }\href {https://cds.cern.ch/record/157249/files/p171.pdf} {\bibfield  {journal} {\bibinfo  {journal} {Part. Accel.}\ }\textbf {\bibinfo {volume} {17}},\ \bibinfo {pages} {171} (\bibinfo {year} {1985})}\BibitemShut {NoStop}%
\bibitem [{\citenamefont {Litos}\ \emph {et~al.}(2014)\citenamefont {Litos} \emph {et~al.}}]{Litos:2014yqa}%
  \BibitemOpen
  \bibfield  {author} {\bibinfo {author} {\bibfnamefont {M.}~\bibnamefont {Litos}} \emph {et~al.},\ }\bibfield  {title} {\bibinfo {title} {{High-efficiency acceleration of an electron beam in a plasma wakefield accelerator}},\ }\href {https://doi.org/10.1038/nature13882} {\bibfield  {journal} {\bibinfo  {journal} {Nature}\ }\textbf {\bibinfo {volume} {515}},\ \bibinfo {pages} {92} (\bibinfo {year} {2014})}\BibitemShut {NoStop}%
\bibitem [{\citenamefont {Corde}\ \emph {et~al.}(2015)\citenamefont {Corde}, \citenamefont {Adli}, \citenamefont {Allen}, \citenamefont {An}, \citenamefont {Clarke}, \citenamefont {Clayton} \emph {et~al.}}]{Corde:2015zxa}%
  \BibitemOpen
  \bibfield  {author} {\bibinfo {author} {\bibfnamefont {S.}~\bibnamefont {Corde}}, \bibinfo {author} {\bibfnamefont {E.}~\bibnamefont {Adli}}, \bibinfo {author} {\bibfnamefont {J.~M.}\ \bibnamefont {Allen}}, \bibinfo {author} {\bibfnamefont {W.}~\bibnamefont {An}}, \bibinfo {author} {\bibfnamefont {C.~I.}\ \bibnamefont {Clarke}}, \bibinfo {author} {\bibfnamefont {C.~E.}\ \bibnamefont {Clayton}}, \emph {et~al.},\ }\bibfield  {title} {\bibinfo {title} {{Multi-gigaelectronvolt acceleration of positrons in a self-loaded plasma wakefield}},\ }\href {https://doi.org/10.1038/nature14890} {\bibfield  {journal} {\bibinfo  {journal} {Nature}\ }\textbf {\bibinfo {volume} {524}},\ \bibinfo {pages} {442} (\bibinfo {year} {2015})}\BibitemShut {NoStop}%
\bibitem [{\citenamefont {Keinigs}\ and\ \citenamefont {Jones}(1987)}]{Keinigs:1986sk}%
  \BibitemOpen
  \bibfield  {author} {\bibinfo {author} {\bibfnamefont {R.}~\bibnamefont {Keinigs}}\ and\ \bibinfo {author} {\bibfnamefont {M.~E.}\ \bibnamefont {Jones}},\ }\bibfield  {title} {\bibinfo {title} {{Two‐dimensional dynamics of the plasma wakefield accelerator}},\ }\href {https://doi.org/10.1063/1.866183} {\bibfield  {journal} {\bibinfo  {journal} {The Physics of Fluids}\ }\textbf {\bibinfo {volume} {30}},\ \bibinfo {pages} {252} (\bibinfo {year} {1987})}\BibitemShut {NoStop}%
\bibitem [{\citenamefont {Rosenzweig}\ \emph {et~al.}(1991)\citenamefont {Rosenzweig}, \citenamefont {Breizman}, \citenamefont {Katsouleas},\ and\ \citenamefont {Su}}]{Rosenzweig:1991yx}%
  \BibitemOpen
  \bibfield  {author} {\bibinfo {author} {\bibfnamefont {J.~B.}\ \bibnamefont {Rosenzweig}}, \bibinfo {author} {\bibfnamefont {B.}~\bibnamefont {Breizman}}, \bibinfo {author} {\bibfnamefont {T.~C.}\ \bibnamefont {Katsouleas}},\ and\ \bibinfo {author} {\bibfnamefont {J.~J.}\ \bibnamefont {Su}},\ }\bibfield  {title} {\bibinfo {title} {{Acceleration and focusing of electrons in two-dimensional nonlinear plasma wake fields}},\ }\href {https://doi.org/10.1103/PhysRevA.44.R6189} {\bibfield  {journal} {\bibinfo  {journal} {Phys. Rev. A}\ }\textbf {\bibinfo {volume} {44}},\ \bibinfo {pages} {R6189} (\bibinfo {year} {1991})}\BibitemShut {NoStop}%
\bibitem [{\citenamefont {Floettmann}(2003)}]{Floettmann:2003}%
  \BibitemOpen
  \bibfield  {author} {\bibinfo {author} {\bibfnamefont {K.}~\bibnamefont {Floettmann}},\ }\bibfield  {title} {\bibinfo {title} {Some basic features of the beam emittance},\ }\href {https://doi.org/10.1103/PhysRevSTAB.6.034202} {\bibfield  {journal} {\bibinfo  {journal} {Phys. Rev. ST Accel. Beams}\ }\textbf {\bibinfo {volume} {6}},\ \bibinfo {pages} {034202} (\bibinfo {year} {2003})}\BibitemShut {NoStop}%
\bibitem [{\citenamefont {Katsouleas}\ \emph {et~al.}(1987)\citenamefont {Katsouleas}, \citenamefont {Wilks}, \citenamefont {Chen}, \citenamefont {Dawson},\ and\ \citenamefont {Su}}]{Katsouleas:1987yd}%
  \BibitemOpen
  \bibfield  {author} {\bibinfo {author} {\bibfnamefont {T.~C.}\ \bibnamefont {Katsouleas}}, \bibinfo {author} {\bibfnamefont {S.}~\bibnamefont {Wilks}}, \bibinfo {author} {\bibfnamefont {P.}~\bibnamefont {Chen}}, \bibinfo {author} {\bibfnamefont {J.~M.}\ \bibnamefont {Dawson}},\ and\ \bibinfo {author} {\bibfnamefont {J.~J.}\ \bibnamefont {Su}},\ }\bibfield  {title} {\bibinfo {title} {{Beam Loading in Plasma Accelerators}},\ }\href {https://s3.cern.ch/inspire-prod-files-3/30e3b8e467eb7298ce19d6bbc95b457f} {\bibfield  {journal} {\bibinfo  {journal} {Part. Accel.}\ }\textbf {\bibinfo {volume} {22}},\ \bibinfo {pages} {81} (\bibinfo {year} {1987})}\BibitemShut {NoStop}%
\bibitem [{\citenamefont {Tzoufras}\ \emph {et~al.}(2008)\citenamefont {Tzoufras}, \citenamefont {Lu}, \citenamefont {Tsung}, \citenamefont {Huang}, \citenamefont {Mori}, \citenamefont {Katsouleas}, \citenamefont {Vieira}, \citenamefont {Fonseca},\ and\ \citenamefont {Silva}}]{tzoufras_2008}%
  \BibitemOpen
  \bibfield  {author} {\bibinfo {author} {\bibfnamefont {M.}~\bibnamefont {Tzoufras}}, \bibinfo {author} {\bibfnamefont {W.}~\bibnamefont {Lu}}, \bibinfo {author} {\bibfnamefont {F.~S.}\ \bibnamefont {Tsung}}, \bibinfo {author} {\bibfnamefont {C.}~\bibnamefont {Huang}}, \bibinfo {author} {\bibfnamefont {W.~B.}\ \bibnamefont {Mori}}, \bibinfo {author} {\bibfnamefont {T.}~\bibnamefont {Katsouleas}}, \bibinfo {author} {\bibfnamefont {J.}~\bibnamefont {Vieira}}, \bibinfo {author} {\bibfnamefont {R.~A.}\ \bibnamefont {Fonseca}},\ and\ \bibinfo {author} {\bibfnamefont {L.~O.}\ \bibnamefont {Silva}},\ }\bibfield  {title} {\bibinfo {title} {Beam loading in the nonlinear regime of plasma-based acceleration},\ }\href {https://doi.org/10.1103/PhysRevLett.101.145002} {\bibfield  {journal} {\bibinfo  {journal} {Phys. Rev. Lett.}\ }\textbf {\bibinfo {volume} {101}},\ \bibinfo {pages} {145002} (\bibinfo {year} {2008})}\BibitemShut {NoStop}%
\bibitem [{\citenamefont {Rosenzweig}\ \emph {et~al.}(1988)\citenamefont {Rosenzweig}, \citenamefont {Cline}, \citenamefont {Cole}, \citenamefont {Figueroa}, \citenamefont {Gai}, \citenamefont {Konecny}, \citenamefont {Norem}, \citenamefont {Schoessow},\ and\ \citenamefont {Simpson}}]{Rosenzweig_1988}%
  \BibitemOpen
  \bibfield  {author} {\bibinfo {author} {\bibfnamefont {J.~B.}\ \bibnamefont {Rosenzweig}}, \bibinfo {author} {\bibfnamefont {D.~B.}\ \bibnamefont {Cline}}, \bibinfo {author} {\bibfnamefont {B.}~\bibnamefont {Cole}}, \bibinfo {author} {\bibfnamefont {H.}~\bibnamefont {Figueroa}}, \bibinfo {author} {\bibfnamefont {W.}~\bibnamefont {Gai}}, \bibinfo {author} {\bibfnamefont {R.}~\bibnamefont {Konecny}}, \bibinfo {author} {\bibfnamefont {J.}~\bibnamefont {Norem}}, \bibinfo {author} {\bibfnamefont {P.}~\bibnamefont {Schoessow}},\ and\ \bibinfo {author} {\bibfnamefont {J.}~\bibnamefont {Simpson}},\ }\bibfield  {title} {\bibinfo {title} {Experimental observation of plasma wake-field acceleration},\ }\href {https://doi.org/10.1103/PhysRevLett.61.98} {\bibfield  {journal} {\bibinfo  {journal} {Phys. Rev. Lett.}\ }\textbf {\bibinfo {volume} {61}},\ \bibinfo {pages} {98} (\bibinfo {year} {1988})}\BibitemShut {NoStop}%
\bibitem [{\citenamefont {Modena}\ \emph {et~al.}(1995)\citenamefont {Modena}, \citenamefont {Najmudin}, \citenamefont {Dangor}, \citenamefont {Clayton}, \citenamefont {Marsh}, \citenamefont {Joshi} \emph {et~al.}}]{Modena1995}%
  \BibitemOpen
  \bibfield  {author} {\bibinfo {author} {\bibfnamefont {A.}~\bibnamefont {Modena}}, \bibinfo {author} {\bibfnamefont {Z.}~\bibnamefont {Najmudin}}, \bibinfo {author} {\bibfnamefont {A.~E.}\ \bibnamefont {Dangor}}, \bibinfo {author} {\bibfnamefont {C.~E.}\ \bibnamefont {Clayton}}, \bibinfo {author} {\bibfnamefont {K.~A.}\ \bibnamefont {Marsh}}, \bibinfo {author} {\bibfnamefont {C.}~\bibnamefont {Joshi}}, \emph {et~al.},\ }\bibfield  {title} {\bibinfo {title} {{Electron acceleration from the breaking of relativistic plasma waves}},\ }\href {https://doi.org/10.1038/377606a0} {\bibfield  {journal} {\bibinfo  {journal} {Nature}\ }\textbf {\bibinfo {volume} {377}},\ \bibinfo {pages} {606} (\bibinfo {year} {1995})}\BibitemShut {NoStop}%
\bibitem [{\citenamefont {Barov}\ \emph {et~al.}(2000)\citenamefont {Barov}, \citenamefont {Rosenzweig}, \citenamefont {Conde}, \citenamefont {Gai},\ and\ \citenamefont {Power}}]{Barov_2000}%
  \BibitemOpen
  \bibfield  {author} {\bibinfo {author} {\bibfnamefont {N.}~\bibnamefont {Barov}}, \bibinfo {author} {\bibfnamefont {J.~B.}\ \bibnamefont {Rosenzweig}}, \bibinfo {author} {\bibfnamefont {M.~E.}\ \bibnamefont {Conde}}, \bibinfo {author} {\bibfnamefont {W.}~\bibnamefont {Gai}},\ and\ \bibinfo {author} {\bibfnamefont {J.~G.}\ \bibnamefont {Power}},\ }\bibfield  {title} {\bibinfo {title} {Observation of plasma wakefield acceleration in the underdense regime},\ }\href {https://doi.org/10.1103/PhysRevSTAB.3.011301} {\bibfield  {journal} {\bibinfo  {journal} {Phys. Rev. ST Accel. Beams}\ }\textbf {\bibinfo {volume} {3}},\ \bibinfo {pages} {011301} (\bibinfo {year} {2000})}\BibitemShut {NoStop}%
\bibitem [{\citenamefont {Blumenfeld}\ \emph {et~al.}(2007)\citenamefont {Blumenfeld} \emph {et~al.}}]{Blumenfeld:2007ph}%
  \BibitemOpen
  \bibfield  {author} {\bibinfo {author} {\bibfnamefont {I.}~\bibnamefont {Blumenfeld}} \emph {et~al.},\ }\bibfield  {title} {\bibinfo {title} {{Energy doubling of 42 GeV electrons in a metre-scale plasma wakefield accelerator}},\ }\href {https://doi.org/10.1038/nature05538} {\bibfield  {journal} {\bibinfo  {journal} {Nature}\ }\textbf {\bibinfo {volume} {445}},\ \bibinfo {pages} {741} (\bibinfo {year} {2007})}\BibitemShut {NoStop}%
\bibitem [{\citenamefont {Lindstr\o{}m}\ \emph {et~al.}(2021)\citenamefont {Lindstr\o{}m} \emph {et~al.}}]{Lindstrom:2021tkb}%
  \BibitemOpen
  \bibfield  {author} {\bibinfo {author} {\bibfnamefont {C.~A.}\ \bibnamefont {Lindstr\o{}m}} \emph {et~al.},\ }\bibfield  {title} {\bibinfo {title} {{Energy-Spread Preservation and High Efficiency in a Plasma-Wakefield Accelerator}},\ }\href {https://doi.org/10.1103/PhysRevLett.126.014801} {\bibfield  {journal} {\bibinfo  {journal} {Phys. Rev. Lett.}\ }\textbf {\bibinfo {volume} {126}},\ \bibinfo {pages} {014801} (\bibinfo {year} {2021})}\BibitemShut {NoStop}%
\bibitem [{\citenamefont {Geddes}\ \emph {et~al.}(2004)\citenamefont {Geddes}, \citenamefont {Toth}, \citenamefont {van Tilborg}, \citenamefont {Esarey}, \citenamefont {Schroeder}, \citenamefont {Bruhwiler}, \citenamefont {Nieter}, \citenamefont {Cary},\ and\ \citenamefont {Leemans}}]{Geddes:2004tb}%
  \BibitemOpen
  \bibfield  {author} {\bibinfo {author} {\bibfnamefont {C.~G.~R.}\ \bibnamefont {Geddes}}, \bibinfo {author} {\bibfnamefont {C.}~\bibnamefont {Toth}}, \bibinfo {author} {\bibfnamefont {J.}~\bibnamefont {van Tilborg}}, \bibinfo {author} {\bibfnamefont {E.}~\bibnamefont {Esarey}}, \bibinfo {author} {\bibfnamefont {C.~B.}\ \bibnamefont {Schroeder}}, \bibinfo {author} {\bibfnamefont {D.}~\bibnamefont {Bruhwiler}}, \bibinfo {author} {\bibfnamefont {C.}~\bibnamefont {Nieter}}, \bibinfo {author} {\bibfnamefont {J.}~\bibnamefont {Cary}},\ and\ \bibinfo {author} {\bibfnamefont {W.~P.}\ \bibnamefont {Leemans}},\ }\bibfield  {title} {\bibinfo {title} {{High-quality electron beams from a laser wakefield accelerator using plasma-channel guiding}},\ }\href {https://doi.org/10.1038/nature02900} {\bibfield  {journal} {\bibinfo  {journal} {Nature}\ }\textbf {\bibinfo {volume} {431}},\ \bibinfo {pages} {538} (\bibinfo {year} {2004})}\BibitemShut {NoStop}%
\bibitem [{\citenamefont {Faure}\ \emph {et~al.}(2004)\citenamefont {Faure}, \citenamefont {Glinec}, \citenamefont {Pukhov}, \citenamefont {Kiselev}, \citenamefont {Gordienko}, \citenamefont {Lefebvre}, \citenamefont {Rousseau}, \citenamefont {Burgy},\ and\ \citenamefont {Malka}}]{Faure:2004tc}%
  \BibitemOpen
  \bibfield  {author} {\bibinfo {author} {\bibfnamefont {J.}~\bibnamefont {Faure}}, \bibinfo {author} {\bibfnamefont {Y.}~\bibnamefont {Glinec}}, \bibinfo {author} {\bibfnamefont {A.}~\bibnamefont {Pukhov}}, \bibinfo {author} {\bibfnamefont {S.}~\bibnamefont {Kiselev}}, \bibinfo {author} {\bibfnamefont {S.}~\bibnamefont {Gordienko}}, \bibinfo {author} {\bibfnamefont {E.}~\bibnamefont {Lefebvre}}, \bibinfo {author} {\bibfnamefont {J.~P.}\ \bibnamefont {Rousseau}}, \bibinfo {author} {\bibfnamefont {F.}~\bibnamefont {Burgy}},\ and\ \bibinfo {author} {\bibfnamefont {V.}~\bibnamefont {Malka}},\ }\bibfield  {title} {\bibinfo {title} {{A laser-plasma accelerator producing monoenergetic electron beams}},\ }\href {https://doi.org/10.1038/nature02963} {\bibfield  {journal} {\bibinfo  {journal} {Nature}\ }\textbf {\bibinfo {volume} {431}},\ \bibinfo {pages} {541} (\bibinfo {year} {2004})}\BibitemShut {NoStop}%
\bibitem [{\citenamefont {Mangles}\ \emph {et~al.}(2004)\citenamefont {Mangles}, \citenamefont {Murphy}, \citenamefont {Najmudin}, \citenamefont {Thomas}, \citenamefont {Collier}, \citenamefont {Dangor} \emph {et~al.}}]{Mangles:2004ta}%
  \BibitemOpen
  \bibfield  {author} {\bibinfo {author} {\bibfnamefont {S.~P.~D.}\ \bibnamefont {Mangles}}, \bibinfo {author} {\bibfnamefont {C.~D.}\ \bibnamefont {Murphy}}, \bibinfo {author} {\bibfnamefont {Z.}~\bibnamefont {Najmudin}}, \bibinfo {author} {\bibfnamefont {A.~G.~R.}\ \bibnamefont {Thomas}}, \bibinfo {author} {\bibfnamefont {J.~L.}\ \bibnamefont {Collier}}, \bibinfo {author} {\bibfnamefont {A.~E.}\ \bibnamefont {Dangor}}, \emph {et~al.},\ }\bibfield  {title} {\bibinfo {title} {{Monoenergetic beams of relativistic electrons from intense laser-plasma interactions}},\ }\href {https://doi.org/10.1038/nature02939} {\bibfield  {journal} {\bibinfo  {journal} {Nature}\ }\textbf {\bibinfo {volume} {431}},\ \bibinfo {pages} {535} (\bibinfo {year} {2004})}\BibitemShut {NoStop}%
\bibitem [{\citenamefont {Gonsalves}\ \emph {et~al.}(2019)\citenamefont {Gonsalves}, \citenamefont {Nakamura}, \citenamefont {Daniels}, \citenamefont {Benedetti}, \citenamefont {Pieronek}, \citenamefont {de~Raadt} \emph {et~al.}}]{Gonsalves_2019}%
  \BibitemOpen
  \bibfield  {author} {\bibinfo {author} {\bibfnamefont {A.~J.}\ \bibnamefont {Gonsalves}}, \bibinfo {author} {\bibfnamefont {K.}~\bibnamefont {Nakamura}}, \bibinfo {author} {\bibfnamefont {J.}~\bibnamefont {Daniels}}, \bibinfo {author} {\bibfnamefont {C.}~\bibnamefont {Benedetti}}, \bibinfo {author} {\bibfnamefont {C.}~\bibnamefont {Pieronek}}, \bibinfo {author} {\bibfnamefont {T.~C.~H.}\ \bibnamefont {de~Raadt}}, \emph {et~al.},\ }\bibfield  {title} {\bibinfo {title} {Petawatt laser guiding and electron beam acceleration to 8 gev in a laser-heated capillary discharge waveguide},\ }\href {https://doi.org/10.1103/PhysRevLett.122.084801} {\bibfield  {journal} {\bibinfo  {journal} {Phys. Rev. Lett.}\ }\textbf {\bibinfo {volume} {122}},\ \bibinfo {pages} {084801} (\bibinfo {year} {2019})}\BibitemShut {NoStop}%
\bibitem [{\citenamefont {Wang}\ \emph {et~al.}(2021)\citenamefont {Wang}, \citenamefont {Feng}, \citenamefont {Ke}, \citenamefont {Yu}, \citenamefont {Xu}, \citenamefont {Qi} \emph {et~al.}}]{wang_2021_fel}%
  \BibitemOpen
  \bibfield  {author} {\bibinfo {author} {\bibfnamefont {W.}~\bibnamefont {Wang}}, \bibinfo {author} {\bibfnamefont {K.}~\bibnamefont {Feng}}, \bibinfo {author} {\bibfnamefont {L.}~\bibnamefont {Ke}}, \bibinfo {author} {\bibfnamefont {C.}~\bibnamefont {Yu}}, \bibinfo {author} {\bibfnamefont {Y.}~\bibnamefont {Xu}}, \bibinfo {author} {\bibfnamefont {R.}~\bibnamefont {Qi}}, \emph {et~al.},\ }\bibfield  {title} {\bibinfo {title} {{Free-electron lasing at 27 nanometres based on a laser wakefield accelerator}},\ }\href {https://doi.org/10.1038/s41586-021-03678-x} {\bibfield  {journal} {\bibinfo  {journal} {Nature}\ }\textbf {\bibinfo {volume} {595}},\ \bibinfo {pages} {516} (\bibinfo {year} {2021})}\BibitemShut {NoStop}%
\bibitem [{\citenamefont {Labat}\ \emph {et~al.}(2023)\citenamefont {Labat}, \citenamefont {Cabadag}, \citenamefont {Ghaith}, \citenamefont {Irman}, \citenamefont {Berlioux},\ and\ \citenamefont {Berteaud}}]{labat_2022}%
  \BibitemOpen
  \bibfield  {author} {\bibinfo {author} {\bibfnamefont {M.}~\bibnamefont {Labat}}, \bibinfo {author} {\bibfnamefont {J.}~\bibnamefont {Cabadag}}, \bibinfo {author} {\bibfnamefont {A.}~\bibnamefont {Ghaith}}, \bibinfo {author} {\bibfnamefont {A.}~\bibnamefont {Irman}}, \bibinfo {author} {\bibfnamefont {A.}~\bibnamefont {Berlioux}},\ and\ \bibinfo {author} {\bibfnamefont {P.~o.}\ \bibnamefont {Berteaud}},\ }\bibfield  {title} {\bibinfo {title} {{Seeded free-electron laser driven by a compact laser plasma accelerator}},\ }\href {https://doi.org/10.1038/s41566-022-01104-w} {\bibfield  {journal} {\bibinfo  {journal} {Nat. Photon.}\ }\textbf {\bibinfo {volume} {17}},\ \bibinfo {pages} {150} (\bibinfo {year} {2023})}\BibitemShut {NoStop}%
\bibitem [{\citenamefont {{Pe{\~n}a}}\ \emph {et~al.}(2023)\citenamefont {{Pe{\~n}a}}, \citenamefont {{Lindstr{\o}m}}, \citenamefont {{Beinortait{\.{e}}}}, \citenamefont {{Bj{\"o}rklund Svensson}}, \citenamefont {{Boulton}}, \citenamefont {{Diederichs}} \emph {et~al.}}]{Pena_2023}%
  \BibitemOpen
  \bibfield  {author} {\bibinfo {author} {\bibfnamefont {F.}~\bibnamefont {{Pe{\~n}a}}}, \bibinfo {author} {\bibfnamefont {C.~A.}\ \bibnamefont {{Lindstr{\o}m}}}, \bibinfo {author} {\bibfnamefont {J.}~\bibnamefont {{Beinortait{\.{e}}}}}, \bibinfo {author} {\bibfnamefont {J.}~\bibnamefont {{Bj{\"o}rklund Svensson}}}, \bibinfo {author} {\bibfnamefont {L.}~\bibnamefont {{Boulton}}}, \bibinfo {author} {\bibfnamefont {S.}~\bibnamefont {{Diederichs}}}, \emph {et~al.},\ }\bibfield  {title} {\bibinfo {title} {{Energy Depletion and Re-Acceleration of Driver Electrons in a Plasma-Wakefield Accelerator}},\ }\href@noop {} {\bibfield  {journal} {\bibinfo  {journal} {arXiv e-prints}\ } (\bibinfo {year} {2023})},\ \Eprint {https://arxiv.org/abs/2305.09581} {arXiv:2305.09581} \BibitemShut {NoStop}%
\bibitem [{\citenamefont {Lindstr\o{}m}(2021)}]{Lindstrom:2021prab}%
  \BibitemOpen
  \bibfield  {author} {\bibinfo {author} {\bibfnamefont {C.~A.}\ \bibnamefont {Lindstr\o{}m}},\ }\bibfield  {title} {\bibinfo {title} {Staging of plasma-wakefield accelerators},\ }\href {https://doi.org/10.1103/PhysRevAccelBeams.24.014801} {\bibfield  {journal} {\bibinfo  {journal} {Phys. Rev. Accel. Beams}\ }\textbf {\bibinfo {volume} {24}},\ \bibinfo {pages} {014801} (\bibinfo {year} {2021})}\BibitemShut {NoStop}%
\bibitem [{\citenamefont {Steinke}\ \emph {et~al.}(2016)\citenamefont {Steinke} \emph {et~al.}}]{Steinke:2016cyx}%
  \BibitemOpen
  \bibfield  {author} {\bibinfo {author} {\bibfnamefont {S.}~\bibnamefont {Steinke}} \emph {et~al.},\ }\bibfield  {title} {\bibinfo {title} {{Multistage coupling of independent laser-plasma accelerators}},\ }\href {https://doi.org/10.1038/nature16525} {\bibfield  {journal} {\bibinfo  {journal} {Nature}\ }\textbf {\bibinfo {volume} {530}},\ \bibinfo {pages} {190} (\bibinfo {year} {2016})}\BibitemShut {NoStop}%
\bibitem [{\citenamefont {Rosenzweig}\ \emph {et~al.}(2005)\citenamefont {Rosenzweig}, \citenamefont {Cook}, \citenamefont {Scott}, \citenamefont {Thompson},\ and\ \citenamefont {Yoder}}]{Rosenzweig_2005}%
  \BibitemOpen
  \bibfield  {author} {\bibinfo {author} {\bibfnamefont {J.~B.}\ \bibnamefont {Rosenzweig}}, \bibinfo {author} {\bibfnamefont {A.~M.}\ \bibnamefont {Cook}}, \bibinfo {author} {\bibfnamefont {A.}~\bibnamefont {Scott}}, \bibinfo {author} {\bibfnamefont {M.~C.}\ \bibnamefont {Thompson}},\ and\ \bibinfo {author} {\bibfnamefont {R.~B.}\ \bibnamefont {Yoder}},\ }\bibfield  {title} {\bibinfo {title} {Effects of ion motion in intense beam-driven plasma wakefield accelerators},\ }\href {https://doi.org/10.1103/PhysRevLett.95.195002} {\bibfield  {journal} {\bibinfo  {journal} {Phys. Rev. Lett.}\ }\textbf {\bibinfo {volume} {95}},\ \bibinfo {pages} {195002} (\bibinfo {year} {2005})}\BibitemShut {NoStop}%
\bibitem [{\citenamefont {An}\ \emph {et~al.}(2017)\citenamefont {An}, \citenamefont {Lu}, \citenamefont {Huang}, \citenamefont {Xu}, \citenamefont {Hogan}, \citenamefont {Joshi},\ and\ \citenamefont {Mori}}]{An_2017}%
  \BibitemOpen
  \bibfield  {author} {\bibinfo {author} {\bibfnamefont {W.}~\bibnamefont {An}}, \bibinfo {author} {\bibfnamefont {W.}~\bibnamefont {Lu}}, \bibinfo {author} {\bibfnamefont {C.}~\bibnamefont {Huang}}, \bibinfo {author} {\bibfnamefont {X.}~\bibnamefont {Xu}}, \bibinfo {author} {\bibfnamefont {M.~J.}\ \bibnamefont {Hogan}}, \bibinfo {author} {\bibfnamefont {C.}~\bibnamefont {Joshi}},\ and\ \bibinfo {author} {\bibfnamefont {W.~B.}\ \bibnamefont {Mori}},\ }\bibfield  {title} {\bibinfo {title} {Ion motion induced emittance growth of matched electron beams in plasma wakefields},\ }\href {https://doi.org/10.1103/PhysRevLett.118.244801} {\bibfield  {journal} {\bibinfo  {journal} {Phys. Rev. Lett.}\ }\textbf {\bibinfo {volume} {118}},\ \bibinfo {pages} {244801} (\bibinfo {year} {2017})}\BibitemShut {NoStop}%
\bibitem [{\citenamefont {Whittum}\ \emph {et~al.}(1991)\citenamefont {Whittum}, \citenamefont {Sharp}, \citenamefont {Yu}, \citenamefont {Lampe},\ and\ \citenamefont {Joyce}}]{Whittum:1990cy}%
  \BibitemOpen
  \bibfield  {author} {\bibinfo {author} {\bibfnamefont {D.~H.}\ \bibnamefont {Whittum}}, \bibinfo {author} {\bibfnamefont {W.~M.}\ \bibnamefont {Sharp}}, \bibinfo {author} {\bibfnamefont {S.~S.}\ \bibnamefont {Yu}}, \bibinfo {author} {\bibfnamefont {M.}~\bibnamefont {Lampe}},\ and\ \bibinfo {author} {\bibfnamefont {G.}~\bibnamefont {Joyce}},\ }\bibfield  {title} {\bibinfo {title} {{Electron-hose instability in the ion-focused regime}},\ }\href {https://doi.org/10.1103/PhysRevLett.67.991} {\bibfield  {journal} {\bibinfo  {journal} {Phys. Rev. Lett.}\ }\textbf {\bibinfo {volume} {67}},\ \bibinfo {pages} {991} (\bibinfo {year} {1991})}\BibitemShut {NoStop}%
\bibitem [{\citenamefont {Lebedev}\ \emph {et~al.}(2017)\citenamefont {Lebedev}, \citenamefont {Burov},\ and\ \citenamefont {Nagaitsev}}]{Lebedev:2017dcs}%
  \BibitemOpen
  \bibfield  {author} {\bibinfo {author} {\bibfnamefont {V.}~\bibnamefont {Lebedev}}, \bibinfo {author} {\bibfnamefont {A.}~\bibnamefont {Burov}},\ and\ \bibinfo {author} {\bibfnamefont {S.}~\bibnamefont {Nagaitsev}},\ }\bibfield  {title} {\bibinfo {title} {{Efficiency versus instability in plasma accelerators}},\ }\href {https://doi.org/10.1103/PhysRevAccelBeams.20.121301} {\bibfield  {journal} {\bibinfo  {journal} {Phys. Rev. Accel. Beams}\ }\textbf {\bibinfo {volume} {20}},\ \bibinfo {pages} {121301} (\bibinfo {year} {2017})}\BibitemShut {NoStop}%
\bibitem [{\citenamefont {Lehe}\ \emph {et~al.}(2017)\citenamefont {Lehe}, \citenamefont {Schroeder}, \citenamefont {Vay}, \citenamefont {Esarey},\ and\ \citenamefont {Leemans}}]{Lehe_2017}%
  \BibitemOpen
  \bibfield  {author} {\bibinfo {author} {\bibfnamefont {R.}~\bibnamefont {Lehe}}, \bibinfo {author} {\bibfnamefont {C.~B.}\ \bibnamefont {Schroeder}}, \bibinfo {author} {\bibfnamefont {J.-L.}\ \bibnamefont {Vay}}, \bibinfo {author} {\bibfnamefont {E.}~\bibnamefont {Esarey}},\ and\ \bibinfo {author} {\bibfnamefont {W.~P.}\ \bibnamefont {Leemans}},\ }\bibfield  {title} {\bibinfo {title} {Saturation of the hosing instability in quasilinear plasma accelerators},\ }\href {https://doi.org/10.1103/PhysRevLett.119.244801} {\bibfield  {journal} {\bibinfo  {journal} {Phys. Rev. Lett.}\ }\textbf {\bibinfo {volume} {119}},\ \bibinfo {pages} {244801} (\bibinfo {year} {2017})}\BibitemShut {NoStop}%
\bibitem [{\citenamefont {Mehrling}\ \emph {et~al.}(2018)\citenamefont {Mehrling}, \citenamefont {Benedetti}, \citenamefont {Schroeder}, \citenamefont {Esarey},\ and\ \citenamefont {Leemans}}]{mehrling_2018}%
  \BibitemOpen
  \bibfield  {author} {\bibinfo {author} {\bibfnamefont {T.~J.}\ \bibnamefont {Mehrling}}, \bibinfo {author} {\bibfnamefont {C.}~\bibnamefont {Benedetti}}, \bibinfo {author} {\bibfnamefont {C.~B.}\ \bibnamefont {Schroeder}}, \bibinfo {author} {\bibfnamefont {E.}~\bibnamefont {Esarey}},\ and\ \bibinfo {author} {\bibfnamefont {W.~P.}\ \bibnamefont {Leemans}},\ }\bibfield  {title} {\bibinfo {title} {Suppression of beam hosing in plasma accelerators with ion motion},\ }\href {https://doi.org/10.1103/PhysRevLett.121.264802} {\bibfield  {journal} {\bibinfo  {journal} {Phys. Rev. Lett.}\ }\textbf {\bibinfo {volume} {121}},\ \bibinfo {pages} {264802} (\bibinfo {year} {2018})}\BibitemShut {NoStop}%
\bibitem [{\citenamefont {Mehrling}\ \emph {et~al.}(2019)\citenamefont {Mehrling}, \citenamefont {Fonseca}, \citenamefont {Martinez de~la Ossa},\ and\ \citenamefont {Vieira}}]{mehrling_2019}%
  \BibitemOpen
  \bibfield  {author} {\bibinfo {author} {\bibfnamefont {T.~J.}\ \bibnamefont {Mehrling}}, \bibinfo {author} {\bibfnamefont {R.~A.}\ \bibnamefont {Fonseca}}, \bibinfo {author} {\bibfnamefont {A.}~\bibnamefont {Martinez de~la Ossa}},\ and\ \bibinfo {author} {\bibfnamefont {J.}~\bibnamefont {Vieira}},\ }\bibfield  {title} {\bibinfo {title} {Mechanisms for the mitigation of the hose instability in plasma-wakefield accelerators},\ }\href {https://doi.org/10.1103/PhysRevAccelBeams.22.031302} {\bibfield  {journal} {\bibinfo  {journal} {Phys. Rev. Accel. Beams}\ }\textbf {\bibinfo {volume} {22}},\ \bibinfo {pages} {031302} (\bibinfo {year} {2019})}\BibitemShut {NoStop}%
\bibitem [{\citenamefont {Vieira}\ \emph {et~al.}(2011)\citenamefont {Vieira}, \citenamefont {Huang}, \citenamefont {Mori},\ and\ \citenamefont {Silva}}]{vieira_2011}%
  \BibitemOpen
  \bibfield  {author} {\bibinfo {author} {\bibfnamefont {J.}~\bibnamefont {Vieira}}, \bibinfo {author} {\bibfnamefont {C.-K.}\ \bibnamefont {Huang}}, \bibinfo {author} {\bibfnamefont {W.~B.}\ \bibnamefont {Mori}},\ and\ \bibinfo {author} {\bibfnamefont {L.~O.}\ \bibnamefont {Silva}},\ }\bibfield  {title} {\bibinfo {title} {Polarized beam conditioning in plasma based acceleration},\ }\href {https://doi.org/10.1103/PhysRevSTAB.14.071303} {\bibfield  {journal} {\bibinfo  {journal} {Phys. Rev. ST Accel. Beams}\ }\textbf {\bibinfo {volume} {14}},\ \bibinfo {pages} {071303} (\bibinfo {year} {2011})}\BibitemShut {NoStop}%
\bibitem [{\citenamefont {Cowley}\ \emph {et~al.}(2017)\citenamefont {Cowley} \emph {et~al.}}]{Cowley:2017skm}%
  \BibitemOpen
  \bibfield  {author} {\bibinfo {author} {\bibfnamefont {J.}~\bibnamefont {Cowley}} \emph {et~al.},\ }\bibfield  {title} {\bibinfo {title} {{Excitation and Control of Plasma Wakefields by Multiple Laser Pulses}},\ }\href {https://doi.org/10.1103/PhysRevLett.119.044802} {\bibfield  {journal} {\bibinfo  {journal} {Phys. Rev. Lett.}\ }\textbf {\bibinfo {volume} {119}},\ \bibinfo {pages} {044802} (\bibinfo {year} {2017})}\BibitemShut {NoStop}%
\bibitem [{\citenamefont {D'Arcy}\ \emph {et~al.}(2022)\citenamefont {D'Arcy} \emph {et~al.}}]{DArcy:2022zwq}%
  \BibitemOpen
  \bibfield  {author} {\bibinfo {author} {\bibfnamefont {R.}~\bibnamefont {D'Arcy}} \emph {et~al.},\ }\bibfield  {title} {\bibinfo {title} {{Recovery time of a plasma-wakefield accelerator}},\ }\href {https://doi.org/10.1038/s41586-021-04348-8} {\bibfield  {journal} {\bibinfo  {journal} {Nature}\ }\textbf {\bibinfo {volume} {603}},\ \bibinfo {pages} {58} (\bibinfo {year} {2022})}\BibitemShut {NoStop}%
\bibitem [{\citenamefont {Courant}\ and\ \citenamefont {Snyder}(1958)}]{COURANT19581}%
  \BibitemOpen
  \bibfield  {author} {\bibinfo {author} {\bibfnamefont {E.~D.}\ \bibnamefont {Courant}}\ and\ \bibinfo {author} {\bibfnamefont {H.~S.}\ \bibnamefont {Snyder}},\ }\bibfield  {title} {\bibinfo {title} {Theory of the alternating-gradient synchrotron},\ }\href {https://doi.org/https://doi.org/10.1016/0003-4916(58)90012-5} {\bibfield  {journal} {\bibinfo  {journal} {Ann. Phys.}\ }\textbf {\bibinfo {volume} {3}},\ \bibinfo {pages} {1} (\bibinfo {year} {1958})}\BibitemShut {NoStop}%
\bibitem [{\citenamefont {Dolgashev}\ \emph {et~al.}(2021)\citenamefont {Dolgashev}, \citenamefont {Faillace}, \citenamefont {Spataro}, \citenamefont {Tantawi},\ and\ \citenamefont {Bonifazi}}]{dolgashev_2021}%
  \BibitemOpen
  \bibfield  {author} {\bibinfo {author} {\bibfnamefont {V.~A.}\ \bibnamefont {Dolgashev}}, \bibinfo {author} {\bibfnamefont {L.}~\bibnamefont {Faillace}}, \bibinfo {author} {\bibfnamefont {B.}~\bibnamefont {Spataro}}, \bibinfo {author} {\bibfnamefont {S.}~\bibnamefont {Tantawi}},\ and\ \bibinfo {author} {\bibfnamefont {R.}~\bibnamefont {Bonifazi}},\ }\bibfield  {title} {\bibinfo {title} {High-gradient rf tests of welded $x$-band accelerating cavities},\ }\href {https://doi.org/10.1103/PhysRevAccelBeams.24.081002} {\bibfield  {journal} {\bibinfo  {journal} {Phys. Rev. Accel. Beams}\ }\textbf {\bibinfo {volume} {24}},\ \bibinfo {pages} {081002} (\bibinfo {year} {2021})}\BibitemShut {NoStop}%
\bibitem [{\citenamefont {Hooker}\ \emph {et~al.}(2014)\citenamefont {Hooker}, \citenamefont {Bartolini}, \citenamefont {Mangles}, \citenamefont {T\"unnermann}, \citenamefont {Corner}, \citenamefont {Limpert}, \citenamefont {Seryi},\ and\ \citenamefont {Walczak}}]{Hooker:2014cza}%
  \BibitemOpen
  \bibfield  {author} {\bibinfo {author} {\bibfnamefont {S.~M.}\ \bibnamefont {Hooker}}, \bibinfo {author} {\bibfnamefont {R.}~\bibnamefont {Bartolini}}, \bibinfo {author} {\bibfnamefont {S.~P.~D.}\ \bibnamefont {Mangles}}, \bibinfo {author} {\bibfnamefont {A.}~\bibnamefont {T\"unnermann}}, \bibinfo {author} {\bibfnamefont {L.}~\bibnamefont {Corner}}, \bibinfo {author} {\bibfnamefont {J.}~\bibnamefont {Limpert}}, \bibinfo {author} {\bibfnamefont {A.}~\bibnamefont {Seryi}},\ and\ \bibinfo {author} {\bibfnamefont {R.}~\bibnamefont {Walczak}},\ }\bibfield  {title} {\bibinfo {title} {{Multi-Pulse Laser Wakefield Acceleration: A New Route to Efficient, High-Repetition-Rate Plasma Accelerators and High Flux Radiation Sources}},\ }\href {https://doi.org/10.1088/0953-4075/47/23/234003} {\bibfield  {journal} {\bibinfo  {journal} {J. Phys. B}\ }\textbf {\bibinfo {volume} {47}},\ \bibinfo {pages} {234003} (\bibinfo {year} {2014})}\BibitemShut {NoStop}%
\bibitem [{\citenamefont {Lotov}(2005)}]{Lotov_2005}%
  \BibitemOpen
  \bibfield  {author} {\bibinfo {author} {\bibfnamefont {K.~V.}\ \bibnamefont {Lotov}},\ }\bibfield  {title} {\bibinfo {title} {{Efficient operating mode of the plasma wakefield accelerator}},\ }\href {https://doi.org/10.1063/1.1889444} {\bibfield  {journal} {\bibinfo  {journal} {Phys. Plasmas}\ }\textbf {\bibinfo {volume} {12}},\ \bibinfo {pages} {053105} (\bibinfo {year} {2005})}\BibitemShut {NoStop}%
\bibitem [{\citenamefont {Su}\ \emph {et~al.}(2023)\citenamefont {Su}, \citenamefont {Larson}, \citenamefont {Dalichaouch}, \citenamefont {Li}, \citenamefont {An}, \citenamefont {Hildebrand} \emph {et~al.}}]{Su_2023}%
  \BibitemOpen
  \bibfield  {author} {\bibinfo {author} {\bibfnamefont {Q.}~\bibnamefont {Su}}, \bibinfo {author} {\bibfnamefont {J.}~\bibnamefont {Larson}}, \bibinfo {author} {\bibfnamefont {T.~N.}\ \bibnamefont {Dalichaouch}}, \bibinfo {author} {\bibfnamefont {F.}~\bibnamefont {Li}}, \bibinfo {author} {\bibfnamefont {W.}~\bibnamefont {An}}, \bibinfo {author} {\bibfnamefont {L.}~\bibnamefont {Hildebrand}}, \emph {et~al.},\ }\bibfield  {title} {\bibinfo {title} {{Optimization of transformer ratio and beam loading in a plasma wakefield accelerator with a structure-exploiting algorithm}},\ }\href {https://doi.org/10.1063/5.0142940} {\bibfield  {journal} {\bibinfo  {journal} {Phys. Plasmas}\ }\textbf {\bibinfo {volume} {30}},\ \bibinfo {pages} {053108} (\bibinfo {year} {2023})}\BibitemShut {NoStop}%
\bibitem [{\citenamefont {Mehrling}\ \emph {et~al.}(2017)\citenamefont {Mehrling}, \citenamefont {Fonseca}, \citenamefont {Martinez de~la Ossa},\ and\ \citenamefont {Vieira}}]{mehrling_2017}%
  \BibitemOpen
  \bibfield  {author} {\bibinfo {author} {\bibfnamefont {T.~J.}\ \bibnamefont {Mehrling}}, \bibinfo {author} {\bibfnamefont {R.~A.}\ \bibnamefont {Fonseca}}, \bibinfo {author} {\bibfnamefont {A.}~\bibnamefont {Martinez de~la Ossa}},\ and\ \bibinfo {author} {\bibfnamefont {J.}~\bibnamefont {Vieira}},\ }\bibfield  {title} {\bibinfo {title} {Mitigation of the hose instability in plasma-wakefield accelerators},\ }\href {https://doi.org/10.1103/PhysRevLett.118.174801} {\bibfield  {journal} {\bibinfo  {journal} {Phys. Rev. Lett.}\ }\textbf {\bibinfo {volume} {118}},\ \bibinfo {pages} {174801} (\bibinfo {year} {2017})}\BibitemShut {NoStop}%
\bibitem [{\citenamefont {Zhou}\ \emph {et~al.}(2007)\citenamefont {Zhou}, \citenamefont {Clayton}, \citenamefont {Huang}, \citenamefont {Joshi}, \citenamefont {Lu}, \citenamefont {Marsh} \emph {et~al.}}]{zhou_2007}%
  \BibitemOpen
  \bibfield  {author} {\bibinfo {author} {\bibfnamefont {M.}~\bibnamefont {Zhou}}, \bibinfo {author} {\bibfnamefont {C.~E.}\ \bibnamefont {Clayton}}, \bibinfo {author} {\bibfnamefont {C.}~\bibnamefont {Huang}}, \bibinfo {author} {\bibfnamefont {C.}~\bibnamefont {Joshi}}, \bibinfo {author} {\bibfnamefont {W.}~\bibnamefont {Lu}}, \bibinfo {author} {\bibfnamefont {K.~A.}\ \bibnamefont {Marsh}}, \emph {et~al.},\ }\bibfield  {title} {\bibinfo {title} {Beam head erosion in self-ionized plasma wakefield accelerators},\ }in\ \href {https://doi.org/10.1109/PAC.2007.4440669} {\emph {\bibinfo {booktitle} {Proceedings of PAC07, Albuquerque, NM, USA}}}\ (\bibinfo {year} {2007})\ p.\ \bibinfo {pages} {3064}\BibitemShut {NoStop}%
\bibitem [{\citenamefont {{Blumenfeld}}(2009)}]{blumenfeld_phd}%
  \BibitemOpen
  \bibfield  {author} {\bibinfo {author} {\bibfnamefont {I.}~\bibnamefont {{Blumenfeld}}},\ }\emph {\bibinfo {title} {{Scaling of the longitudinal electric fields and transformer ratio in a non-linear Plasma Wakefield Accelerator}}},\ \href {https://www.proquest.com/openview/3d032ec976c1f2443d2e92a423b11d0a/1?pq-origsite=gscholar&cbl=18750} {Ph.D. thesis},\ \bibinfo  {school} {Stanford University, California} (\bibinfo {year} {2009})\BibitemShut {NoStop}%
\bibitem [{\citenamefont {Lu}\ \emph {et~al.}(2007)\citenamefont {Lu}, \citenamefont {Tzoufras}, \citenamefont {Joshi}, \citenamefont {Tsung}, \citenamefont {Mori}, \citenamefont {Vieria}, \citenamefont {Fonseca},\ and\ \citenamefont {Silva}}]{Lu_2007_laserdep}%
  \BibitemOpen
  \bibfield  {author} {\bibinfo {author} {\bibfnamefont {W.}~\bibnamefont {Lu}}, \bibinfo {author} {\bibfnamefont {M.}~\bibnamefont {Tzoufras}}, \bibinfo {author} {\bibfnamefont {C.}~\bibnamefont {Joshi}}, \bibinfo {author} {\bibfnamefont {F.~S.}\ \bibnamefont {Tsung}}, \bibinfo {author} {\bibfnamefont {W.~B.}\ \bibnamefont {Mori}}, \bibinfo {author} {\bibfnamefont {J.}~\bibnamefont {Vieria}}, \bibinfo {author} {\bibfnamefont {R.~A.}\ \bibnamefont {Fonseca}},\ and\ \bibinfo {author} {\bibfnamefont {L.~O.}\ \bibnamefont {Silva}},\ }\bibfield  {title} {\bibinfo {title} {{Generating multi-GeV electron bunches using single stage laser wakefield acceleration in a 3D nonlinear regime}},\ }\href {https://doi.org/10.1103/PhysRevSTAB.10.061301} {\bibfield  {journal} {\bibinfo  {journal} {Phys. Rev. ST Accel. Beams}\ }\textbf {\bibinfo {volume} {10}},\ \bibinfo {pages} {061301} (\bibinfo {year} {2007})}\BibitemShut {NoStop}%
\bibitem [{\citenamefont {Shadwick}\ \emph {et~al.}(2009)\citenamefont {Shadwick}, \citenamefont {Schroeder},\ and\ \citenamefont {Esarey}}]{shadwick2009}%
  \BibitemOpen
  \bibfield  {author} {\bibinfo {author} {\bibfnamefont {B.~A.}\ \bibnamefont {Shadwick}}, \bibinfo {author} {\bibfnamefont {C.~B.}\ \bibnamefont {Schroeder}},\ and\ \bibinfo {author} {\bibfnamefont {E.}~\bibnamefont {Esarey}},\ }\bibfield  {title} {\bibinfo {title} {{Nonlinear laser energy depletion in laser-plasma acceleratorsa)}},\ }\href {https://doi.org/10.1063/1.3124185} {\bibfield  {journal} {\bibinfo  {journal} {Phys. Plasmas}\ }\textbf {\bibinfo {volume} {16}},\ \bibinfo {pages} {056704} (\bibinfo {year} {2009})}\BibitemShut {NoStop}%
\bibitem [{\citenamefont {Schulte}(1996)}]{schulte_phd}%
  \BibitemOpen
  \bibfield  {author} {\bibinfo {author} {\bibfnamefont {D.}~\bibnamefont {Schulte}},\ }\emph {\bibinfo {title} {Study of Electromagnetic and Hadronic Background in the Interaction Region of the TESLA Collider}},\ \href {https://cds.cern.ch/record/331845?ln=en} {Ph.D. thesis},\ \bibinfo  {school} {University of Hamburg} (\bibinfo {year} {1996})\BibitemShut {NoStop}%
\bibitem [{\citenamefont {Schulte}(2016)}]{Schulte:2016ijt}%
  \BibitemOpen
  \bibfield  {author} {\bibinfo {author} {\bibfnamefont {D.}~\bibnamefont {Schulte}},\ }\bibfield  {title} {\bibinfo {title} {{Beam-Beam Effects in Linear Colliders}},\ }\href {https://icfa-usa.jlab.org/archive/newsletter/icfa_bd_nl_69.pdf} {\bibfield  {journal} {\bibinfo  {journal} {ICFA Beam Dyn. Newslett.}\ }\textbf {\bibinfo {volume} {69}},\ \bibinfo {pages} {237} (\bibinfo {year} {2016})}\BibitemShut {NoStop}%
\bibitem [{\citenamefont {{Lindstr{\o}m}}\ and\ \citenamefont {{Th{\'e}venet}}(2022)}]{Lindstrom_2022}%
  \BibitemOpen
  \bibfield  {author} {\bibinfo {author} {\bibfnamefont {C.~A.}\ \bibnamefont {{Lindstr{\o}m}}}\ and\ \bibinfo {author} {\bibfnamefont {M.}~\bibnamefont {{Th{\'e}venet}}},\ }\bibfield  {title} {\bibinfo {title} {{Emittance preservation in advanced accelerators}},\ }\href {https://doi.org/10.1088/1748-0221/17/05/P05016} {\bibfield  {journal} {\bibinfo  {journal} {Journal of Instrumentation}\ }\textbf {\bibinfo {volume} {17}}\bibinfo  {number} { (5)},\ \bibinfo {eid} {P05016}}\BibitemShut {NoStop}%
\bibitem [{\citenamefont {Raimondi}\ and\ \citenamefont {Seryi}(2001)}]{Raimondi:2001cx}%
  \BibitemOpen
\bibfield  {number} {  }\bibfield  {author} {\bibinfo {author} {\bibfnamefont {P.}~\bibnamefont {Raimondi}}\ and\ \bibinfo {author} {\bibfnamefont {A.}~\bibnamefont {Seryi}},\ }\bibfield  {title} {\bibinfo {title} {{A Novel final focus design for future linear colliders}},\ }\href {https://doi.org/10.1103/PhysRevLett.86.3779} {\bibfield  {journal} {\bibinfo  {journal} {Phys. Rev. Lett.}\ }\textbf {\bibinfo {volume} {86}},\ \bibinfo {pages} {3779} (\bibinfo {year} {2001})}\BibitemShut {NoStop}%
\bibitem [{\citenamefont {Migliorati}\ \emph {et~al.}(2013)\citenamefont {Migliorati}, \citenamefont {Bacci}, \citenamefont {Benedetti}, \citenamefont {Chiadroni}, \citenamefont {Ferrario}, \citenamefont {Mostacci}, \citenamefont {Palumbo}, \citenamefont {Rossi}, \citenamefont {Serafini},\ and\ \citenamefont {Antici}}]{migliorati_2013}%
  \BibitemOpen
  \bibfield  {author} {\bibinfo {author} {\bibfnamefont {M.}~\bibnamefont {Migliorati}}, \bibinfo {author} {\bibfnamefont {A.}~\bibnamefont {Bacci}}, \bibinfo {author} {\bibfnamefont {C.}~\bibnamefont {Benedetti}}, \bibinfo {author} {\bibfnamefont {E.}~\bibnamefont {Chiadroni}}, \bibinfo {author} {\bibfnamefont {M.}~\bibnamefont {Ferrario}}, \bibinfo {author} {\bibfnamefont {A.}~\bibnamefont {Mostacci}}, \bibinfo {author} {\bibfnamefont {L.}~\bibnamefont {Palumbo}}, \bibinfo {author} {\bibfnamefont {A.~R.}\ \bibnamefont {Rossi}}, \bibinfo {author} {\bibfnamefont {L.}~\bibnamefont {Serafini}},\ and\ \bibinfo {author} {\bibfnamefont {P.}~\bibnamefont {Antici}},\ }\bibfield  {title} {\bibinfo {title} {Intrinsic normalized emittance growth in laser-driven electron accelerators},\ }\href {https://doi.org/10.1103/PhysRevSTAB.16.011302} {\bibfield  {journal} {\bibinfo  {journal} {Phys. Rev. ST Accel. Beams}\ }\textbf {\bibinfo {volume} {16}},\ \bibinfo {pages} {011302} (\bibinfo {year} {2013})}\BibitemShut {NoStop}%
\bibitem [{\citenamefont {D'Arcy}\ \emph {et~al.}(2019)\citenamefont {D'Arcy}, \citenamefont {Wesch}, \citenamefont {Aschikhin}, \citenamefont {Bohlen}, \citenamefont {Behrens}, \citenamefont {Garland} \emph {et~al.}}]{darcy_2019}%
  \BibitemOpen
  \bibfield  {author} {\bibinfo {author} {\bibfnamefont {R.}~\bibnamefont {D'Arcy}}, \bibinfo {author} {\bibfnamefont {S.}~\bibnamefont {Wesch}}, \bibinfo {author} {\bibfnamefont {A.}~\bibnamefont {Aschikhin}}, \bibinfo {author} {\bibfnamefont {S.}~\bibnamefont {Bohlen}}, \bibinfo {author} {\bibfnamefont {C.}~\bibnamefont {Behrens}}, \bibinfo {author} {\bibfnamefont {M.~J.}\ \bibnamefont {Garland}}, \emph {et~al.},\ }\bibfield  {title} {\bibinfo {title} {Tunable plasma-based energy dechirper},\ }\href {https://doi.org/10.1103/PhysRevLett.122.034801} {\bibfield  {journal} {\bibinfo  {journal} {Phys. Rev. Lett.}\ }\textbf {\bibinfo {volume} {122}},\ \bibinfo {pages} {034801} (\bibinfo {year} {2019})}\BibitemShut {NoStop}%
\bibitem [{\citenamefont {Shpakov}\ \emph {et~al.}(2019)\citenamefont {Shpakov}, \citenamefont {Anania}, \citenamefont {Bellaveglia}, \citenamefont {Biagioni}, \citenamefont {Bisesto}, \citenamefont {Cardelli} \emph {et~al.}}]{shpakov_2019}%
  \BibitemOpen
  \bibfield  {author} {\bibinfo {author} {\bibfnamefont {V.}~\bibnamefont {Shpakov}}, \bibinfo {author} {\bibfnamefont {M.~P.}\ \bibnamefont {Anania}}, \bibinfo {author} {\bibfnamefont {M.}~\bibnamefont {Bellaveglia}}, \bibinfo {author} {\bibfnamefont {A.}~\bibnamefont {Biagioni}}, \bibinfo {author} {\bibfnamefont {F.}~\bibnamefont {Bisesto}}, \bibinfo {author} {\bibfnamefont {F.}~\bibnamefont {Cardelli}}, \emph {et~al.},\ }\bibfield  {title} {\bibinfo {title} {Longitudinal phase-space manipulation with beam-driven plasma wakefields},\ }\href {https://doi.org/10.1103/PhysRevLett.122.114801} {\bibfield  {journal} {\bibinfo  {journal} {Phys. Rev. Lett.}\ }\textbf {\bibinfo {volume} {122}},\ \bibinfo {pages} {114801} (\bibinfo {year} {2019})}\BibitemShut {NoStop}%
\bibitem [{\citenamefont {Wu}\ \emph {et~al.}(2019{\natexlab{a}})\citenamefont {Wu}, \citenamefont {Hua}, \citenamefont {Zhou}, \citenamefont {Zhang}, \citenamefont {Liu}, \citenamefont {Peng} \emph {et~al.}}]{Wu_2019}%
  \BibitemOpen
  \bibfield  {author} {\bibinfo {author} {\bibfnamefont {Y.~P.}\ \bibnamefont {Wu}}, \bibinfo {author} {\bibfnamefont {J.~F.}\ \bibnamefont {Hua}}, \bibinfo {author} {\bibfnamefont {Z.}~\bibnamefont {Zhou}}, \bibinfo {author} {\bibfnamefont {J.}~\bibnamefont {Zhang}}, \bibinfo {author} {\bibfnamefont {S.}~\bibnamefont {Liu}}, \bibinfo {author} {\bibfnamefont {B.}~\bibnamefont {Peng}}, \emph {et~al.},\ }\bibfield  {title} {\bibinfo {title} {Phase space dynamics of a plasma wakefield dechirper for energy spread reduction},\ }\href {https://doi.org/10.1103/PhysRevLett.122.204804} {\bibfield  {journal} {\bibinfo  {journal} {Phys. Rev. Lett.}\ }\textbf {\bibinfo {volume} {122}},\ \bibinfo {pages} {204804} (\bibinfo {year} {2019}{\natexlab{a}})}\BibitemShut {NoStop}%
\bibitem [{\citenamefont {Mane}\ \emph {et~al.}(2005)\citenamefont {Mane}, \citenamefont {Shatunov},\ and\ \citenamefont {Yokoya}}]{Mane_2005}%
  \BibitemOpen
  \bibfield  {author} {\bibinfo {author} {\bibfnamefont {S.~R.}\ \bibnamefont {Mane}}, \bibinfo {author} {\bibfnamefont {Y.~M.}\ \bibnamefont {Shatunov}},\ and\ \bibinfo {author} {\bibfnamefont {K.}~\bibnamefont {Yokoya}},\ }\bibfield  {title} {\bibinfo {title} {Spin-polarized charged particle beams in high-energy accelerators},\ }\href {https://doi.org/10.1088/0034-4885/68/9/R01} {\bibfield  {journal} {\bibinfo  {journal} {Rep. Prog. Phys.}\ }\textbf {\bibinfo {volume} {68}},\ \bibinfo {pages} {1997} (\bibinfo {year} {2005})}\BibitemShut {NoStop}%
\bibitem [{\citenamefont {Clendenin}(1994)}]{Clendenin:1993pk}%
  \BibitemOpen
  \bibfield  {author} {\bibinfo {author} {\bibfnamefont {J.}~\bibnamefont {Clendenin}},\ }\bibfield  {title} {\bibinfo {title} {{Polarized electron beams for linear colliders}},\ }\href {https://doi.org/10.1016/0168-9002(94)91275-0} {\bibfield  {journal} {\bibinfo  {journal} {Nucl. Instrum. Meth. A}\ }\textbf {\bibinfo {volume} {340}},\ \bibinfo {pages} {1} (\bibinfo {year} {1994})}\BibitemShut {NoStop}%
\bibitem [{\citenamefont {Lau}(1989)}]{Lau:1989}%
  \BibitemOpen
  \bibfield  {author} {\bibinfo {author} {\bibfnamefont {Y.~Y.}\ \bibnamefont {Lau}},\ }\bibfield  {title} {\bibinfo {title} {Classification of beam breakup instabilities in linear accelerators},\ }\href {https://doi.org/10.1103/PhysRevLett.63.1141} {\bibfield  {journal} {\bibinfo  {journal} {Phys. Rev. Lett.}\ }\textbf {\bibinfo {volume} {63}},\ \bibinfo {pages} {1141} (\bibinfo {year} {1989})}\BibitemShut {NoStop}%
\bibitem [{\citenamefont {Lebedev}\ \emph {et~al.}(2016)\citenamefont {Lebedev}, \citenamefont {Burov},\ and\ \citenamefont {Nagaitsev}}]{Lebedev:2016azx}%
  \BibitemOpen
  \bibfield  {author} {\bibinfo {author} {\bibfnamefont {V.}~\bibnamefont {Lebedev}}, \bibinfo {author} {\bibfnamefont {A.}~\bibnamefont {Burov}},\ and\ \bibinfo {author} {\bibfnamefont {S.}~\bibnamefont {Nagaitsev}},\ }\bibfield  {title} {\bibinfo {title} {{Luminosity Limitations of Linear Colliders Based on Plasma Acceleration}},\ }\href {https://doi.org/10.1142/S1793626816300097} {\bibfield  {journal} {\bibinfo  {journal} {Rev. Accel. Sci. Tech.}\ }\textbf {\bibinfo {volume} {09}},\ \bibinfo {pages} {187} (\bibinfo {year} {2016})}\BibitemShut {NoStop}%
\bibitem [{\citenamefont {Lee}\ \emph {et~al.}(2001)\citenamefont {Lee}, \citenamefont {Katsouleas}, \citenamefont {Hemker}, \citenamefont {Dodd},\ and\ \citenamefont {Mori}}]{Lee:2001bb}%
  \BibitemOpen
  \bibfield  {author} {\bibinfo {author} {\bibfnamefont {S.}~\bibnamefont {Lee}}, \bibinfo {author} {\bibfnamefont {T.~C.}\ \bibnamefont {Katsouleas}}, \bibinfo {author} {\bibfnamefont {R.~G.}\ \bibnamefont {Hemker}}, \bibinfo {author} {\bibfnamefont {E.~S.}\ \bibnamefont {Dodd}},\ and\ \bibinfo {author} {\bibfnamefont {W.~B.}\ \bibnamefont {Mori}},\ }\bibfield  {title} {\bibinfo {title} {{Plasma-wakefield acceleration of a positron beam}},\ }\href {https://doi.org/10.1103/PhysRevE.64.045501} {\bibfield  {journal} {\bibinfo  {journal} {Phys. Rev. E}\ }\textbf {\bibinfo {volume} {64}},\ \bibinfo {pages} {045501} (\bibinfo {year} {2001})}\BibitemShut {NoStop}%
\bibitem [{\citenamefont {Tajima}(1983)}]{Tajima:1983egt}%
  \BibitemOpen
  \bibfield  {author} {\bibinfo {author} {\bibfnamefont {T.}~\bibnamefont {Tajima}},\ }\bibfield  {title} {\bibinfo {title} {{Laser accelerator for ultrahigh energies}},\ }\href {https://inspirehep.net/literature/263522} {\bibfield  {journal} {\bibinfo  {journal} {Conf. Proc. C}\ }\textbf {\bibinfo {volume} {830811}},\ \bibinfo {pages} {470} (\bibinfo {year} {1983})}\BibitemShut {NoStop}%
\bibitem [{\citenamefont {Ng}\ \emph {et~al.}(2001)\citenamefont {Ng} \emph {et~al.}}]{Ng:2001is}%
  \BibitemOpen
  \bibfield  {author} {\bibinfo {author} {\bibfnamefont {J.~S.~T.}\ \bibnamefont {Ng}} \emph {et~al.},\ }\bibfield  {title} {\bibinfo {title} {{Observation of plasma focusing of a 28.5-GeV positron beam}},\ }\href {https://doi.org/10.1103/PhysRevLett.87.244801} {\bibfield  {journal} {\bibinfo  {journal} {Phys. Rev. Lett.}\ }\textbf {\bibinfo {volume} {87}},\ \bibinfo {pages} {244801} (\bibinfo {year} {2001})}\BibitemShut {NoStop}%
\bibitem [{\citenamefont {Marsh}\ \emph {et~al.}(2003)\citenamefont {Marsh} \emph {et~al.}}]{Marsh:2003dnx}%
  \BibitemOpen
  \bibfield  {author} {\bibinfo {author} {\bibfnamefont {K.~A.}\ \bibnamefont {Marsh}} \emph {et~al.},\ }\bibfield  {title} {\bibinfo {title} {Positron beam propagation in a meter long plasma channel},\ }in\ \href {https://doi.org/10.1109/PAC.2003.1289023} {\emph {\bibinfo {booktitle} {Proceedings of the 2003 Particle Accelerator Conference, Portland, OR, USA}}},\ Vol.~\bibinfo {volume} {1}\ (\bibinfo {year} {2003})\ p.\ \bibinfo {pages} {731}\BibitemShut {NoStop}%
\bibitem [{\citenamefont {Blue}\ \emph {et~al.}(2003)\citenamefont {Blue} \emph {et~al.}}]{Blue:2003nk}%
  \BibitemOpen
  \bibfield  {author} {\bibinfo {author} {\bibfnamefont {B.~E.}\ \bibnamefont {Blue}} \emph {et~al.},\ }\bibfield  {title} {\bibinfo {title} {{Plasma wake field acceleration of an intense positron beam}},\ }\href {https://doi.org/10.1103/PhysRevLett.90.214801} {\bibfield  {journal} {\bibinfo  {journal} {Phys. Rev. Lett.}\ }\textbf {\bibinfo {volume} {90}},\ \bibinfo {pages} {214801} (\bibinfo {year} {2003})}\BibitemShut {NoStop}%
\bibitem [{\citenamefont {Hogan}\ \emph {et~al.}(2003)\citenamefont {Hogan} \emph {et~al.}}]{Hogan:2003bs}%
  \BibitemOpen
  \bibfield  {author} {\bibinfo {author} {\bibfnamefont {M.~J.}\ \bibnamefont {Hogan}} \emph {et~al.},\ }\bibfield  {title} {\bibinfo {title} {{Ultrarelativistic positron beam transport through meter scale plasmas}},\ }\href {https://doi.org/10.1103/PhysRevLett.90.205002} {\bibfield  {journal} {\bibinfo  {journal} {Phys. Rev. Lett.}\ }\textbf {\bibinfo {volume} {90}},\ \bibinfo {pages} {205002} (\bibinfo {year} {2003})}\BibitemShut {NoStop}%
\bibitem [{\citenamefont {Muggli}\ \emph {et~al.}(2008)\citenamefont {Muggli} \emph {et~al.}}]{Muggli:2008zzb}%
  \BibitemOpen
  \bibfield  {author} {\bibinfo {author} {\bibfnamefont {P.}~\bibnamefont {Muggli}} \emph {et~al.},\ }\bibfield  {title} {\bibinfo {title} {{Halo Formation and Emittance Growth of Positron Beams in Plasmas}},\ }\href {https://doi.org/10.1103/PhysRevLett.101.055001} {\bibfield  {journal} {\bibinfo  {journal} {Phys. Rev. Lett.}\ }\textbf {\bibinfo {volume} {101}},\ \bibinfo {pages} {055001} (\bibinfo {year} {2008})}\BibitemShut {NoStop}%
\bibitem [{\citenamefont {Gessner}\ \emph {et~al.}(2016{\natexlab{a}})\citenamefont {Gessner}, \citenamefont {Adli}, \citenamefont {Allen}, \citenamefont {An}, \citenamefont {Clarke} \emph {et~al.}}]{Gessner:2016bqz}%
  \BibitemOpen
  \bibfield  {author} {\bibinfo {author} {\bibfnamefont {S.}~\bibnamefont {Gessner}}, \bibinfo {author} {\bibfnamefont {E.}~\bibnamefont {Adli}}, \bibinfo {author} {\bibfnamefont {J.~M.}\ \bibnamefont {Allen}}, \bibinfo {author} {\bibfnamefont {W.}~\bibnamefont {An}}, \bibinfo {author} {\bibfnamefont {C.~E.}\ \bibnamefont {Clarke}, \bibfnamefont {C.~I.~Clayton}}, \emph {et~al.},\ }\bibfield  {title} {\bibinfo {title} {{Demonstration of a positron beam-driven hollow channel plasma wakefield accelerator}},\ }\href {https://doi.org/10.1038/ncomms11785} {\bibfield  {journal} {\bibinfo  {journal} {Nat. Commun.}\ }\textbf {\bibinfo {volume} {7}},\ \bibinfo {pages} {11785} (\bibinfo {year} {2016}{\natexlab{a}})}\BibitemShut {NoStop}%
\bibitem [{\citenamefont {Doche}\ \emph {et~al.}(2017)\citenamefont {Doche}, \citenamefont {Beekman}, \citenamefont {Corde}, \citenamefont {Allen}, \citenamefont {Clarke}, \citenamefont {Frederico} \emph {et~al.}}]{Doche:2017jhd}%
  \BibitemOpen
  \bibfield  {author} {\bibinfo {author} {\bibfnamefont {A.}~\bibnamefont {Doche}}, \bibinfo {author} {\bibfnamefont {C.}~\bibnamefont {Beekman}}, \bibinfo {author} {\bibfnamefont {S.}~\bibnamefont {Corde}}, \bibinfo {author} {\bibfnamefont {J.~M.}\ \bibnamefont {Allen}}, \bibinfo {author} {\bibfnamefont {C.~I.}\ \bibnamefont {Clarke}}, \bibinfo {author} {\bibfnamefont {J.}~\bibnamefont {Frederico}}, \emph {et~al.},\ }\bibfield  {title} {\bibinfo {title} {{Acceleration of a trailing positron bunch in a plasma wakefield accelerator}},\ }\href {https://doi.org/10.1038/s41598-017-14524-4} {\bibfield  {journal} {\bibinfo  {journal} {Sci. Rep.}\ }\textbf {\bibinfo {volume} {7}},\ \bibinfo {pages} {14180} (\bibinfo {year} {2017})}\BibitemShut {NoStop}%
\bibitem [{\citenamefont {Lindstr\o{}m}\ \emph {et~al.}(2018)\citenamefont {Lindstr\o{}m} \emph {et~al.}}]{Lindstrom:2018hhy}%
  \BibitemOpen
  \bibfield  {author} {\bibinfo {author} {\bibfnamefont {C.~A.}\ \bibnamefont {Lindstr\o{}m}} \emph {et~al.},\ }\bibfield  {title} {\bibinfo {title} {{Measurement of transverse wakefields induced by a misaligned positron bunch in a hollow channel plasma accelerator}},\ }\href {https://doi.org/10.1103/PhysRevLett.120.124802} {\bibfield  {journal} {\bibinfo  {journal} {Phys. Rev. Lett.}\ }\textbf {\bibinfo {volume} {120}},\ \bibinfo {pages} {124802} (\bibinfo {year} {2018})}\BibitemShut {NoStop}%
\bibitem [{\citenamefont {Gessner}\ \emph {et~al.}(2023)\citenamefont {Gessner}, \citenamefont {Adli}, \citenamefont {Allen}, \citenamefont {An}, \citenamefont {Clarke}, \citenamefont {Clayton} \emph {et~al.}}]{Gessner:2023arxiv}%
  \BibitemOpen
  \bibfield  {author} {\bibinfo {author} {\bibfnamefont {S.}~\bibnamefont {Gessner}}, \bibinfo {author} {\bibfnamefont {E.}~\bibnamefont {Adli}}, \bibinfo {author} {\bibfnamefont {J.~M.}\ \bibnamefont {Allen}}, \bibinfo {author} {\bibfnamefont {W.}~\bibnamefont {An}}, \bibinfo {author} {\bibfnamefont {C.~I.}\ \bibnamefont {Clarke}}, \bibinfo {author} {\bibfnamefont {C.~E.}\ \bibnamefont {Clayton}}, \emph {et~al.},\ }\bibfield  {title} {\bibinfo {title} {{Acceleration of a Positron Bunch in a Hollow Channel Plasma}},\ }\href@noop {} {\bibfield  {journal} {\bibinfo  {journal} {arXiv e-prints}\ } (\bibinfo {year} {2023})},\ \Eprint {https://arxiv.org/abs/2304.01700} {arXiv:2304.01700} \BibitemShut {NoStop}%
\bibitem [{\citenamefont {Seeman}(1991)}]{Seeman:1991wf}%
  \BibitemOpen
  \bibfield  {author} {\bibinfo {author} {\bibfnamefont {J.~T.}\ \bibnamefont {Seeman}},\ }\bibfield  {title} {\bibinfo {title} {{The Stanford Linear Collider}},\ }\href {https://doi.org/10.1146/annurev.ns.41.120191.002133} {\bibfield  {journal} {\bibinfo  {journal} {Ann. Rev. Nucl. Part. Sci.}\ }\textbf {\bibinfo {volume} {41}},\ \bibinfo {pages} {389} (\bibinfo {year} {1991})}\BibitemShut {NoStop}%
\bibitem [{\citenamefont {Burke}(1992)}]{Burke:1991xr}%
  \BibitemOpen
  \bibfield  {author} {\bibinfo {author} {\bibfnamefont {D.}~\bibnamefont {Burke}},\ }\bibfield  {title} {\bibinfo {title} {{The final focus test beam project}},\ }\href {https://doi.org/10.1063/1.41971} {\bibfield  {journal} {\bibinfo  {journal} {AIP Conference Proceedings}\ }\textbf {\bibinfo {volume} {249}},\ \bibinfo {pages} {2082} (\bibinfo {year} {1992})}\BibitemShut {NoStop}%
\bibitem [{\citenamefont {Berndt}\ \emph {et~al.}(1991)\citenamefont {Berndt} \emph {et~al.}}]{Berndt:1991ug}%
  \BibitemOpen
  \bibfield  {author} {\bibinfo {author} {\bibfnamefont {M.}~\bibnamefont {Berndt}} \emph {et~al.},\ }\href {https://www-public.slac.stanford.edu/sciDoc/docMeta.aspx?slacPubNumber=slac-r-376} {\emph {\bibinfo {title} {{Final focus test beam: Project design report}}}},\ \bibinfo {type} {Tech. Rep.}\ (\bibinfo {year} {1991})\BibitemShut {NoStop}%
\bibitem [{\citenamefont {Betz}\ \emph {et~al.}(1991)\citenamefont {Betz}, \citenamefont {Chen}, \citenamefont {Cline}, \citenamefont {Gundersen}, \citenamefont {Joshi}, \citenamefont {Katsouleas} \emph {et~al.}}]{Betz1991}%
  \BibitemOpen
  \bibfield  {author} {\bibinfo {author} {\bibfnamefont {D.}~\bibnamefont {Betz}}, \bibinfo {author} {\bibfnamefont {P.}~\bibnamefont {Chen}}, \bibinfo {author} {\bibfnamefont {D.}~\bibnamefont {Cline}}, \bibinfo {author} {\bibfnamefont {M.}~\bibnamefont {Gundersen}}, \bibinfo {author} {\bibfnamefont {C.}~\bibnamefont {Joshi}}, \bibinfo {author} {\bibfnamefont {T.}~\bibnamefont {Katsouleas}}, \emph {et~al.},\ }\bibfield  {title} {\bibinfo {title} {Plasma lenses for slac final focus test facility},\ }in\ \href {https://doi.org/10.1109/PAC.1991.164382} {\emph {\bibinfo {booktitle} {Conference Record of the 1991 IEEE Particle Accelerator Conference, San Francisco, CA, USA}}},\ Vol.~\bibinfo {volume} {1}\ (\bibinfo {year} {1991})\ p.\ \bibinfo {pages} {619}\BibitemShut {NoStop}%
\bibitem [{\citenamefont {Chen}\ \emph {et~al.}(1998{\natexlab{a}})\citenamefont {Chen}, \citenamefont {Cline}, \citenamefont {Craddock}, \citenamefont {Decker}, \citenamefont {Iverson}, \citenamefont {Katsouleas} \emph {et~al.}}]{Chen:1998ut}%
  \BibitemOpen
  \bibfield  {author} {\bibinfo {author} {\bibfnamefont {P.}~\bibnamefont {Chen}}, \bibinfo {author} {\bibfnamefont {D.}~\bibnamefont {Cline}}, \bibinfo {author} {\bibfnamefont {W.}~\bibnamefont {Craddock}}, \bibinfo {author} {\bibfnamefont {F.}~\bibnamefont {Decker}}, \bibinfo {author} {\bibfnamefont {R.}~\bibnamefont {Iverson}}, \bibinfo {author} {\bibfnamefont {T.}~\bibnamefont {Katsouleas}}, \emph {et~al.},\ }\bibfield  {title} {\bibinfo {title} {{Plasma lens experiment at the final focus test beam}},\ }\href {https://doi.org/10.1016/S0168-9002(98)00153-3} {\bibfield  {journal} {\bibinfo  {journal} {Nucl. Instrum. Methods Phys. Res. A}\ }\textbf {\bibinfo {volume} {410}},\ \bibinfo {pages} {407} (\bibinfo {year} {1998}{\natexlab{a}})}\BibitemShut {NoStop}%
\bibitem [{\citenamefont {Assmann}\ \emph {et~al.}(1998)\citenamefont {Assmann}, \citenamefont {Chen}, \citenamefont {Decker}, \citenamefont {Iverson}, \citenamefont {Raimondi}, \citenamefont {Raubenheimer} \emph {et~al.}}]{ASSMANN1998396}%
  \BibitemOpen
  \bibfield  {author} {\bibinfo {author} {\bibfnamefont {R.}~\bibnamefont {Assmann}}, \bibinfo {author} {\bibfnamefont {P.}~\bibnamefont {Chen}}, \bibinfo {author} {\bibfnamefont {F.}~\bibnamefont {Decker}}, \bibinfo {author} {\bibfnamefont {R.}~\bibnamefont {Iverson}}, \bibinfo {author} {\bibfnamefont {P.}~\bibnamefont {Raimondi}}, \bibinfo {author} {\bibfnamefont {T.}~\bibnamefont {Raubenheimer}}, \emph {et~al.},\ }\bibfield  {title} {\bibinfo {title} {{Proposal for a one GeV plasma wakefield acceleration experiment at SLAC}},\ }\href {https://doi.org/https://doi.org/10.1016/S0168-9002(98)00169-7} {\bibfield  {journal} {\bibinfo  {journal} {Nucl. Instrum. Methods Phys. Res. A}\ }\textbf {\bibinfo {volume} {410}},\ \bibinfo {pages} {396} (\bibinfo {year} {1998})}\BibitemShut {NoStop}%
\bibitem [{\citenamefont {Hogan}\ \emph {et~al.}(2000)\citenamefont {Hogan} \emph {et~al.}}]{Hogan:2000xq}%
  \BibitemOpen
  \bibfield  {author} {\bibinfo {author} {\bibfnamefont {M.~J.}\ \bibnamefont {Hogan}} \emph {et~al.},\ }\bibfield  {title} {\bibinfo {title} {{E-157: A 1.4 meter long plasma wake field acceleration experiment using a 30-GeV electron beam from the Stanford Linear Accelerator Center linac}},\ }\href {https://doi.org/10.1063/1.874059} {\bibfield  {journal} {\bibinfo  {journal} {Phys. Plasmas}\ }\textbf {\bibinfo {volume} {7}},\ \bibinfo {pages} {2241} (\bibinfo {year} {2000})}\BibitemShut {NoStop}%
\bibitem [{\citenamefont {Baird}\ \emph {et~al.}(2000)\citenamefont {Baird}, \citenamefont {Decker}, \citenamefont {Hogan}, \citenamefont {Iverson}, \citenamefont {Raimondi}, \citenamefont {H.} \emph {et~al.}}]{Baird2000}%
  \BibitemOpen
  \bibfield  {author} {\bibinfo {author} {\bibfnamefont {K.}~\bibnamefont {Baird}}, \bibinfo {author} {\bibfnamefont {F.-J.}\ \bibnamefont {Decker}}, \bibinfo {author} {\bibfnamefont {M.~J.}\ \bibnamefont {Hogan}}, \bibinfo {author} {\bibfnamefont {R.~H.}\ \bibnamefont {Iverson}}, \bibinfo {author} {\bibfnamefont {P.}~\bibnamefont {Raimondi}}, \bibinfo {author} {\bibfnamefont {S.~R.}\ \bibnamefont {H.}}, \emph {et~al.},\ }\href {https://www.slac.stanford.edu/pubs/slactns/tn04/slac-tn-05-083.pdf} {\emph {\bibinfo {title} {{E-162: Positron and Electron Dynamics in a Plasma Wakefield Accelerator}}}},\ \bibinfo {type} {Tech. Rep.}\ (\bibinfo {year} {2000})\BibitemShut {NoStop}%
\bibitem [{\citenamefont {Chen}\ \emph {et~al.}(1998{\natexlab{b}})\citenamefont {Chen}, \citenamefont {Cline}, \citenamefont {Craddock}, \citenamefont {Decker}, \citenamefont {Iverson}, \citenamefont {Katsouleas} \emph {et~al.}}]{CHEN1998407}%
  \BibitemOpen
  \bibfield  {author} {\bibinfo {author} {\bibfnamefont {P.}~\bibnamefont {Chen}}, \bibinfo {author} {\bibfnamefont {D.}~\bibnamefont {Cline}}, \bibinfo {author} {\bibfnamefont {W.}~\bibnamefont {Craddock}}, \bibinfo {author} {\bibfnamefont {F.}~\bibnamefont {Decker}}, \bibinfo {author} {\bibfnamefont {R.}~\bibnamefont {Iverson}}, \bibinfo {author} {\bibfnamefont {T.}~\bibnamefont {Katsouleas}}, \emph {et~al.},\ }\bibfield  {title} {\bibinfo {title} {Plasma lens experiment at the final focus test beam},\ }\href {https://doi.org/https://doi.org/10.1016/S0168-9002(98)00153-3} {\bibfield  {journal} {\bibinfo  {journal} {Nucl. Instrum. Methods Phys. Res. A}\ }\textbf {\bibinfo {volume} {410}},\ \bibinfo {pages} {407} (\bibinfo {year} {1998}{\natexlab{b}})}\BibitemShut {NoStop}%
\bibitem [{\citenamefont {Hogan}\ \emph {et~al.}(2010)\citenamefont {Hogan} \emph {et~al.}}]{Hogan:2010zz}%
  \BibitemOpen
  \bibfield  {author} {\bibinfo {author} {\bibfnamefont {M.~J.}\ \bibnamefont {Hogan}} \emph {et~al.},\ }\bibfield  {title} {\bibinfo {title} {{Plasma wakefield acceleration experiments at FACET}},\ }\href {https://doi.org/10.1088/1367-2630/12/5/055030} {\bibfield  {journal} {\bibinfo  {journal} {New J. Phys.}\ }\textbf {\bibinfo {volume} {12}},\ \bibinfo {pages} {055030} (\bibinfo {year} {2010})}\BibitemShut {NoStop}%
\bibitem [{\citenamefont {Clarke}\ \emph {et~al.}(2011)\citenamefont {Clarke} \emph {et~al.}}]{Clarke:2011za}%
  \BibitemOpen
  \bibfield  {author} {\bibinfo {author} {\bibfnamefont {C.~I.}\ \bibnamefont {Clarke}} \emph {et~al.},\ }\bibfield  {title} {\bibinfo {title} {{FACET: The New User Facility at SLAC}},\ }in\ \href {https://accelconf.web.cern.ch/ipac2011/papers/weoab02.pdf} {\emph {\bibinfo {booktitle} {{Proceedings of IPAC2011, San Sebastián, Spain}}}}\ (\bibinfo {year} {2011})\ p.\ \bibinfo {pages} {1953}\BibitemShut {NoStop}%
\bibitem [{\citenamefont {Lu}\ \emph {et~al.}(2005)\citenamefont {Lu}, \citenamefont {Huang}, \citenamefont {Zhou}, \citenamefont {Mori},\ and\ \citenamefont {Katsouleas}}]{Lu_2005}%
  \BibitemOpen
  \bibfield  {author} {\bibinfo {author} {\bibfnamefont {W.}~\bibnamefont {Lu}}, \bibinfo {author} {\bibfnamefont {C.}~\bibnamefont {Huang}}, \bibinfo {author} {\bibfnamefont {M.~M.}\ \bibnamefont {Zhou}}, \bibinfo {author} {\bibfnamefont {W.~B.}\ \bibnamefont {Mori}},\ and\ \bibinfo {author} {\bibfnamefont {T.}~\bibnamefont {Katsouleas}},\ }\bibfield  {title} {\bibinfo {title} {{Limits of linear plasma wakefield theory for electron or positron beams}},\ }\href {https://doi.org/10.1063/1.1905587} {\bibfield  {journal} {\bibinfo  {journal} {Phys. Plasmas}\ }\textbf {\bibinfo {volume} {12}},\ \bibinfo {pages} {063101} (\bibinfo {year} {2005})}\BibitemShut {NoStop}%
\bibitem [{\citenamefont {{Zhou}}\ \emph {et~al.}(2006)\citenamefont {{Zhou}}, \citenamefont {{He}},\ and\ \citenamefont {{Yu}}}]{Zhou_2006}%
  \BibitemOpen
  \bibfield  {author} {\bibinfo {author} {\bibfnamefont {C.~T.}\ \bibnamefont {{Zhou}}}, \bibinfo {author} {\bibfnamefont {X.~T.}\ \bibnamefont {{He}}},\ and\ \bibinfo {author} {\bibfnamefont {M.~Y.}\ \bibnamefont {{Yu}}},\ }\bibfield  {title} {\bibinfo {title} {{A comparison of ultrarelativistic electron- and positron-bunch propagation in plasmas}},\ }\href {https://doi.org/10.1063/1.2345580} {\bibfield  {journal} {\bibinfo  {journal} {Phys. Plasmas}\ }\textbf {\bibinfo {volume} {13}},\ \bibinfo {eid} {092109} (\bibinfo {year} {2006})}\BibitemShut {NoStop}%
\bibitem [{\citenamefont {{Lotov}}(2007)}]{Lotov_2007}%
  \BibitemOpen
  \bibfield  {author} {\bibinfo {author} {\bibfnamefont {K.~V.}\ \bibnamefont {{Lotov}}},\ }\bibfield  {title} {\bibinfo {title} {{Acceleration of positrons by electron beam-driven wakefields in a plasma}},\ }\href {https://doi.org/10.1063/1.2434793} {\bibfield  {journal} {\bibinfo  {journal} {Phys. Plasmas}\ }\textbf {\bibinfo {volume} {14}},\ \bibinfo {eid} {023101} (\bibinfo {year} {2007})}\BibitemShut {NoStop}%
\bibitem [{\citenamefont {An}\ \emph {et~al.}(2010)\citenamefont {An}, \citenamefont {Huang}, \citenamefont {Lu}, \citenamefont {Mori},\ and\ \citenamefont {Katsouleas}}]{An:2010exa}%
  \BibitemOpen
  \bibfield  {author} {\bibinfo {author} {\bibfnamefont {W.}~\bibnamefont {An}}, \bibinfo {author} {\bibfnamefont {C.}~\bibnamefont {Huang}}, \bibinfo {author} {\bibfnamefont {W.}~\bibnamefont {Lu}}, \bibinfo {author} {\bibfnamefont {W.}~\bibnamefont {Mori}},\ and\ \bibinfo {author} {\bibfnamefont {T.}~\bibnamefont {Katsouleas}},\ }\bibfield  {title} {\bibinfo {title} {{Positron Acceleration by Using a Particle Beam-Driven Wake Field in Plasma}},\ }in\ \href {https://accelconf.web.cern.ch/pac2009/papers/we6rfp093.pdf} {\emph {\bibinfo {booktitle} {{Proceedings of PAC2009, Vancouver, BC, Canada}}}}\ (\bibinfo {year} {2010})\ p.\ \bibinfo {pages} {3013}\BibitemShut {NoStop}%
\bibitem [{\citenamefont {Li}\ \emph {et~al.}(2010)\citenamefont {Li}, \citenamefont {Muggli},\ and\ \citenamefont {Martins}}]{Li:2010moa}%
  \BibitemOpen
  \bibfield  {author} {\bibinfo {author} {\bibfnamefont {X.}~\bibnamefont {Li}}, \bibinfo {author} {\bibfnamefont {P.}~\bibnamefont {Muggli}},\ and\ \bibinfo {author} {\bibfnamefont {S.}~\bibnamefont {Martins}},\ }\bibfield  {title} {\bibinfo {title} {{Simulations of Positron Beams Propagating in Plasma}},\ }in\ \href {https://accelconf.web.cern.ch/pac2009/papers/fr5rfp025.pdf} {\emph {\bibinfo {booktitle} {{Proceedings of PAC2009, Vancouver, BC, Canada}}}}\ (\bibinfo {year} {2010})\ p.\ \bibinfo {pages} {4586}\BibitemShut {NoStop}%
\bibitem [{\citenamefont {Gessner}\ \emph {et~al.}(2012)\citenamefont {Gessner}, \citenamefont {Adli}, \citenamefont {Corde}, \citenamefont {England}, \citenamefont {Frederico}, \citenamefont {Hogan} \emph {et~al.}}]{Gessner:2012zz}%
  \BibitemOpen
  \bibfield  {author} {\bibinfo {author} {\bibfnamefont {S.~J.}\ \bibnamefont {Gessner}}, \bibinfo {author} {\bibfnamefont {E.}~\bibnamefont {Adli}}, \bibinfo {author} {\bibfnamefont {S.}~\bibnamefont {Corde}}, \bibinfo {author} {\bibfnamefont {J.}~\bibnamefont {England}}, \bibinfo {author} {\bibfnamefont {J.}~\bibnamefont {Frederico}}, \bibinfo {author} {\bibfnamefont {M.~J.}\ \bibnamefont {Hogan}}, \emph {et~al.},\ }\bibfield  {title} {\bibinfo {title} {{Positron PWFA Simulations for FACET}},\ }in\ \href {https://accelconf.web.cern.ch/ipac2012/papers/weppp056.pdf} {\emph {\bibinfo {booktitle} {{Proceedings of IPAC2012, New Orleans, Louisiana, USA}}}}\ (\bibinfo {year} {2012})\ p.\ \bibinfo {pages} {2834}\BibitemShut {NoStop}%
\bibitem [{\citenamefont {Fujii}\ \emph {et~al.}(2019)\citenamefont {Fujii}, \citenamefont {Marsh}, \citenamefont {An}, \citenamefont {Corde}, \citenamefont {Hogan}, \citenamefont {Yakimenko},\ and\ \citenamefont {Joshi}}]{Fujii:2019qxb}%
  \BibitemOpen
  \bibfield  {author} {\bibinfo {author} {\bibfnamefont {H.}~\bibnamefont {Fujii}}, \bibinfo {author} {\bibfnamefont {K.~A.}\ \bibnamefont {Marsh}}, \bibinfo {author} {\bibfnamefont {W.}~\bibnamefont {An}}, \bibinfo {author} {\bibfnamefont {S.}~\bibnamefont {Corde}}, \bibinfo {author} {\bibfnamefont {M.~J.}\ \bibnamefont {Hogan}}, \bibinfo {author} {\bibfnamefont {V.}~\bibnamefont {Yakimenko}},\ and\ \bibinfo {author} {\bibfnamefont {C.}~\bibnamefont {Joshi}},\ }\bibfield  {title} {\bibinfo {title} {{Positron beam extraction from an electron-beam-driven plasma wakefield accelerator}},\ }\href {https://doi.org/10.1103/PhysRevAccelBeams.22.091301} {\bibfield  {journal} {\bibinfo  {journal} {Phys. Rev. Accel. Beams}\ }\textbf {\bibinfo {volume} {22}},\ \bibinfo {pages} {091301} (\bibinfo {year} {2019})}\BibitemShut {NoStop}%
\bibitem [{\citenamefont {Marsh}\ \emph {et~al.}(2005)\citenamefont {Marsh}, \citenamefont {Clayton}, \citenamefont {Johnson}, \citenamefont {Huang}, \citenamefont {Joshi}, \citenamefont {Lu} \emph {et~al.}}]{marsh_2005}%
  \BibitemOpen
  \bibfield  {author} {\bibinfo {author} {\bibfnamefont {K.~A.}\ \bibnamefont {Marsh}}, \bibinfo {author} {\bibfnamefont {C.~E.}\ \bibnamefont {Clayton}}, \bibinfo {author} {\bibfnamefont {D.~K.}\ \bibnamefont {Johnson}}, \bibinfo {author} {\bibfnamefont {C.}~\bibnamefont {Huang}}, \bibinfo {author} {\bibfnamefont {C.}~\bibnamefont {Joshi}}, \bibinfo {author} {\bibfnamefont {W.}~\bibnamefont {Lu}}, \emph {et~al.},\ }\bibfield  {title} {\bibinfo {title} {Beam matching to a plasma wake field accelerator using a ramped density profile at the plasma boundary},\ }in\ \href {https://doi.org/10.1109/PAC.2005.1591234} {\emph {\bibinfo {booktitle} {Proceedings of PAC2005, Knoxville, TN, USA}}}\ (\bibinfo {year} {2005})\ p.\ \bibinfo {pages} {2702}\BibitemShut {NoStop}%
\bibitem [{\citenamefont {Dornmair}\ \emph {et~al.}(2015)\citenamefont {Dornmair}, \citenamefont {Floettmann},\ and\ \citenamefont {Maier}}]{Dornmair_2015}%
  \BibitemOpen
  \bibfield  {author} {\bibinfo {author} {\bibfnamefont {I.}~\bibnamefont {Dornmair}}, \bibinfo {author} {\bibfnamefont {K.}~\bibnamefont {Floettmann}},\ and\ \bibinfo {author} {\bibfnamefont {A.~R.}\ \bibnamefont {Maier}},\ }\bibfield  {title} {\bibinfo {title} {Emittance conservation by tailored focusing profiles in a plasma accelerator},\ }\href {https://doi.org/10.1103/PhysRevSTAB.18.041302} {\bibfield  {journal} {\bibinfo  {journal} {Phys. Rev. ST Accel. Beams}\ }\textbf {\bibinfo {volume} {18}},\ \bibinfo {pages} {041302} (\bibinfo {year} {2015})}\BibitemShut {NoStop}%
\bibitem [{\citenamefont {Xu}\ \emph {et~al.}(2016)\citenamefont {Xu} \emph {et~al.}}]{Xu:2016kia}%
  \BibitemOpen
  \bibfield  {author} {\bibinfo {author} {\bibfnamefont {X.~L.}\ \bibnamefont {Xu}} \emph {et~al.},\ }\bibfield  {title} {\bibinfo {title} {{Physics of Phase Space Matching for Staging Plasma and Traditional Accelerator Components Using Longitudinally Tailored Plasma Profiles}},\ }\href {https://doi.org/10.1103/PhysRevLett.116.124801} {\bibfield  {journal} {\bibinfo  {journal} {Phys. Rev. Lett.}\ }\textbf {\bibinfo {volume} {116}},\ \bibinfo {pages} {124801} (\bibinfo {year} {2016})}\BibitemShut {NoStop}%
\bibitem [{\citenamefont {Ariniello}\ \emph {et~al.}(2019)\citenamefont {Ariniello}, \citenamefont {Doss}, \citenamefont {Hunt-Stone}, \citenamefont {Cary},\ and\ \citenamefont {Litos}}]{ariniello_2019}%
  \BibitemOpen
  \bibfield  {author} {\bibinfo {author} {\bibfnamefont {R.}~\bibnamefont {Ariniello}}, \bibinfo {author} {\bibfnamefont {C.~E.}\ \bibnamefont {Doss}}, \bibinfo {author} {\bibfnamefont {K.}~\bibnamefont {Hunt-Stone}}, \bibinfo {author} {\bibfnamefont {J.~R.}\ \bibnamefont {Cary}},\ and\ \bibinfo {author} {\bibfnamefont {M.~D.}\ \bibnamefont {Litos}},\ }\bibfield  {title} {\bibinfo {title} {Transverse beam dynamics in a plasma density ramp},\ }\href {https://doi.org/10.1103/PhysRevAccelBeams.22.041304} {\bibfield  {journal} {\bibinfo  {journal} {Phys. Rev. Accel. Beams}\ }\textbf {\bibinfo {volume} {22}},\ \bibinfo {pages} {041304} (\bibinfo {year} {2019})}\BibitemShut {NoStop}%
\bibitem [{\citenamefont {Litos}\ \emph {et~al.}(2019)\citenamefont {Litos}, \citenamefont {Ariniello}, \citenamefont {Doss}, \citenamefont {Hunt-Stone},\ and\ \citenamefont {Cary}}]{Litos:2019tzi}%
  \BibitemOpen
  \bibfield  {author} {\bibinfo {author} {\bibfnamefont {M.~D.}\ \bibnamefont {Litos}}, \bibinfo {author} {\bibfnamefont {R.}~\bibnamefont {Ariniello}}, \bibinfo {author} {\bibfnamefont {C.~E.}\ \bibnamefont {Doss}}, \bibinfo {author} {\bibfnamefont {K.}~\bibnamefont {Hunt-Stone}},\ and\ \bibinfo {author} {\bibfnamefont {J.~R.}\ \bibnamefont {Cary}},\ }\bibfield  {title} {\bibinfo {title} {{Beam emittance preservation using Gaussian density ramps in a beam-driven plasma wakefield accelerator}},\ }\href {https://doi.org/10.1098/rsta.2018.0181} {\bibfield  {journal} {\bibinfo  {journal} {Phil. Trans. Roy. Soc. Lond. A}\ }\textbf {\bibinfo {volume} {377}},\ \bibinfo {pages} {20180181} (\bibinfo {year} {2019})}\BibitemShut {NoStop}%
\bibitem [{\citenamefont {Li}\ \emph {et~al.}(2019{\natexlab{a}})\citenamefont {Li}, \citenamefont {Chanc\'e},\ and\ \citenamefont {Nghiem}}]{Li_2019b}%
  \BibitemOpen
  \bibfield  {author} {\bibinfo {author} {\bibfnamefont {X.}~\bibnamefont {Li}}, \bibinfo {author} {\bibfnamefont {A.}~\bibnamefont {Chanc\'e}},\ and\ \bibinfo {author} {\bibfnamefont {P.~A.~P.}\ \bibnamefont {Nghiem}},\ }\bibfield  {title} {\bibinfo {title} {Preserving emittance by matching out and matching in plasma wakefield acceleration stage},\ }\href {https://doi.org/10.1103/PhysRevAccelBeams.22.021304} {\bibfield  {journal} {\bibinfo  {journal} {Phys. Rev. Accel. Beams}\ }\textbf {\bibinfo {volume} {22}},\ \bibinfo {pages} {021304} (\bibinfo {year} {2019}{\natexlab{a}})}\BibitemShut {NoStop}%
\bibitem [{\citenamefont {Zhao}\ \emph {et~al.}(2020)\citenamefont {Zhao}, \citenamefont {An}, \citenamefont {Xu}, \citenamefont {Li}, \citenamefont {Hildebrand}, \citenamefont {Hogan}, \citenamefont {Yakimenko}, \citenamefont {Joshi},\ and\ \citenamefont {Mori}}]{Zhao_2020}%
  \BibitemOpen
  \bibfield  {author} {\bibinfo {author} {\bibfnamefont {Y.}~\bibnamefont {Zhao}}, \bibinfo {author} {\bibfnamefont {W.}~\bibnamefont {An}}, \bibinfo {author} {\bibfnamefont {X.}~\bibnamefont {Xu}}, \bibinfo {author} {\bibfnamefont {F.}~\bibnamefont {Li}}, \bibinfo {author} {\bibfnamefont {L.}~\bibnamefont {Hildebrand}}, \bibinfo {author} {\bibfnamefont {M.~J.}\ \bibnamefont {Hogan}}, \bibinfo {author} {\bibfnamefont {V.}~\bibnamefont {Yakimenko}}, \bibinfo {author} {\bibfnamefont {C.}~\bibnamefont {Joshi}},\ and\ \bibinfo {author} {\bibfnamefont {W.~B.}\ \bibnamefont {Mori}},\ }\bibfield  {title} {\bibinfo {title} {Emittance preservation through density ramp matching sections in a plasma wakefield accelerator},\ }\href {https://doi.org/10.1103/PhysRevAccelBeams.23.011302} {\bibfield  {journal} {\bibinfo  {journal} {Phys. Rev. Accel. Beams}\ }\textbf {\bibinfo {volume} {23}},\ \bibinfo {pages} {011302} (\bibinfo {year} {2020})}\BibitemShut {NoStop}%
\bibitem [{\citenamefont {Kimura}\ \emph {et~al.}(2011)\citenamefont {Kimura}, \citenamefont {Milchberg}, \citenamefont {Muggli}, \citenamefont {Li},\ and\ \citenamefont {Mori}}]{Kimura:2011zz}%
  \BibitemOpen
  \bibfield  {author} {\bibinfo {author} {\bibfnamefont {W.~D.}\ \bibnamefont {Kimura}}, \bibinfo {author} {\bibfnamefont {H.~M.}\ \bibnamefont {Milchberg}}, \bibinfo {author} {\bibfnamefont {P.}~\bibnamefont {Muggli}}, \bibinfo {author} {\bibfnamefont {X.}~\bibnamefont {Li}},\ and\ \bibinfo {author} {\bibfnamefont {W.~B.}\ \bibnamefont {Mori}},\ }\bibfield  {title} {\bibinfo {title} {{Hollow plasma channel for positron plasma wakefield acceleration}},\ }\href {https://doi.org/10.1103/PhysRevSTAB.14.041301} {\bibfield  {journal} {\bibinfo  {journal} {Phys. Rev. ST Accel. Beams}\ }\textbf {\bibinfo {volume} {14}},\ \bibinfo {pages} {041301} (\bibinfo {year} {2011})}\BibitemShut {NoStop}%
\bibitem [{\citenamefont {Barnes}\ \emph {et~al.}(1987)\citenamefont {Barnes}, \citenamefont {Kurki-Suonio},\ and\ \citenamefont {Tajima}}]{Barnes:1987}%
  \BibitemOpen
  \bibfield  {author} {\bibinfo {author} {\bibfnamefont {D.~C.}\ \bibnamefont {Barnes}}, \bibinfo {author} {\bibfnamefont {T.}~\bibnamefont {Kurki-Suonio}},\ and\ \bibinfo {author} {\bibfnamefont {T.}~\bibnamefont {Tajima}},\ }\bibfield  {title} {\bibinfo {title} {Laser self-trapping for the plasma fiber accelerator},\ }\href {https://doi.org/10.1109/TPS.1987.4316678} {\bibfield  {journal} {\bibinfo  {journal} {IEEE T. Plasma Sci.}\ }\textbf {\bibinfo {volume} {15}},\ \bibinfo {pages} {154} (\bibinfo {year} {1987})}\BibitemShut {NoStop}%
\bibitem [{\citenamefont {Katsouleas}\ \emph {et~al.}(1992)\citenamefont {Katsouleas}, \citenamefont {Chiou}, \citenamefont {Decker}, \citenamefont {Mori}, \citenamefont {Wurtele}, \citenamefont {Shvets},\ and\ \citenamefont {Su}}]{Katsouleas:1992yn}%
  \BibitemOpen
  \bibfield  {author} {\bibinfo {author} {\bibfnamefont {T.~C.}\ \bibnamefont {Katsouleas}}, \bibinfo {author} {\bibfnamefont {T.~C.}\ \bibnamefont {Chiou}}, \bibinfo {author} {\bibfnamefont {C.}~\bibnamefont {Decker}}, \bibinfo {author} {\bibfnamefont {W.~B.}\ \bibnamefont {Mori}}, \bibinfo {author} {\bibfnamefont {J.~S.}\ \bibnamefont {Wurtele}}, \bibinfo {author} {\bibfnamefont {G.}~\bibnamefont {Shvets}},\ and\ \bibinfo {author} {\bibfnamefont {J.~J.}\ \bibnamefont {Su}},\ }\bibfield  {title} {\bibinfo {title} {{Laser wakefield acceleration \& optical guiding in a hollow plasma channel}},\ }\href {https://doi.org/10.1063/1.44067} {\bibfield  {journal} {\bibinfo  {journal} {AIP Conf. Proc.}\ }\textbf {\bibinfo {volume} {279}},\ \bibinfo {pages} {480} (\bibinfo {year} {1992})}\BibitemShut {NoStop}%
\bibitem [{\citenamefont {Chiou}\ \emph {et~al.}(1995)\citenamefont {Chiou}, \citenamefont {Katsouleas}, \citenamefont {Decker}, \citenamefont {Mori}, \citenamefont {Wurtele}, \citenamefont {Shvets},\ and\ \citenamefont {Su}}]{Chiou_1995}%
  \BibitemOpen
  \bibfield  {author} {\bibinfo {author} {\bibfnamefont {T.~C.}\ \bibnamefont {Chiou}}, \bibinfo {author} {\bibfnamefont {T.}~\bibnamefont {Katsouleas}}, \bibinfo {author} {\bibfnamefont {C.}~\bibnamefont {Decker}}, \bibinfo {author} {\bibfnamefont {W.~B.}\ \bibnamefont {Mori}}, \bibinfo {author} {\bibfnamefont {J.~S.}\ \bibnamefont {Wurtele}}, \bibinfo {author} {\bibfnamefont {G.}~\bibnamefont {Shvets}},\ and\ \bibinfo {author} {\bibfnamefont {J.~J.}\ \bibnamefont {Su}},\ }\bibfield  {title} {\bibinfo {title} {Laser wake-field acceleration and optical guiding in a hollow plasma channel},\ }\href {https://doi.org/10.1063/1.871107} {\bibfield  {journal} {\bibinfo  {journal} {Phys. Plasmas}\ }\textbf {\bibinfo {volume} {2}},\ \bibinfo {pages} {310} (\bibinfo {year} {1995})}\BibitemShut {NoStop}%
\bibitem [{\citenamefont {Chiou}\ and\ \citenamefont {Katsouleas}(1998)}]{chiou_1998}%
  \BibitemOpen
  \bibfield  {author} {\bibinfo {author} {\bibfnamefont {T.~C.}\ \bibnamefont {Chiou}}\ and\ \bibinfo {author} {\bibfnamefont {T.}~\bibnamefont {Katsouleas}},\ }\bibfield  {title} {\bibinfo {title} {High beam quality and efficiency in plasma-based accelerators},\ }\href {https://doi.org/10.1103/PhysRevLett.81.3411} {\bibfield  {journal} {\bibinfo  {journal} {Phys. Rev. Lett.}\ }\textbf {\bibinfo {volume} {81}},\ \bibinfo {pages} {3411} (\bibinfo {year} {1998})}\BibitemShut {NoStop}%
\bibitem [{\citenamefont {Schroeder}\ \emph {et~al.}(1999)\citenamefont {Schroeder}, \citenamefont {Wurtele},\ and\ \citenamefont {Whittum}}]{Schroeder:1999cb}%
  \BibitemOpen
  \bibfield  {author} {\bibinfo {author} {\bibfnamefont {C.~B.}\ \bibnamefont {Schroeder}}, \bibinfo {author} {\bibfnamefont {J.~S.}\ \bibnamefont {Wurtele}},\ and\ \bibinfo {author} {\bibfnamefont {D.~H.}\ \bibnamefont {Whittum}},\ }\bibfield  {title} {\bibinfo {title} {{Multimode analysis of the hollow plasma channel wake field Accelerator}},\ }\href {https://doi.org/10.1103/PhysRevLett.82.1177} {\bibfield  {journal} {\bibinfo  {journal} {Phys. Rev. Lett.}\ }\textbf {\bibinfo {volume} {82}},\ \bibinfo {pages} {1177} (\bibinfo {year} {1999})}\BibitemShut {NoStop}%
\bibitem [{\citenamefont {Lee}\ \emph {et~al.}(2002)\citenamefont {Lee} \emph {et~al.}}]{Lee:2002bh}%
  \BibitemOpen
  \bibfield  {author} {\bibinfo {author} {\bibfnamefont {S.}~\bibnamefont {Lee}} \emph {et~al.},\ }\bibfield  {title} {\bibinfo {title} {{Energy doubler for a linear collider}},\ }\href {https://doi.org/10.1103/PhysRevSTAB.5.011001} {\bibfield  {journal} {\bibinfo  {journal} {Phys. Rev. ST Accel. Beams}\ }\textbf {\bibinfo {volume} {5}},\ \bibinfo {pages} {011001} (\bibinfo {year} {2002})}\BibitemShut {NoStop}%
\bibitem [{\citenamefont {Raubenheimer}(2004)}]{Raubenheimer_2004}%
  \BibitemOpen
  \bibfield  {author} {\bibinfo {author} {\bibfnamefont {T.~O.}\ \bibnamefont {Raubenheimer}},\ }\bibfield  {title} {\bibinfo {title} {{An Afterburner at the ILC: The Collider Viewpoint}},\ }\href {https://doi.org/10.1063/1.1842536} {\bibfield  {journal} {\bibinfo  {journal} {AIP Conf. Proc.}\ }\textbf {\bibinfo {volume} {737}},\ \bibinfo {pages} {86} (\bibinfo {year} {2004})}\BibitemShut {NoStop}%
\bibitem [{\citenamefont {Kirby}\ \emph {et~al.}(2010)\citenamefont {Kirby}, \citenamefont {Blumenfeld}, \citenamefont {Hogan}, \citenamefont {Siemann}, \citenamefont {Walz}, \citenamefont {Davidson},\ and\ \citenamefont {Huang}}]{Kirby:2009zza}%
  \BibitemOpen
  \bibfield  {author} {\bibinfo {author} {\bibfnamefont {N.}~\bibnamefont {Kirby}}, \bibinfo {author} {\bibfnamefont {I.}~\bibnamefont {Blumenfeld}}, \bibinfo {author} {\bibfnamefont {M.~J.}\ \bibnamefont {Hogan}}, \bibinfo {author} {\bibfnamefont {R.~H.}\ \bibnamefont {Siemann}}, \bibinfo {author} {\bibfnamefont {D.~R.}\ \bibnamefont {Walz}}, \bibinfo {author} {\bibfnamefont {A.~W.}\ \bibnamefont {Davidson}},\ and\ \bibinfo {author} {\bibfnamefont {C.}~\bibnamefont {Huang}},\ }\bibfield  {title} {\bibinfo {title} {{Investigation of a gas jet-produced hollow plasma wakefield accelerator}},\ }in\ \href {https://accelconf.web.cern.ch/pac2009/papers/fr5rfp017.pdf} {\emph {\bibinfo {booktitle} {{Proceedings of PAC2009, Vancouver, BC, Canada}}}}\ (\bibinfo {year} {2010})\ p.\ \bibinfo {pages} {4566}\BibitemShut {NoStop}%
\bibitem [{\citenamefont {Fan}\ \emph {et~al.}(2000)\citenamefont {Fan}, \citenamefont {Parra}, \citenamefont {Alexeev}, \citenamefont {Kim}, \citenamefont {Milchberg}, \citenamefont {Margolin},\ and\ \citenamefont {Pyatnitskii}}]{Fan_2000}%
  \BibitemOpen
  \bibfield  {author} {\bibinfo {author} {\bibfnamefont {J.}~\bibnamefont {Fan}}, \bibinfo {author} {\bibfnamefont {E.}~\bibnamefont {Parra}}, \bibinfo {author} {\bibfnamefont {I.}~\bibnamefont {Alexeev}}, \bibinfo {author} {\bibfnamefont {K.~Y.}\ \bibnamefont {Kim}}, \bibinfo {author} {\bibfnamefont {H.~M.}\ \bibnamefont {Milchberg}}, \bibinfo {author} {\bibfnamefont {L.~Y.}\ \bibnamefont {Margolin}},\ and\ \bibinfo {author} {\bibfnamefont {L.~N.}\ \bibnamefont {Pyatnitskii}},\ }\bibfield  {title} {\bibinfo {title} {Tubular plasma generation with a high-power hollow bessel beam},\ }\href {https://doi.org/10.1103/PhysRevE.62.R7603} {\bibfield  {journal} {\bibinfo  {journal} {Phys. Rev. E}\ }\textbf {\bibinfo {volume} {62}},\ \bibinfo {pages} {R7603} (\bibinfo {year} {2000})}\BibitemShut {NoStop}%
\bibitem [{\citenamefont {Andreev}\ \emph {et~al.}(1996)\citenamefont {Andreev}, \citenamefont {Bychkov}, \citenamefont {Kotlyar}, \citenamefont {Margolin}, \citenamefont {Pyatnitskii},\ and\ \citenamefont {Serafimovich}}]{Andreev_1996}%
  \BibitemOpen
  \bibfield  {author} {\bibinfo {author} {\bibfnamefont {N.~E.}\ \bibnamefont {Andreev}}, \bibinfo {author} {\bibfnamefont {S.~S.}\ \bibnamefont {Bychkov}}, \bibinfo {author} {\bibfnamefont {V.~V.}\ \bibnamefont {Kotlyar}}, \bibinfo {author} {\bibfnamefont {L.~Y.}\ \bibnamefont {Margolin}}, \bibinfo {author} {\bibfnamefont {L.~N.}\ \bibnamefont {Pyatnitskii}},\ and\ \bibinfo {author} {\bibfnamefont {P.~G.}\ \bibnamefont {Serafimovich}},\ }\bibfield  {title} {\bibinfo {title} {Formation of high-power hollow bessel light beams},\ }\href {https://doi.org/10.1070/QE1996v026n02ABEH000607} {\bibfield  {journal} {\bibinfo  {journal} {Quantum Electron.}\ }\textbf {\bibinfo {volume} {26}},\ \bibinfo {pages} {126} (\bibinfo {year} {1996})}\BibitemShut {NoStop}%
\bibitem [{\citenamefont {Gessner}\ \emph {et~al.}(2016{\natexlab{b}})\citenamefont {Gessner}, \citenamefont {Allen}, \citenamefont {Clarke}, \citenamefont {Delahaye}, \citenamefont {Frederico}, \citenamefont {Green} \emph {et~al.}}]{Gessner:2016uof}%
  \BibitemOpen
  \bibfield  {author} {\bibinfo {author} {\bibfnamefont {S.}~\bibnamefont {Gessner}}, \bibinfo {author} {\bibfnamefont {J.~M.}\ \bibnamefont {Allen}}, \bibinfo {author} {\bibfnamefont {C.~I.}\ \bibnamefont {Clarke}}, \bibinfo {author} {\bibfnamefont {J.-P.}\ \bibnamefont {Delahaye}}, \bibinfo {author} {\bibfnamefont {J.~T.}\ \bibnamefont {Frederico}}, \bibinfo {author} {\bibfnamefont {S.}~\bibnamefont {Green}}, \emph {et~al.},\ }\bibfield  {title} {\bibinfo {title} {{Demonstration of the Hollow Channel Plasma Wakefield Accelerator}},\ }in\ \href {https://doi.org/10.18429/JACoW-IPAC2016-THPPA01} {\emph {\bibinfo {booktitle} {{Proceedings of IPAC2016, Busan, Korea}}}}\ (\bibinfo {year} {2016})\ p.\ \bibinfo {pages} {3202}\BibitemShut {NoStop}%
\bibitem [{\citenamefont {Gessner}(2016)}]{gessner_phd}%
  \BibitemOpen
  \bibfield  {author} {\bibinfo {author} {\bibfnamefont {S.~J.}\ \bibnamefont {Gessner}},\ }\emph {\bibinfo {title} {Demonstration of the hollow channel plasma wakefield accelerator}},\ \href {https://doi.org/10.2172/1340170} {Ph.D. thesis},\ \bibinfo  {school} {Stanford University} (\bibinfo {year} {2016})\BibitemShut {NoStop}%
\bibitem [{\citenamefont {Penn}\ \emph {et~al.}(2017)\citenamefont {Penn}, \citenamefont {Vay}, \citenamefont {Lehe}, \citenamefont {Schroeder},\ and\ \citenamefont {Esarey}}]{Penn:2017uyh}%
  \BibitemOpen
  \bibfield  {author} {\bibinfo {author} {\bibfnamefont {G.}~\bibnamefont {Penn}}, \bibinfo {author} {\bibfnamefont {J.~L.}\ \bibnamefont {Vay}}, \bibinfo {author} {\bibfnamefont {R.}~\bibnamefont {Lehe}}, \bibinfo {author} {\bibfnamefont {C.}~\bibnamefont {Schroeder}},\ and\ \bibinfo {author} {\bibfnamefont {E.}~\bibnamefont {Esarey}},\ }\bibfield  {title} {\bibinfo {title} {{Beam breakup studies in a hollow plasma channel}},\ }\href {https://doi.org/10.1063/1.4975856} {\bibfield  {journal} {\bibinfo  {journal} {AIP Conf. Proc.}\ }\textbf {\bibinfo {volume} {1812}},\ \bibinfo {pages} {040009} (\bibinfo {year} {2017})}\BibitemShut {NoStop}%
\bibitem [{\citenamefont {Yi}\ \emph {et~al.}(2013)\citenamefont {Yi} \emph {et~al.}}]{Yi:2013upa}%
  \BibitemOpen
  \bibfield  {author} {\bibinfo {author} {\bibfnamefont {L.}~\bibnamefont {Yi}} \emph {et~al.},\ }\bibfield  {title} {\bibinfo {title} {{Scheme for proton-driven plasma-wakefield acceleration of positively charged particles in a hollow plasma channel}},\ }\href {https://doi.org/10.1103/PhysRevSTAB.16.071301} {\bibfield  {journal} {\bibinfo  {journal} {Phys. Rev. ST Accel. Beams}\ }\textbf {\bibinfo {volume} {16}},\ \bibinfo {pages} {071301} (\bibinfo {year} {2013})}\BibitemShut {NoStop}%
\bibitem [{\citenamefont {Yi}\ \emph {et~al.}(2014)\citenamefont {Yi}, \citenamefont {Shen}, \citenamefont {Ji}, \citenamefont {Lotov}, \citenamefont {Sosedkin}, \citenamefont {Zhang} \emph {et~al.}}]{Yi:2014gta}%
  \BibitemOpen
  \bibfield  {author} {\bibinfo {author} {\bibfnamefont {L.}~\bibnamefont {Yi}}, \bibinfo {author} {\bibfnamefont {B.}~\bibnamefont {Shen}}, \bibinfo {author} {\bibfnamefont {L.}~\bibnamefont {Ji}}, \bibinfo {author} {\bibfnamefont {K.}~\bibnamefont {Lotov}}, \bibinfo {author} {\bibfnamefont {A.}~\bibnamefont {Sosedkin}}, \bibinfo {author} {\bibfnamefont {X.}~\bibnamefont {Zhang}}, \emph {et~al.},\ }\bibfield  {title} {\bibinfo {title} {{Positron acceleration in a hollow plasma channel up to TeV regime}},\ }\href {https://doi.org/10.1038/srep04171} {\bibfield  {journal} {\bibinfo  {journal} {Sci. Rep.}\ }\textbf {\bibinfo {volume} {4}},\ \bibinfo {pages} {4171} (\bibinfo {year} {2014})}\BibitemShut {NoStop}%
\bibitem [{\citenamefont {Li}\ \emph {et~al.}(2019{\natexlab{b}})\citenamefont {Li}, \citenamefont {Xia}, \citenamefont {Lotov}, \citenamefont {Sosedkin},\ and\ \citenamefont {Zhao}}]{Li_2019}%
  \BibitemOpen
  \bibfield  {author} {\bibinfo {author} {\bibfnamefont {Y.}~\bibnamefont {Li}}, \bibinfo {author} {\bibfnamefont {G.}~\bibnamefont {Xia}}, \bibinfo {author} {\bibfnamefont {K.~V.}\ \bibnamefont {Lotov}}, \bibinfo {author} {\bibfnamefont {A.~P.}\ \bibnamefont {Sosedkin}},\ and\ \bibinfo {author} {\bibfnamefont {Y.}~\bibnamefont {Zhao}},\ }\bibfield  {title} {\bibinfo {title} {High-quality positrons from a multi-proton bunch driven hollow plasma wakefield accelerator},\ }\href {https://doi.org/10.1088/1361-6587/aaf121} {\bibfield  {journal} {\bibinfo  {journal} {Plasma Phys. Control. Fusion}\ }\textbf {\bibinfo {volume} {61}},\ \bibinfo {pages} {025012} (\bibinfo {year} {2019}{\natexlab{b}})}\BibitemShut {NoStop}%
\bibitem [{\citenamefont {Amorim}\ \emph {et~al.}(2016)\citenamefont {Amorim}, \citenamefont {Vieira}, \citenamefont {Fonseca},\ and\ \citenamefont {Silva}}]{Amorim:2016prc}%
  \BibitemOpen
  \bibfield  {author} {\bibinfo {author} {\bibfnamefont {L.~D.}\ \bibnamefont {Amorim}}, \bibinfo {author} {\bibfnamefont {J.}~\bibnamefont {Vieira}}, \bibinfo {author} {\bibfnamefont {R.~A.}\ \bibnamefont {Fonseca}},\ and\ \bibinfo {author} {\bibfnamefont {L.~O.}\ \bibnamefont {Silva}},\ }\bibfield  {title} {\bibinfo {title} {{Positron plasma wakefield acceleration in a self-driven hollow channel}},\ }\href {https://doi.org/10.1063/1.4965644} {\bibfield  {journal} {\bibinfo  {journal} {AIP Conf. Proc.}\ }\textbf {\bibinfo {volume} {1777}},\ \bibinfo {pages} {070001} (\bibinfo {year} {2016})}\BibitemShut {NoStop}%
\bibitem [{\citenamefont {Golovanov}\ \emph {et~al.}(2017)\citenamefont {Golovanov}, \citenamefont {Kostyukov}, \citenamefont {Thomas},\ and\ \citenamefont {Pukhov}}]{Golovanov_2017}%
  \BibitemOpen
  \bibfield  {author} {\bibinfo {author} {\bibfnamefont {A.~A.}\ \bibnamefont {Golovanov}}, \bibinfo {author} {\bibfnamefont {I.~Y.}\ \bibnamefont {Kostyukov}}, \bibinfo {author} {\bibfnamefont {J.}~\bibnamefont {Thomas}},\ and\ \bibinfo {author} {\bibfnamefont {A.}~\bibnamefont {Pukhov}},\ }\bibfield  {title} {\bibinfo {title} {{Analytic model for electromagnetic fields in the bubble regime of plasma wakefield in non-uniform plasmas}},\ }\href {https://doi.org/10.1063/1.4996856} {\bibfield  {journal} {\bibinfo  {journal} {Phys. Plasmas}\ }\textbf {\bibinfo {volume} {24}},\ \bibinfo {pages} {103104} (\bibinfo {year} {2017})}\BibitemShut {NoStop}%
\bibitem [{\citenamefont {Wu}\ \emph {et~al.}(2019{\natexlab{b}})\citenamefont {Wu}, \citenamefont {Hua}, \citenamefont {Pai}, \citenamefont {An}, \citenamefont {Zhou}, \citenamefont {Zhang} \emph {et~al.}}]{Wu_2019_hc}%
  \BibitemOpen
  \bibfield  {author} {\bibinfo {author} {\bibfnamefont {Y.}~\bibnamefont {Wu}}, \bibinfo {author} {\bibfnamefont {J.}~\bibnamefont {Hua}}, \bibinfo {author} {\bibfnamefont {C.-H.}\ \bibnamefont {Pai}}, \bibinfo {author} {\bibfnamefont {W.}~\bibnamefont {An}}, \bibinfo {author} {\bibfnamefont {Z.}~\bibnamefont {Zhou}}, \bibinfo {author} {\bibfnamefont {J.}~\bibnamefont {Zhang}}, \emph {et~al.},\ }\bibfield  {title} {\bibinfo {title} {Near-ideal dechirper for plasma-based electron and positron acceleration using a hollow channel plasma},\ }\href {https://doi.org/10.1103/PhysRevApplied.12.064011} {\bibfield  {journal} {\bibinfo  {journal} {Phys. Rev. Appl.}\ }\textbf {\bibinfo {volume} {12}},\ \bibinfo {pages} {064011} (\bibinfo {year} {2019}{\natexlab{b}})}\BibitemShut {NoStop}%
\bibitem [{\citenamefont {Wu}\ \emph {et~al.}(2023)\citenamefont {Wu}, \citenamefont {Zhou}, \citenamefont {Du}, \citenamefont {Hua}, \citenamefont {Lu}, \citenamefont {Mori},\ and\ \citenamefont {Joshi}}]{Wu_2023}%
  \BibitemOpen
  \bibfield  {author} {\bibinfo {author} {\bibfnamefont {Y.}~\bibnamefont {Wu}}, \bibinfo {author} {\bibfnamefont {Z.}~\bibnamefont {Zhou}}, \bibinfo {author} {\bibfnamefont {Y.}~\bibnamefont {Du}}, \bibinfo {author} {\bibfnamefont {J.}~\bibnamefont {Hua}}, \bibinfo {author} {\bibfnamefont {W.}~\bibnamefont {Lu}}, \bibinfo {author} {\bibfnamefont {W.~B.}\ \bibnamefont {Mori}},\ and\ \bibinfo {author} {\bibfnamefont {C.}~\bibnamefont {Joshi}},\ }\bibfield  {title} {\bibinfo {title} {Linearization of an electron beam's longitudinal phase space using a hollow-channel plasma},\ }\href {https://doi.org/10.1103/PhysRevApplied.19.064013} {\bibfield  {journal} {\bibinfo  {journal} {Phys. Rev. Appl.}\ }\textbf {\bibinfo {volume} {19}},\ \bibinfo {pages} {064013} (\bibinfo {year} {2023})}\BibitemShut {NoStop}%
\bibitem [{\citenamefont {Yakimenko}\ \emph {et~al.}(2019)\citenamefont {Yakimenko}, \citenamefont {Alsberg}, \citenamefont {Bong}, \citenamefont {Bouchard}, \citenamefont {Clarke}, \citenamefont {Emma} \emph {et~al.}}]{facet_ii}%
  \BibitemOpen
  \bibfield  {author} {\bibinfo {author} {\bibfnamefont {V.}~\bibnamefont {Yakimenko}}, \bibinfo {author} {\bibfnamefont {L.}~\bibnamefont {Alsberg}}, \bibinfo {author} {\bibfnamefont {E.}~\bibnamefont {Bong}}, \bibinfo {author} {\bibfnamefont {G.}~\bibnamefont {Bouchard}}, \bibinfo {author} {\bibfnamefont {C.}~\bibnamefont {Clarke}}, \bibinfo {author} {\bibfnamefont {C.}~\bibnamefont {Emma}}, \emph {et~al.},\ }\bibfield  {title} {\bibinfo {title} {{FACET-II} facility for advanced accelerator experimental tests},\ }\href {https://doi.org/10.1103/PhysRevAccelBeams.22.101301} {\bibfield  {journal} {\bibinfo  {journal} {Phys. Rev. Accel. Beams}\ }\textbf {\bibinfo {volume} {22}},\ \bibinfo {pages} {101301} (\bibinfo {year} {2019})}\BibitemShut {NoStop}%
\bibitem [{\citenamefont {Gahn}\ \emph {et~al.}(2002)\citenamefont {Gahn}, \citenamefont {Tsakiris}, \citenamefont {Pretzler}, \citenamefont {Witte}, \citenamefont {Thirolf}, \citenamefont {Habs}, \citenamefont {Delfin},\ and\ \citenamefont {Wahlström}}]{lwfa_gahn}%
  \BibitemOpen
  \bibfield  {author} {\bibinfo {author} {\bibfnamefont {C.}~\bibnamefont {Gahn}}, \bibinfo {author} {\bibfnamefont {G.~D.}\ \bibnamefont {Tsakiris}}, \bibinfo {author} {\bibfnamefont {G.}~\bibnamefont {Pretzler}}, \bibinfo {author} {\bibfnamefont {K.~J.}\ \bibnamefont {Witte}}, \bibinfo {author} {\bibfnamefont {P.}~\bibnamefont {Thirolf}}, \bibinfo {author} {\bibfnamefont {D.}~\bibnamefont {Habs}}, \bibinfo {author} {\bibfnamefont {C.}~\bibnamefont {Delfin}},\ and\ \bibinfo {author} {\bibfnamefont {C.-G.}\ \bibnamefont {Wahlström}},\ }\bibfield  {title} {\bibinfo {title} {{Generation of MeV electrons and positrons with femtosecond pulses from a table-top laser system}},\ }\href {https://doi.org/10.1063/1.1446879} {\bibfield  {journal} {\bibinfo  {journal} {Phys. Plasmas}\ }\textbf {\bibinfo {volume} {9}},\ \bibinfo {pages} {987} (\bibinfo {year} {2002})}\BibitemShut {NoStop}%
\bibitem [{\citenamefont {Chen}\ \emph {et~al.}(2009)\citenamefont {Chen}, \citenamefont {Wilks}, \citenamefont {Bonlie}, \citenamefont {Liang}, \citenamefont {Myatt}, \citenamefont {Price}, \citenamefont {Meyerhofer},\ and\ \citenamefont {Beiersdorfer}}]{lwfa_Chen}%
  \BibitemOpen
  \bibfield  {author} {\bibinfo {author} {\bibfnamefont {H.}~\bibnamefont {Chen}}, \bibinfo {author} {\bibfnamefont {S.~C.}\ \bibnamefont {Wilks}}, \bibinfo {author} {\bibfnamefont {J.~D.}\ \bibnamefont {Bonlie}}, \bibinfo {author} {\bibfnamefont {E.~P.}\ \bibnamefont {Liang}}, \bibinfo {author} {\bibfnamefont {J.}~\bibnamefont {Myatt}}, \bibinfo {author} {\bibfnamefont {D.~F.}\ \bibnamefont {Price}}, \bibinfo {author} {\bibfnamefont {D.~D.}\ \bibnamefont {Meyerhofer}},\ and\ \bibinfo {author} {\bibfnamefont {P.}~\bibnamefont {Beiersdorfer}},\ }\bibfield  {title} {\bibinfo {title} {Relativistic positron creation using ultraintense short pulse lasers},\ }\href {https://doi.org/10.1103/PhysRevLett.102.105001} {\bibfield  {journal} {\bibinfo  {journal} {Phys. Rev. Lett.}\ }\textbf {\bibinfo {volume} {102}},\ \bibinfo {pages} {105001} (\bibinfo {year} {2009})}\BibitemShut {NoStop}%
\bibitem [{\citenamefont {Chen}\ \emph {et~al.}(2010)\citenamefont {Chen}, \citenamefont {Wilks}, \citenamefont {Meyerhofer}, \citenamefont {Bonlie}, \citenamefont {Chen}, \citenamefont {Chen} \emph {et~al.}}]{lwfa_Chen2}%
  \BibitemOpen
  \bibfield  {author} {\bibinfo {author} {\bibfnamefont {H.}~\bibnamefont {Chen}}, \bibinfo {author} {\bibfnamefont {S.~C.}\ \bibnamefont {Wilks}}, \bibinfo {author} {\bibfnamefont {D.~D.}\ \bibnamefont {Meyerhofer}}, \bibinfo {author} {\bibfnamefont {J.}~\bibnamefont {Bonlie}}, \bibinfo {author} {\bibfnamefont {C.~D.}\ \bibnamefont {Chen}}, \bibinfo {author} {\bibfnamefont {S.~N.}\ \bibnamefont {Chen}}, \emph {et~al.},\ }\bibfield  {title} {\bibinfo {title} {Relativistic quasimonoenergetic positron jets from intense laser-solid interactions},\ }\href {https://doi.org/10.1103/PhysRevLett.105.015003} {\bibfield  {journal} {\bibinfo  {journal} {Phys. Rev. Lett.}\ }\textbf {\bibinfo {volume} {105}},\ \bibinfo {pages} {015003} (\bibinfo {year} {2010})}\BibitemShut {NoStop}%
\bibitem [{\citenamefont {Sarri}\ \emph {et~al.}(2013)\citenamefont {Sarri}, \citenamefont {Schumaker}, \citenamefont {Di~Piazza}, \citenamefont {Vargas}, \citenamefont {Dromey}, \citenamefont {Dieckmann} \emph {et~al.}}]{lwfa_Sarri}%
  \BibitemOpen
  \bibfield  {author} {\bibinfo {author} {\bibfnamefont {G.}~\bibnamefont {Sarri}}, \bibinfo {author} {\bibfnamefont {W.}~\bibnamefont {Schumaker}}, \bibinfo {author} {\bibfnamefont {A.}~\bibnamefont {Di~Piazza}}, \bibinfo {author} {\bibfnamefont {M.}~\bibnamefont {Vargas}}, \bibinfo {author} {\bibfnamefont {B.}~\bibnamefont {Dromey}}, \bibinfo {author} {\bibfnamefont {M.~E.}\ \bibnamefont {Dieckmann}}, \emph {et~al.},\ }\bibfield  {title} {\bibinfo {title} {Table-top laser-based source of femtosecond, collimated, ultrarelativistic positron beams},\ }\href {https://doi.org/10.1103/PhysRevLett.110.255002} {\bibfield  {journal} {\bibinfo  {journal} {Phys. Rev. Lett.}\ }\textbf {\bibinfo {volume} {110}},\ \bibinfo {pages} {255002} (\bibinfo {year} {2013})}\BibitemShut {NoStop}%
\bibitem [{\citenamefont {Wang}\ \emph {et~al.}(2006)\citenamefont {Wang}, \citenamefont {Ischebeck}, \citenamefont {Joshi}, \citenamefont {Muggli},\ and\ \citenamefont {Katsouleas}}]{wang_2006}%
  \BibitemOpen
  \bibfield  {author} {\bibinfo {author} {\bibfnamefont {X.}~\bibnamefont {Wang}}, \bibinfo {author} {\bibfnamefont {R.}~\bibnamefont {Ischebeck}}, \bibinfo {author} {\bibfnamefont {C.}~\bibnamefont {Joshi}}, \bibinfo {author} {\bibfnamefont {P.}~\bibnamefont {Muggli}},\ and\ \bibinfo {author} {\bibfnamefont {T.}~\bibnamefont {Katsouleas}},\ }\bibfield  {title} {\bibinfo {title} {Novel module for plasma wakefield acceleration of a positron beam},\ }\href {https://doi.org/10.1063/1.2409185} {\bibfield  {journal} {\bibinfo  {journal} {AIP Conf. Proc.}\ }\textbf {\bibinfo {volume} {877}},\ \bibinfo {pages} {568} (\bibinfo {year} {2006})}\BibitemShut {NoStop}%
\bibitem [{\citenamefont {Wang}\ \emph {et~al.}(2008)\citenamefont {Wang}, \citenamefont {Ischebeck}, \citenamefont {Muggli}, \citenamefont {Katsouleas}, \citenamefont {Joshi}, \citenamefont {Mori},\ and\ \citenamefont {Hogan}}]{wang_2008}%
  \BibitemOpen
  \bibfield  {author} {\bibinfo {author} {\bibfnamefont {X.}~\bibnamefont {Wang}}, \bibinfo {author} {\bibfnamefont {R.}~\bibnamefont {Ischebeck}}, \bibinfo {author} {\bibfnamefont {P.}~\bibnamefont {Muggli}}, \bibinfo {author} {\bibfnamefont {T.}~\bibnamefont {Katsouleas}}, \bibinfo {author} {\bibfnamefont {C.}~\bibnamefont {Joshi}}, \bibinfo {author} {\bibfnamefont {W.~B.}\ \bibnamefont {Mori}},\ and\ \bibinfo {author} {\bibfnamefont {M.~J.}\ \bibnamefont {Hogan}},\ }\bibfield  {title} {\bibinfo {title} {Positron injection and acceleration on the wake driven by an electron beam in a foil-and-gas plasma},\ }\href {https://doi.org/10.1103/PhysRevLett.101.124801} {\bibfield  {journal} {\bibinfo  {journal} {Phys. Rev. Lett.}\ }\textbf {\bibinfo {volume} {101}},\ \bibinfo {pages} {124801} (\bibinfo {year} {2008})}\BibitemShut {NoStop}%
\bibitem [{\citenamefont {Wang}\ \emph {et~al.}(2009)\citenamefont {Wang}, \citenamefont {Muggli}, \citenamefont {Katsouleas}, \citenamefont {Joshi}, \citenamefont {Mori}, \citenamefont {Ischebeck},\ and\ \citenamefont {Hogan}}]{wang_2009}%
  \BibitemOpen
  \bibfield  {author} {\bibinfo {author} {\bibfnamefont {X.}~\bibnamefont {Wang}}, \bibinfo {author} {\bibfnamefont {P.}~\bibnamefont {Muggli}}, \bibinfo {author} {\bibfnamefont {T.}~\bibnamefont {Katsouleas}}, \bibinfo {author} {\bibfnamefont {C.}~\bibnamefont {Joshi}}, \bibinfo {author} {\bibfnamefont {W.~B.}\ \bibnamefont {Mori}}, \bibinfo {author} {\bibfnamefont {R.}~\bibnamefont {Ischebeck}},\ and\ \bibinfo {author} {\bibfnamefont {M.~J.}\ \bibnamefont {Hogan}},\ }\bibfield  {title} {\bibinfo {title} {Optimization of positron trapping and acceleration in an electron-beam-driven plasma wakefield accelerator},\ }\href {https://doi.org/10.1103/PhysRevSTAB.12.051303} {\bibfield  {journal} {\bibinfo  {journal} {Phys. Rev. ST Accel. Beams}\ }\textbf {\bibinfo {volume} {12}},\ \bibinfo {pages} {051303} (\bibinfo {year} {2009})}\BibitemShut {NoStop}%
\bibitem [{\citenamefont {Firouzjaei}\ and\ \citenamefont {Shokri}(2017)}]{Firouzjaei:2017kfh}%
  \BibitemOpen
  \bibfield  {author} {\bibinfo {author} {\bibfnamefont {A.~S.}\ \bibnamefont {Firouzjaei}}\ and\ \bibinfo {author} {\bibfnamefont {B.}~\bibnamefont {Shokri}},\ }\bibfield  {title} {\bibinfo {title} {{Trapping and acceleration of hollow electron and positron bunch in a quasi-linear donut wakefield}},\ }\href {https://doi.org/10.1063/1.4973598} {\bibfield  {journal} {\bibinfo  {journal} {Phys. Plasmas}\ }\textbf {\bibinfo {volume} {24}},\ \bibinfo {pages} {013107} (\bibinfo {year} {2017})}\BibitemShut {NoStop}%
\bibitem [{\citenamefont {Sahai}(2018)}]{Sahai:2018ksu}%
  \BibitemOpen
  \bibfield  {author} {\bibinfo {author} {\bibfnamefont {A.~A.}\ \bibnamefont {Sahai}},\ }\bibfield  {title} {\bibinfo {title} {{Quasimonoenergetic laser plasma positron accelerator using particle-shower plasma-wave interactions}},\ }\href {https://doi.org/10.1103/PhysRevAccelBeams.21.081301} {\bibfield  {journal} {\bibinfo  {journal} {Phys. Rev. Accel. Beams}\ }\textbf {\bibinfo {volume} {21}},\ \bibinfo {pages} {081301} (\bibinfo {year} {2018})}\BibitemShut {NoStop}%
\bibitem [{\citenamefont {Xu}\ \emph {et~al.}(2020)\citenamefont {Xu} \emph {et~al.}}]{Xu:2019zov}%
  \BibitemOpen
  \bibfield  {author} {\bibinfo {author} {\bibfnamefont {Z.~Y.}\ \bibnamefont {Xu}} \emph {et~al.},\ }\bibfield  {title} {\bibinfo {title} {{New injection and acceleration scheme of positrons in the laser-plasma bubble regime}},\ }\href {https://doi.org/10.1103/PhysRevAccelBeams.23.091301} {\bibfield  {journal} {\bibinfo  {journal} {Phys. Rev. Accel. Beams}\ }\textbf {\bibinfo {volume} {23}},\ \bibinfo {pages} {091301} (\bibinfo {year} {2020})}\BibitemShut {NoStop}%
\bibitem [{\citenamefont {Martinez}\ \emph {et~al.}(2023)\citenamefont {Martinez}, \citenamefont {Barbosa},\ and\ \citenamefont {Vranic}}]{Martinez:2022idh}%
  \BibitemOpen
  \bibfield  {author} {\bibinfo {author} {\bibfnamefont {B.}~\bibnamefont {Martinez}}, \bibinfo {author} {\bibfnamefont {B.}~\bibnamefont {Barbosa}},\ and\ \bibinfo {author} {\bibfnamefont {M.}~\bibnamefont {Vranic}},\ }\bibfield  {title} {\bibinfo {title} {Creation and direct laser acceleration of positrons in a single stage},\ }\href {https://doi.org/10.1103/PhysRevAccelBeams.26.011301} {\bibfield  {journal} {\bibinfo  {journal} {Phys. Rev. Accel. Beams}\ }\textbf {\bibinfo {volume} {26}},\ \bibinfo {pages} {011301} (\bibinfo {year} {2023})}\BibitemShut {NoStop}%
\bibitem [{\citenamefont {Liu}\ \emph {et~al.}(2022)\citenamefont {Liu}, \citenamefont {Xue}, \citenamefont {Wan}, \citenamefont {Chen}, \citenamefont {Li}, \citenamefont {Liu}, \citenamefont {Weng}, \citenamefont {Sheng},\ and\ \citenamefont {Zhang}}]{Liu_2022b}%
  \BibitemOpen
  \bibfield  {author} {\bibinfo {author} {\bibfnamefont {W.-Y.}\ \bibnamefont {Liu}}, \bibinfo {author} {\bibfnamefont {K.}~\bibnamefont {Xue}}, \bibinfo {author} {\bibfnamefont {F.}~\bibnamefont {Wan}}, \bibinfo {author} {\bibfnamefont {M.}~\bibnamefont {Chen}}, \bibinfo {author} {\bibfnamefont {J.-X.}\ \bibnamefont {Li}}, \bibinfo {author} {\bibfnamefont {F.}~\bibnamefont {Liu}}, \bibinfo {author} {\bibfnamefont {S.-M.}\ \bibnamefont {Weng}}, \bibinfo {author} {\bibfnamefont {Z.-M.}\ \bibnamefont {Sheng}},\ and\ \bibinfo {author} {\bibfnamefont {J.}~\bibnamefont {Zhang}},\ }\bibfield  {title} {\bibinfo {title} {Trapping and acceleration of spin-polarized positrons from $\ensuremath{\gamma}$ photon splitting in wakefields},\ }\href {https://doi.org/10.1103/PhysRevResearch.4.L022028} {\bibfield  {journal} {\bibinfo  {journal} {Phys. Rev. Res.}\ }\textbf {\bibinfo {volume} {4}},\ \bibinfo {pages} {L022028} (\bibinfo {year} {2022})}\BibitemShut {NoStop}%
\bibitem [{\citenamefont {Amorim}\ \emph {et~al.}(2023)\citenamefont {Amorim}, \citenamefont {Benedetti}, \citenamefont {Bulanov}, \citenamefont {Terzani}, \citenamefont {Huebl}, \citenamefont {Schroeder}, \citenamefont {Vay},\ and\ \citenamefont {Esarey}}]{Amorim:2023bof}%
  \BibitemOpen
  \bibfield  {author} {\bibinfo {author} {\bibfnamefont {L.~D.}\ \bibnamefont {Amorim}}, \bibinfo {author} {\bibfnamefont {C.}~\bibnamefont {Benedetti}}, \bibinfo {author} {\bibfnamefont {S.~S.}\ \bibnamefont {Bulanov}}, \bibinfo {author} {\bibfnamefont {D.}~\bibnamefont {Terzani}}, \bibinfo {author} {\bibfnamefont {A.}~\bibnamefont {Huebl}}, \bibinfo {author} {\bibfnamefont {C.~B.}\ \bibnamefont {Schroeder}}, \bibinfo {author} {\bibfnamefont {J.-L.}\ \bibnamefont {Vay}},\ and\ \bibinfo {author} {\bibfnamefont {E.}~\bibnamefont {Esarey}},\ }\bibfield  {title} {\bibinfo {title} {{Design study for a compact, two-stage, laser-plasma-based source of positron beams}},\ }\href {https://doi.org/10.1088/1361-6587/ace3f1} {\bibfield  {journal} {\bibinfo  {journal} {Plasma Phys. Control. Fusion}\ }\textbf {\bibinfo {volume} {65}},\ \bibinfo {pages} {085016} (\bibinfo {year} {2023})}\BibitemShut {NoStop}%
\bibitem [{\citenamefont {Sugimoto}\ \emph {et~al.}(2023)\citenamefont {Sugimoto}, \citenamefont {He}, \citenamefont {Iwata}, \citenamefont {Yeh}, \citenamefont {Tangtartharakul}, \citenamefont {Arefiev},\ and\ \citenamefont {Sentoku}}]{Sugimoto_2023}%
  \BibitemOpen
  \bibfield  {author} {\bibinfo {author} {\bibfnamefont {K.}~\bibnamefont {Sugimoto}}, \bibinfo {author} {\bibfnamefont {Y.}~\bibnamefont {He}}, \bibinfo {author} {\bibfnamefont {N.}~\bibnamefont {Iwata}}, \bibinfo {author} {\bibfnamefont {I.-L.}\ \bibnamefont {Yeh}}, \bibinfo {author} {\bibfnamefont {K.}~\bibnamefont {Tangtartharakul}}, \bibinfo {author} {\bibfnamefont {A.}~\bibnamefont {Arefiev}},\ and\ \bibinfo {author} {\bibfnamefont {Y.}~\bibnamefont {Sentoku}},\ }\bibfield  {title} {\bibinfo {title} {Positron generation and acceleration in a self-organized photon collider enabled by an ultraintense laser pulse},\ }\href {https://doi.org/10.1103/PhysRevLett.131.065102} {\bibfield  {journal} {\bibinfo  {journal} {Phys. Rev. Lett.}\ }\textbf {\bibinfo {volume} {131}},\ \bibinfo {pages} {065102} (\bibinfo {year} {2023})}\BibitemShut {NoStop}%
\bibitem [{\citenamefont {Hessami}\ and\ \citenamefont {Gessner}(2023)}]{hessami2023}%
  \BibitemOpen
  \bibfield  {author} {\bibinfo {author} {\bibfnamefont {R.}~\bibnamefont {Hessami}}\ and\ \bibinfo {author} {\bibfnamefont {S.}~\bibnamefont {Gessner}},\ }\bibfield  {title} {\bibinfo {title} {Compact source of positron beams with small thermal emittance},\ }\href {https://doi.org/10.1103/PhysRevAccelBeams.26.123402} {\bibfield  {journal} {\bibinfo  {journal} {Phys. Rev. Accel. Beams}\ }\textbf {\bibinfo {volume} {26}},\ \bibinfo {pages} {123402} (\bibinfo {year} {2023})}\BibitemShut {NoStop}%
\bibitem [{\citenamefont {Esarey}\ \emph {et~al.}(1996)\citenamefont {Esarey}, \citenamefont {Sprangle}, \citenamefont {Krall},\ and\ \citenamefont {Ting}}]{Esarey:1996}%
  \BibitemOpen
  \bibfield  {author} {\bibinfo {author} {\bibfnamefont {E.}~\bibnamefont {Esarey}}, \bibinfo {author} {\bibfnamefont {P.}~\bibnamefont {Sprangle}}, \bibinfo {author} {\bibfnamefont {J.}~\bibnamefont {Krall}},\ and\ \bibinfo {author} {\bibfnamefont {A.}~\bibnamefont {Ting}},\ }\bibfield  {title} {\bibinfo {title} {Overview of plasma-based accelerator concepts},\ }\href {https://doi.org/10.1109/27.509991} {\bibfield  {journal} {\bibinfo  {journal} {IEEE T. Plasma Sci.}\ }\textbf {\bibinfo {volume} {24}},\ \bibinfo {pages} {252} (\bibinfo {year} {1996})}\BibitemShut {NoStop}%
\bibitem [{\citenamefont {Hue}\ \emph {et~al.}(2021)\citenamefont {Hue}, \citenamefont {Cao}, \citenamefont {Andriyash}, \citenamefont {Knetsch}, \citenamefont {Hogan}, \citenamefont {Adli}, \citenamefont {Gessner},\ and\ \citenamefont {Corde}}]{hue_2021}%
  \BibitemOpen
  \bibfield  {author} {\bibinfo {author} {\bibfnamefont {C.~S.}\ \bibnamefont {Hue}}, \bibinfo {author} {\bibfnamefont {G.~J.}\ \bibnamefont {Cao}}, \bibinfo {author} {\bibfnamefont {I.~A.}\ \bibnamefont {Andriyash}}, \bibinfo {author} {\bibfnamefont {A.}~\bibnamefont {Knetsch}}, \bibinfo {author} {\bibfnamefont {M.~J.}\ \bibnamefont {Hogan}}, \bibinfo {author} {\bibfnamefont {E.}~\bibnamefont {Adli}}, \bibinfo {author} {\bibfnamefont {S.}~\bibnamefont {Gessner}},\ and\ \bibinfo {author} {\bibfnamefont {S.}~\bibnamefont {Corde}},\ }\bibfield  {title} {\bibinfo {title} {{Efficiency and beam quality for positron acceleration in loaded plasma wakefields}},\ }\href {https://doi.org/10.1103/PhysRevResearch.3.043063} {\bibfield  {journal} {\bibinfo  {journal} {Phys. Rev. Research}\ }\textbf {\bibinfo {volume} {3}},\ \bibinfo {eid} {043063} (\bibinfo {year} {2021})}\BibitemShut {NoStop}%
\bibitem [{\citenamefont {Cormier-Michel}\ \emph {et~al.}(2011)\citenamefont {Cormier-Michel}, \citenamefont {Esarey}, \citenamefont {Geddes}, \citenamefont {Schroeder}, \citenamefont {Paul}, \citenamefont {Mullowney}, \citenamefont {Cary},\ and\ \citenamefont {Leemans}}]{cormier2011}%
  \BibitemOpen
  \bibfield  {author} {\bibinfo {author} {\bibfnamefont {E.}~\bibnamefont {Cormier-Michel}}, \bibinfo {author} {\bibfnamefont {E.}~\bibnamefont {Esarey}}, \bibinfo {author} {\bibfnamefont {C.~G.~R.}\ \bibnamefont {Geddes}}, \bibinfo {author} {\bibfnamefont {C.~B.}\ \bibnamefont {Schroeder}}, \bibinfo {author} {\bibfnamefont {K.}~\bibnamefont {Paul}}, \bibinfo {author} {\bibfnamefont {P.~J.}\ \bibnamefont {Mullowney}}, \bibinfo {author} {\bibfnamefont {J.~R.}\ \bibnamefont {Cary}},\ and\ \bibinfo {author} {\bibfnamefont {W.~P.}\ \bibnamefont {Leemans}},\ }\bibfield  {title} {\bibinfo {title} {Control of focusing fields in laser-plasma accelerators using higher-order modes},\ }\href {https://doi.org/10.1103/PhysRevSTAB.14.031303} {\bibfield  {journal} {\bibinfo  {journal} {Phys. Rev. ST Accel. Beams}\ }\textbf {\bibinfo {volume} {14}},\ \bibinfo {pages} {031303} (\bibinfo {year} {2011})}\BibitemShut {NoStop}%
\bibitem [{\citenamefont {Hue}(2020)}]{huephd_2020}%
  \BibitemOpen
  \bibfield  {author} {\bibinfo {author} {\bibfnamefont {C.}~\bibnamefont {Hue}},\ }\emph {\bibinfo {title} {{Efficiency and beam quality in a loaded quasi-linear plasma wakefield positron accelerator}}},\ \href {https://theses.hal.science/tel-03592844} {\bibinfo {type} {{PhD thesis}}},\ \bibinfo  {school} {{Institut Polytechnique de Paris}} (\bibinfo {year} {2020})\BibitemShut {NoStop}%
\bibitem [{\citenamefont {Rosenzweig}\ \emph {et~al.}(2004)\citenamefont {Rosenzweig}, \citenamefont {Barov}, \citenamefont {Thompson},\ and\ \citenamefont {Yoder}}]{Rosenzweig_2004}%
  \BibitemOpen
  \bibfield  {author} {\bibinfo {author} {\bibfnamefont {J.~B.}\ \bibnamefont {Rosenzweig}}, \bibinfo {author} {\bibfnamefont {N.}~\bibnamefont {Barov}}, \bibinfo {author} {\bibfnamefont {M.~C.}\ \bibnamefont {Thompson}},\ and\ \bibinfo {author} {\bibfnamefont {R.~B.}\ \bibnamefont {Yoder}},\ }\bibfield  {title} {\bibinfo {title} {Energy loss of a high charge bunched electron beam in plasma: Simulations, scaling, and accelerating wakefields},\ }\href {https://doi.org/10.1103/PhysRevSTAB.7.061302} {\bibfield  {journal} {\bibinfo  {journal} {Phys. Rev. ST Accel. Beams}\ }\textbf {\bibinfo {volume} {7}},\ \bibinfo {pages} {061302} (\bibinfo {year} {2004})}\BibitemShut {NoStop}%
\bibitem [{\citenamefont {Barov}\ \emph {et~al.}(2004)\citenamefont {Barov}, \citenamefont {Rosenzweig}, \citenamefont {Thompson},\ and\ \citenamefont {Yoder}}]{Barov_2004}%
  \BibitemOpen
  \bibfield  {author} {\bibinfo {author} {\bibfnamefont {N.}~\bibnamefont {Barov}}, \bibinfo {author} {\bibfnamefont {J.~B.}\ \bibnamefont {Rosenzweig}}, \bibinfo {author} {\bibfnamefont {M.~C.}\ \bibnamefont {Thompson}},\ and\ \bibinfo {author} {\bibfnamefont {R.~B.}\ \bibnamefont {Yoder}},\ }\bibfield  {title} {\bibinfo {title} {Energy loss of a high-charge bunched electron beam in plasma: Analysis},\ }\href {https://doi.org/10.1103/PhysRevSTAB.7.061301} {\bibfield  {journal} {\bibinfo  {journal} {Phys. Rev. ST Accel. Beams}\ }\textbf {\bibinfo {volume} {7}},\ \bibinfo {pages} {061301} (\bibinfo {year} {2004})}\BibitemShut {NoStop}%
\bibitem [{\citenamefont {Cao}\ \emph {et~al.}(2022)\citenamefont {Cao}, \citenamefont {Adli}, \citenamefont {Corde},\ and\ \citenamefont {Gessner}}]{Cao:2022zkb}%
  \BibitemOpen
  \bibfield  {author} {\bibinfo {author} {\bibfnamefont {J.}~\bibnamefont {Cao}}, \bibinfo {author} {\bibfnamefont {E.}~\bibnamefont {Adli}}, \bibinfo {author} {\bibfnamefont {S.}~\bibnamefont {Corde}},\ and\ \bibinfo {author} {\bibfnamefont {S.}~\bibnamefont {Gessner}},\ }\bibfield  {title} {\bibinfo {title} {{Positron Acceleration in Linear, Moderately Non-Linear and Non-Linear Plasma Wakefields}},\ }in\ \href {https://doi.org/10.18429/JACoW-NAPAC2022-WEYD3} {\emph {\bibinfo {booktitle} {Proceedings of NAPAC2022, Albuquerque, NM, USA}}}\ (\bibinfo {year} {2022})\ p.\ \bibinfo {pages} {560}\BibitemShut {NoStop}%
\bibitem [{\citenamefont {Muggli}\ \emph {et~al.}(2012)\citenamefont {Muggli}, \citenamefont {Allen}, \citenamefont {Fang}, \citenamefont {Yakimenko}, \citenamefont {Babzien}, \citenamefont {Kusche}, \citenamefont {Fedurin}, \citenamefont {Vieira}, \citenamefont {Martins},\ and\ \citenamefont {Silva}}]{Muggli_2012}%
  \BibitemOpen
  \bibfield  {author} {\bibinfo {author} {\bibfnamefont {P.}~\bibnamefont {Muggli}}, \bibinfo {author} {\bibfnamefont {B.}~\bibnamefont {Allen}}, \bibinfo {author} {\bibfnamefont {Y.}~\bibnamefont {Fang}}, \bibinfo {author} {\bibfnamefont {V.}~\bibnamefont {Yakimenko}}, \bibinfo {author} {\bibfnamefont {M.}~\bibnamefont {Babzien}}, \bibinfo {author} {\bibfnamefont {K.}~\bibnamefont {Kusche}}, \bibinfo {author} {\bibfnamefont {M.}~\bibnamefont {Fedurin}}, \bibinfo {author} {\bibfnamefont {J.}~\bibnamefont {Vieira}}, \bibinfo {author} {\bibfnamefont {J.}~\bibnamefont {Martins}},\ and\ \bibinfo {author} {\bibfnamefont {L.}~\bibnamefont {Silva}},\ }\bibfield  {title} {\bibinfo {title} {{Three regimes of relativistic beam - plasma interaction}},\ }\href {https://doi.org/10.1063/1.4773764} {\bibfield  {journal} {\bibinfo  {journal} {AIP Conf. Proc.}\ }\textbf {\bibinfo {volume} {1507}},\ \bibinfo {pages} {594} (\bibinfo {year} {2012})}\BibitemShut {NoStop}%
\bibitem [{\citenamefont {Londrillo}\ \emph {et~al.}(2014)\citenamefont {Londrillo}, \citenamefont {Gatti},\ and\ \citenamefont {Ferrario}}]{LONDRILLO2014236}%
  \BibitemOpen
  \bibfield  {author} {\bibinfo {author} {\bibfnamefont {P.}~\bibnamefont {Londrillo}}, \bibinfo {author} {\bibfnamefont {C.}~\bibnamefont {Gatti}},\ and\ \bibinfo {author} {\bibfnamefont {M.}~\bibnamefont {Ferrario}},\ }\bibfield  {title} {\bibinfo {title} {Numerical investigation of beam-driven pwfa in quasi-nonlinear regime},\ }\href {https://doi.org/https://doi.org/10.1016/j.nima.2013.10.028} {\bibfield  {journal} {\bibinfo  {journal} {Nucl. Instrum. Methods Phys. Res. A}\ }\textbf {\bibinfo {volume} {740}},\ \bibinfo {pages} {236} (\bibinfo {year} {2014})}\BibitemShut {NoStop}%
\bibitem [{\citenamefont {Marocchino}\ \emph {et~al.}(2016)\citenamefont {Marocchino}, \citenamefont {Massimo}, \citenamefont {Rossi}, \citenamefont {Chiadroni},\ and\ \citenamefont {Ferrario}}]{Marocchino:2016tfe}%
  \BibitemOpen
  \bibfield  {author} {\bibinfo {author} {\bibfnamefont {A.}~\bibnamefont {Marocchino}}, \bibinfo {author} {\bibfnamefont {F.}~\bibnamefont {Massimo}}, \bibinfo {author} {\bibfnamefont {A.~R.}\ \bibnamefont {Rossi}}, \bibinfo {author} {\bibfnamefont {E.}~\bibnamefont {Chiadroni}},\ and\ \bibinfo {author} {\bibfnamefont {M.}~\bibnamefont {Ferrario}},\ }\bibfield  {title} {\bibinfo {title} {{Efficient modeling of plasma wakefield acceleration in quasi-non-linear-regimes with the hybrid code Architect}},\ }\href {https://doi.org/10.1016/j.nima.2016.03.005} {\bibfield  {journal} {\bibinfo  {journal} {Nucl. Instrum. Meth. A}\ }\textbf {\bibinfo {volume} {829}},\ \bibinfo {pages} {386} (\bibinfo {year} {2016})}\BibitemShut {NoStop}%
\bibitem [{\citenamefont {Liu}\ \emph {et~al.}(2023)\citenamefont {Liu}, \citenamefont {Zhu}, \citenamefont {Chen}, \citenamefont {Weng}, \citenamefont {He}, \citenamefont {Sheng},\ and\ \citenamefont {Zhang}}]{Liu_2022a}%
  \BibitemOpen
  \bibfield  {author} {\bibinfo {author} {\bibfnamefont {W.-Y.}\ \bibnamefont {Liu}}, \bibinfo {author} {\bibfnamefont {X.-L.}\ \bibnamefont {Zhu}}, \bibinfo {author} {\bibfnamefont {M.}~\bibnamefont {Chen}}, \bibinfo {author} {\bibfnamefont {S.-M.}\ \bibnamefont {Weng}}, \bibinfo {author} {\bibfnamefont {F.}~\bibnamefont {He}}, \bibinfo {author} {\bibfnamefont {Z.-M.}\ \bibnamefont {Sheng}},\ and\ \bibinfo {author} {\bibfnamefont {J.}~\bibnamefont {Zhang}},\ }\href {https://doi.org/10.1103/PhysRevApplied.19.044048} {\bibfield  {journal} {\bibinfo  {journal} {Phys. Rev. Appl.}\ }\textbf {\bibinfo {volume} {19}},\ \bibinfo {pages} {044048} (\bibinfo {year} {2023})}\BibitemShut {NoStop}%
\bibitem [{\citenamefont {Zhou}\ \emph {et~al.}(2022{\natexlab{a}})\citenamefont {Zhou}, \citenamefont {An}, \citenamefont {Ding}, \citenamefont {Hua}, \citenamefont {Mori}, \citenamefont {Joshi},\ and\ \citenamefont {Lu}}]{zhou2022positron}%
  \BibitemOpen
  \bibfield  {author} {\bibinfo {author} {\bibfnamefont {S.}~\bibnamefont {Zhou}}, \bibinfo {author} {\bibfnamefont {W.}~\bibnamefont {An}}, \bibinfo {author} {\bibfnamefont {S.}~\bibnamefont {Ding}}, \bibinfo {author} {\bibfnamefont {J.}~\bibnamefont {Hua}}, \bibinfo {author} {\bibfnamefont {W.~B.}\ \bibnamefont {Mori}}, \bibinfo {author} {\bibfnamefont {C.}~\bibnamefont {Joshi}},\ and\ \bibinfo {author} {\bibfnamefont {W.}~\bibnamefont {Lu}},\ }\bibfield  {title} {\bibinfo {title} {{Positron beam loading and acceleration in the blowout regime of plasma wakefield accelerator}},\ }\href@noop {} {\bibfield  {journal} {\bibinfo  {journal} {arXiv e-prints}\ } (\bibinfo {year} {2022}{\natexlab{a}})},\ \Eprint {https://arxiv.org/abs/2211.07962} {arXiv:2211.07962} \BibitemShut {NoStop}%
\bibitem [{\citenamefont {{Wang}}\ \emph {et~al.}(2021)\citenamefont {{Wang}}, \citenamefont {{Khudik}},\ and\ \citenamefont {{Shvets}}}]{wang_2021}%
  \BibitemOpen
  \bibfield  {author} {\bibinfo {author} {\bibfnamefont {T.}~\bibnamefont {{Wang}}}, \bibinfo {author} {\bibfnamefont {V.}~\bibnamefont {{Khudik}}},\ and\ \bibinfo {author} {\bibfnamefont {G.}~\bibnamefont {{Shvets}}},\ }\bibfield  {title} {\bibinfo {title} {{Positron Acceleration in an Elongated Bubble Regime}},\ }\href@noop {} {\bibfield  {journal} {\bibinfo  {journal} {arXiv e-prints}\ } (\bibinfo {year} {2021})},\ \Eprint {https://arxiv.org/abs/2110.10290} {arXiv:2110.10290} \BibitemShut {NoStop}%
\bibitem [{\citenamefont {Vieira}\ and\ \citenamefont {Mendon\ifmmode~\mbox{\c{c}}\else \c{c}\fi{}a}(2014)}]{vieira_2014}%
  \BibitemOpen
  \bibfield  {author} {\bibinfo {author} {\bibfnamefont {J.}~\bibnamefont {Vieira}}\ and\ \bibinfo {author} {\bibfnamefont {J.~T.}\ \bibnamefont {Mendon\ifmmode~\mbox{\c{c}}\else \c{c}\fi{}a}},\ }\bibfield  {title} {\bibinfo {title} {Nonlinear laser driven donut wakefields for positron and electron acceleration},\ }\href {https://doi.org/10.1103/PhysRevLett.112.215001} {\bibfield  {journal} {\bibinfo  {journal} {Phys. Rev. Lett.}\ }\textbf {\bibinfo {volume} {112}},\ \bibinfo {pages} {215001} (\bibinfo {year} {2014})}\BibitemShut {NoStop}%
\bibitem [{\citenamefont {Mendonça}\ and\ \citenamefont {Vieira}(2014)}]{mendonca_2014}%
  \BibitemOpen
  \bibfield  {author} {\bibinfo {author} {\bibfnamefont {J.~T.}\ \bibnamefont {Mendonça}}\ and\ \bibinfo {author} {\bibfnamefont {J.}~\bibnamefont {Vieira}},\ }\bibfield  {title} {\bibinfo {title} {Donut wakefields generated by intense laser pulses with orbital angular momentum},\ }\href {https://doi.org/10.1063/1.4868967} {\bibfield  {journal} {\bibinfo  {journal} {Phys. Plasmas}\ }\textbf {\bibinfo {volume} {21}},\ \bibinfo {pages} {033107} (\bibinfo {year} {2014})}\BibitemShut {NoStop}%
\bibitem [{\citenamefont {Wang}\ \emph {et~al.}(2020)\citenamefont {Wang}, \citenamefont {Jiang}, \citenamefont {Dong}, \citenamefont {Lu}, \citenamefont {Li}, \citenamefont {Xu} \emph {et~al.}}]{Wang_2020}%
  \BibitemOpen
  \bibfield  {author} {\bibinfo {author} {\bibfnamefont {W.~P.}\ \bibnamefont {Wang}}, \bibinfo {author} {\bibfnamefont {C.}~\bibnamefont {Jiang}}, \bibinfo {author} {\bibfnamefont {H.}~\bibnamefont {Dong}}, \bibinfo {author} {\bibfnamefont {X.~M.}\ \bibnamefont {Lu}}, \bibinfo {author} {\bibfnamefont {J.~F.}\ \bibnamefont {Li}}, \bibinfo {author} {\bibfnamefont {R.~J.}\ \bibnamefont {Xu}}, \emph {et~al.},\ }\bibfield  {title} {\bibinfo {title} {Hollow plasma acceleration driven by a relativistic reflected hollow laser},\ }\href {https://doi.org/10.1103/PhysRevLett.125.034801} {\bibfield  {journal} {\bibinfo  {journal} {Phys. Rev. Lett.}\ }\textbf {\bibinfo {volume} {125}},\ \bibinfo {pages} {034801} (\bibinfo {year} {2020})}\BibitemShut {NoStop}%
\bibitem [{\citenamefont {Jain}\ \emph {et~al.}(2015)\citenamefont {Jain}, \citenamefont {Antonsen},\ and\ \citenamefont {Palastro}}]{jain_2015}%
  \BibitemOpen
  \bibfield  {author} {\bibinfo {author} {\bibfnamefont {N.}~\bibnamefont {Jain}}, \bibinfo {author} {\bibfnamefont {T.~M.}\ \bibnamefont {Antonsen}},\ and\ \bibinfo {author} {\bibfnamefont {J.~P.}\ \bibnamefont {Palastro}},\ }\bibfield  {title} {\bibinfo {title} {Positron acceleration by plasma wakefields driven by a hollow electron beam},\ }\href {https://doi.org/10.1103/PhysRevLett.115.195001} {\bibfield  {journal} {\bibinfo  {journal} {Phys. Rev. Lett.}\ }\textbf {\bibinfo {volume} {115}},\ \bibinfo {pages} {195001} (\bibinfo {year} {2015})}\BibitemShut {NoStop}%
\bibitem [{\citenamefont {Vieira}\ \emph {et~al.}(2016)\citenamefont {Vieira}, \citenamefont {Mendonça},\ and\ \citenamefont {Silva}}]{vieira_2016}%
  \BibitemOpen
  \bibfield  {author} {\bibinfo {author} {\bibfnamefont {J.}~\bibnamefont {Vieira}}, \bibinfo {author} {\bibfnamefont {J.~T.}\ \bibnamefont {Mendonça}},\ and\ \bibinfo {author} {\bibfnamefont {L.~O.}\ \bibnamefont {Silva}},\ }\bibfield  {title} {\bibinfo {title} {Positron acceleration in non-linear beam driven plasma wakefields},\ }\href {https://doi.org/10.1063/1.4965655} {\bibfield  {journal} {\bibinfo  {journal} {AIP Conf. Proc.}\ }\textbf {\bibinfo {volume} {1777}},\ \bibinfo {pages} {070012} (\bibinfo {year} {2016})}\BibitemShut {NoStop}%
\bibitem [{\citenamefont {Silva}\ and\ \citenamefont {Vieira}(2023)}]{Silva_2023}%
  \BibitemOpen
  \bibfield  {author} {\bibinfo {author} {\bibfnamefont {T.}~\bibnamefont {Silva}}\ and\ \bibinfo {author} {\bibfnamefont {J.}~\bibnamefont {Vieira}},\ }\bibfield  {title} {\bibinfo {title} {Positron acceleration in plasma waves driven by non-neutral fireball beams},\ }\href {https://doi.org/10.1103/PhysRevAccelBeams.26.091301} {\bibfield  {journal} {\bibinfo  {journal} {Phys. Rev. Accel. Beams}\ }\textbf {\bibinfo {volume} {26}},\ \bibinfo {pages} {091301} (\bibinfo {year} {2023})}\BibitemShut {NoStop}%
\bibitem [{\citenamefont {Yu}\ \emph {et~al.}(2014)\citenamefont {Yu}, \citenamefont {Schroeder}, \citenamefont {Li}, \citenamefont {Benedetti}, \citenamefont {Chen}, \citenamefont {Weng}, \citenamefont {Sheng},\ and\ \citenamefont {Esarey}}]{Yu_2014}%
  \BibitemOpen
  \bibfield  {author} {\bibinfo {author} {\bibfnamefont {L.-L.}\ \bibnamefont {Yu}}, \bibinfo {author} {\bibfnamefont {C.~B.}\ \bibnamefont {Schroeder}}, \bibinfo {author} {\bibfnamefont {F.-Y.}\ \bibnamefont {Li}}, \bibinfo {author} {\bibfnamefont {C.}~\bibnamefont {Benedetti}}, \bibinfo {author} {\bibfnamefont {M.}~\bibnamefont {Chen}}, \bibinfo {author} {\bibfnamefont {S.-M.}\ \bibnamefont {Weng}}, \bibinfo {author} {\bibfnamefont {Z.-M.}\ \bibnamefont {Sheng}},\ and\ \bibinfo {author} {\bibfnamefont {E.}~\bibnamefont {Esarey}},\ }\bibfield  {title} {\bibinfo {title} {{Control of focusing fields for positron acceleration in nonlinear plasma wakes using multiple laser modes}},\ }\href {https://doi.org/10.1063/1.4903536} {\bibfield  {journal} {\bibinfo  {journal} {Phys. Plasmas}\ }\textbf {\bibinfo {volume} {21}},\ \bibinfo {pages} {120702} (\bibinfo {year} {2014})}\BibitemShut {NoStop}%
\bibitem [{\citenamefont {Pathak}\ \emph {et~al.}(2016)\citenamefont {Pathak}, \citenamefont {Zhidkov}, \citenamefont {Nakanii}, \citenamefont {Masuda}, \citenamefont {Hosokai},\ and\ \citenamefont {Kodama}}]{Pathak_2016}%
  \BibitemOpen
  \bibfield  {author} {\bibinfo {author} {\bibfnamefont {N.}~\bibnamefont {Pathak}}, \bibinfo {author} {\bibfnamefont {A.}~\bibnamefont {Zhidkov}}, \bibinfo {author} {\bibfnamefont {N.}~\bibnamefont {Nakanii}}, \bibinfo {author} {\bibfnamefont {S.}~\bibnamefont {Masuda}}, \bibinfo {author} {\bibfnamefont {T.}~\bibnamefont {Hosokai}},\ and\ \bibinfo {author} {\bibfnamefont {R.}~\bibnamefont {Kodama}},\ }\bibfield  {title} {\bibinfo {title} {{Breaking symmetry in propagation of radially and azimuthally polarized high power laser pulses in underdense plasma}},\ }\href {https://doi.org/10.1063/1.4942942} {\bibfield  {journal} {\bibinfo  {journal} {Phys. Plasmas}\ }\textbf {\bibinfo {volume} {23}},\ \bibinfo {pages} {033102} (\bibinfo {year} {2016})}\BibitemShut {NoStop}%
\bibitem [{\citenamefont {Jain}(2019)}]{Jain_2019}%
  \BibitemOpen
  \bibfield  {author} {\bibinfo {author} {\bibfnamefont {N.}~\bibnamefont {Jain}},\ }\bibfield  {title} {\bibinfo {title} {{Evolution of ultra-relativistic hollow electron beams during their propagation in plasmas}},\ }\href {https://doi.org/10.1063/1.5065375} {\bibfield  {journal} {\bibinfo  {journal} {Phys. Plasmas}\ }\textbf {\bibinfo {volume} {26}},\ \bibinfo {pages} {023107} (\bibinfo {year} {2019})}\BibitemShut {NoStop}%
\bibitem [{\citenamefont {Su}\ \emph {et~al.}(1987)\citenamefont {Su}, \citenamefont {Katsouleas}, \citenamefont {Dawson}, \citenamefont {Chen}, \citenamefont {Jones},\ and\ \citenamefont {Keinigs}}]{Su_1987}%
  \BibitemOpen
  \bibfield  {author} {\bibinfo {author} {\bibfnamefont {J.~J.}\ \bibnamefont {Su}}, \bibinfo {author} {\bibfnamefont {T.}~\bibnamefont {Katsouleas}}, \bibinfo {author} {\bibfnamefont {J.~M.}\ \bibnamefont {Dawson}}, \bibinfo {author} {\bibfnamefont {P.}~\bibnamefont {Chen}}, \bibinfo {author} {\bibfnamefont {M.}~\bibnamefont {Jones}},\ and\ \bibinfo {author} {\bibfnamefont {R.}~\bibnamefont {Keinigs}},\ }\bibfield  {title} {\bibinfo {title} {Stability of the driving bunch in the plasma wakefield accelerator},\ }\href {https://doi.org/10.1109/TPS.1987.4316684} {\bibfield  {journal} {\bibinfo  {journal} {IEEE T. Plasma Sci.}\ }\textbf {\bibinfo {volume} {15}},\ \bibinfo {pages} {192} (\bibinfo {year} {1987})}\BibitemShut {NoStop}%
\bibitem [{\citenamefont {Allen}\ \emph {et~al.}(1992)\citenamefont {Allen}, \citenamefont {Beijersbergen}, \citenamefont {Spreeuw},\ and\ \citenamefont {Woerdman}}]{Allen_1992}%
  \BibitemOpen
  \bibfield  {author} {\bibinfo {author} {\bibfnamefont {L.}~\bibnamefont {Allen}}, \bibinfo {author} {\bibfnamefont {M.~W.}\ \bibnamefont {Beijersbergen}}, \bibinfo {author} {\bibfnamefont {R.~J.~C.}\ \bibnamefont {Spreeuw}},\ and\ \bibinfo {author} {\bibfnamefont {J.~P.}\ \bibnamefont {Woerdman}},\ }\bibfield  {title} {\bibinfo {title} {Orbital angular momentum of light and the transformation of laguerre-gaussian laser modes},\ }\href {https://doi.org/10.1103/PhysRevA.45.8185} {\bibfield  {journal} {\bibinfo  {journal} {Phys. Rev. A}\ }\textbf {\bibinfo {volume} {45}},\ \bibinfo {pages} {8185} (\bibinfo {year} {1992})}\BibitemShut {NoStop}%
\bibitem [{\citenamefont {Nakanii}\ \emph {et~al.}(2016)\citenamefont {Nakanii}, \citenamefont {Hosokai}, \citenamefont {Pathak}, \citenamefont {Masuda}, \citenamefont {Zhidkov}, \citenamefont {Nakahara} \emph {et~al.}}]{Nakanii_2016}%
  \BibitemOpen
  \bibfield  {author} {\bibinfo {author} {\bibfnamefont {N.}~\bibnamefont {Nakanii}}, \bibinfo {author} {\bibfnamefont {T.}~\bibnamefont {Hosokai}}, \bibinfo {author} {\bibfnamefont {N.~C.}\ \bibnamefont {Pathak}}, \bibinfo {author} {\bibfnamefont {S.}~\bibnamefont {Masuda}}, \bibinfo {author} {\bibfnamefont {A.~G.}\ \bibnamefont {Zhidkov}}, \bibinfo {author} {\bibfnamefont {H.}~\bibnamefont {Nakahara}}, \emph {et~al.},\ }\bibfield  {title} {\bibinfo {title} {Decomposition of powerful axisymmetrically polarized laser pulses in underdense plasma},\ }\href {https://doi.org/10.1103/PhysRevE.94.063205} {\bibfield  {journal} {\bibinfo  {journal} {Phys. Rev. E}\ }\textbf {\bibinfo {volume} {94}},\ \bibinfo {pages} {063205} (\bibinfo {year} {2016})}\BibitemShut {NoStop}%
\bibitem [{\citenamefont {Bubley}\ \emph {et~al.}(2002)\citenamefont {Bubley}, \citenamefont {Goncharov}, \citenamefont {Ivanov}, \citenamefont {Konstantinov}, \citenamefont {Konstantinov}, \citenamefont {Kryuchkov} \emph {et~al.}}]{bubley_2002}%
  \BibitemOpen
  \bibfield  {author} {\bibinfo {author} {\bibfnamefont {A.}~\bibnamefont {Bubley}}, \bibinfo {author} {\bibfnamefont {A.}~\bibnamefont {Goncharov}}, \bibinfo {author} {\bibfnamefont {A.}~\bibnamefont {Ivanov}}, \bibinfo {author} {\bibfnamefont {E.}~\bibnamefont {Konstantinov}}, \bibinfo {author} {\bibfnamefont {S.}~\bibnamefont {Konstantinov}}, \bibinfo {author} {\bibfnamefont {A.}~\bibnamefont {Kryuchkov}}, \emph {et~al.},\ }\bibfield  {title} {\bibinfo {title} {{The electron gun with variable beam profile for optimization of electron cooling}},\ }in\ \href {https://accelconf.web.cern.ch/e02/papers/wepri049.pdf} {\emph {\bibinfo {booktitle} {{Proceedings of EPAC2002, Paris, France}}}}\ (\bibinfo {year} {2002})\ p.\ \bibinfo {pages} {1356}\BibitemShut {NoStop}%
\bibitem [{\citenamefont {Bubley}\ \emph {et~al.}(2006)\citenamefont {Bubley}, \citenamefont {Panasyuk}, \citenamefont {Parkhomchuk},\ and\ \citenamefont {Reva}}]{Bubley:2006ia}%
  \BibitemOpen
  \bibfield  {author} {\bibinfo {author} {\bibfnamefont {A.~V.}\ \bibnamefont {Bubley}}, \bibinfo {author} {\bibfnamefont {V.~M.}\ \bibnamefont {Panasyuk}}, \bibinfo {author} {\bibfnamefont {V.~V.}\ \bibnamefont {Parkhomchuk}},\ and\ \bibinfo {author} {\bibfnamefont {V.~B.}\ \bibnamefont {Reva}},\ }\bibfield  {title} {\bibinfo {title} {{Measurements of the profile of an intense electron beam}},\ }\href {https://doi.org/10.1134/S0020441206010106} {\bibfield  {journal} {\bibinfo  {journal} {Instrum. Exp. Tech.}\ }\textbf {\bibinfo {volume} {49}},\ \bibinfo {pages} {83} (\bibinfo {year} {2006})}\BibitemShut {NoStop}%
\bibitem [{\citenamefont {Diederichs}\ \emph {et~al.}(2019)\citenamefont {Diederichs}, \citenamefont {Mehrling}, \citenamefont {Benedetti}, \citenamefont {Schroeder}, \citenamefont {Knetsch}, \citenamefont {Esarey},\ and\ \citenamefont {Osterhoff}}]{Diederichs:2019wnl}%
  \BibitemOpen
  \bibfield  {author} {\bibinfo {author} {\bibfnamefont {S.}~\bibnamefont {Diederichs}}, \bibinfo {author} {\bibfnamefont {T.~J.}\ \bibnamefont {Mehrling}}, \bibinfo {author} {\bibfnamefont {C.}~\bibnamefont {Benedetti}}, \bibinfo {author} {\bibfnamefont {C.~B.}\ \bibnamefont {Schroeder}}, \bibinfo {author} {\bibfnamefont {A.}~\bibnamefont {Knetsch}}, \bibinfo {author} {\bibfnamefont {E.}~\bibnamefont {Esarey}},\ and\ \bibinfo {author} {\bibfnamefont {J.}~\bibnamefont {Osterhoff}},\ }\bibfield  {title} {\bibinfo {title} {{Positron transport and acceleration in beam-driven plasma wakefield accelerators using plasma columns}},\ }\href {https://doi.org/10.1103/PhysRevAccelBeams.22.081301} {\bibfield  {journal} {\bibinfo  {journal} {Phys. Rev. Accel. Beams}\ }\textbf {\bibinfo {volume} {22}},\ \bibinfo {pages} {081301} (\bibinfo {year} {2019})}\BibitemShut {NoStop}%
\bibitem [{\citenamefont {Diederichs}\ \emph {et~al.}(2020)\citenamefont {Diederichs}, \citenamefont {Benedetti}, \citenamefont {Esarey}, \citenamefont {Osterhoff},\ and\ \citenamefont {Schroeder}}]{Diederichs:2020hri}%
  \BibitemOpen
  \bibfield  {author} {\bibinfo {author} {\bibfnamefont {S.}~\bibnamefont {Diederichs}}, \bibinfo {author} {\bibfnamefont {C.}~\bibnamefont {Benedetti}}, \bibinfo {author} {\bibfnamefont {E.}~\bibnamefont {Esarey}}, \bibinfo {author} {\bibfnamefont {J.}~\bibnamefont {Osterhoff}},\ and\ \bibinfo {author} {\bibfnamefont {C.~B.}\ \bibnamefont {Schroeder}},\ }\bibfield  {title} {\bibinfo {title} {{High-quality positron acceleration in beam-driven plasma accelerators}},\ }\href {https://doi.org/10.1103/PhysRevAccelBeams.23.121301} {\bibfield  {journal} {\bibinfo  {journal} {Phys. Rev. Accel. Beams}\ }\textbf {\bibinfo {volume} {23}},\ \bibinfo {pages} {121301} (\bibinfo {year} {2020})}\BibitemShut {NoStop}%
\bibitem [{\citenamefont {Lotov}(2017)}]{lotov_2017}%
  \BibitemOpen
  \bibfield  {author} {\bibinfo {author} {\bibfnamefont {K.~V.}\ \bibnamefont {Lotov}},\ }\bibfield  {title} {\bibinfo {title} {Radial equilibrium of relativistic particle bunches in plasma wakefield accelerators},\ }\href {https://doi.org/10.1063/1.4977058} {\bibfield  {journal} {\bibinfo  {journal} {Phys. Plasmas}\ }\textbf {\bibinfo {volume} {24}},\ \bibinfo {pages} {023119} (\bibinfo {year} {2017})}\BibitemShut {NoStop}%
\bibitem [{\citenamefont {{Diederichs}}\ \emph {et~al.}(2023)\citenamefont {{Diederichs}}, \citenamefont {{Benedetti}}, \citenamefont {{Esarey}}, \citenamefont {{Sinn}}, \citenamefont {{Osterhoff}}, \citenamefont {{Schroeder}},\ and\ \citenamefont {{Th{\'e}venet}}}]{diederichs2023_temp2}%
  \BibitemOpen
  \bibfield  {author} {\bibinfo {author} {\bibfnamefont {S.}~\bibnamefont {{Diederichs}}}, \bibinfo {author} {\bibfnamefont {C.}~\bibnamefont {{Benedetti}}}, \bibinfo {author} {\bibfnamefont {E.}~\bibnamefont {{Esarey}}}, \bibinfo {author} {\bibfnamefont {A.}~\bibnamefont {{Sinn}}}, \bibinfo {author} {\bibfnamefont {J.}~\bibnamefont {{Osterhoff}}}, \bibinfo {author} {\bibfnamefont {C.~B.}\ \bibnamefont {{Schroeder}}},\ and\ \bibinfo {author} {\bibfnamefont {M.}~\bibnamefont {{Th{\'e}venet}}},\ }\bibfield  {title} {\bibinfo {title} {{Emittance-preserving acceleration of high-quality positron beams using warm plasma filaments}},\ }\bibfield  {journal} {\bibinfo  {journal} {arXiv e-prints}\ }\href {https://doi.org/10.48550/arXiv.2311.07402} {10.48550/arXiv.2311.07402} (\bibinfo {year} {2023})\BibitemShut {NoStop}%
\bibitem [{\citenamefont {Diederichs}\ \emph {et~al.}(2022{\natexlab{a}})\citenamefont {Diederichs}, \citenamefont {Benedetti}, \citenamefont {Esarey}, \citenamefont {Th\'evenet}, \citenamefont {Osterhoff},\ and\ \citenamefont {Schroeder}}]{Diederichs:2022pjj}%
  \BibitemOpen
  \bibfield  {author} {\bibinfo {author} {\bibfnamefont {S.}~\bibnamefont {Diederichs}}, \bibinfo {author} {\bibfnamefont {C.}~\bibnamefont {Benedetti}}, \bibinfo {author} {\bibfnamefont {E.}~\bibnamefont {Esarey}}, \bibinfo {author} {\bibfnamefont {M.}~\bibnamefont {Th\'evenet}}, \bibinfo {author} {\bibfnamefont {J.}~\bibnamefont {Osterhoff}},\ and\ \bibinfo {author} {\bibfnamefont {C.~B.}\ \bibnamefont {Schroeder}},\ }\bibfield  {title} {\bibinfo {title} {{Stable electron beam propagation in a plasma column}},\ }\href {https://doi.org/10.1063/5.0087807} {\bibfield  {journal} {\bibinfo  {journal} {Phys. Plasmas}\ }\textbf {\bibinfo {volume} {29}},\ \bibinfo {pages} {043101} (\bibinfo {year} {2022}{\natexlab{a}})}\BibitemShut {NoStop}%
\bibitem [{\citenamefont {Diederichs}\ \emph {et~al.}(2022{\natexlab{b}})\citenamefont {Diederichs}, \citenamefont {Benedetti}, \citenamefont {Th\'evenet}, \citenamefont {Esarey}, \citenamefont {Osterhoff},\ and\ \citenamefont {Schroeder}}]{Diederichs:2022yfd}%
  \BibitemOpen
  \bibfield  {author} {\bibinfo {author} {\bibfnamefont {S.}~\bibnamefont {Diederichs}}, \bibinfo {author} {\bibfnamefont {C.}~\bibnamefont {Benedetti}}, \bibinfo {author} {\bibfnamefont {M.}~\bibnamefont {Th\'evenet}}, \bibinfo {author} {\bibfnamefont {E.}~\bibnamefont {Esarey}}, \bibinfo {author} {\bibfnamefont {J.}~\bibnamefont {Osterhoff}},\ and\ \bibinfo {author} {\bibfnamefont {C.~B.}\ \bibnamefont {Schroeder}},\ }\bibfield  {title} {\bibinfo {title} {{Self-stabilizing positron acceleration in a plasma column}},\ }\href {https://doi.org/10.1103/PhysRevAccelBeams.25.091304} {\bibfield  {journal} {\bibinfo  {journal} {Phys. Rev. Accel. Beams}\ }\textbf {\bibinfo {volume} {25}},\ \bibinfo {pages} {091304} (\bibinfo {year} {2022}{\natexlab{b}})}\BibitemShut {NoStop}%
\bibitem [{\citenamefont {Green}\ \emph {et~al.}(2014)\citenamefont {Green}, \citenamefont {Adli}, \citenamefont {Clarke}, \citenamefont {Corde}, \citenamefont {Edstrom}, \citenamefont {Fisher} \emph {et~al.}}]{Green_2014}%
  \BibitemOpen
  \bibfield  {author} {\bibinfo {author} {\bibfnamefont {S.~Z.}\ \bibnamefont {Green}}, \bibinfo {author} {\bibfnamefont {E.}~\bibnamefont {Adli}}, \bibinfo {author} {\bibfnamefont {C.~I.}\ \bibnamefont {Clarke}}, \bibinfo {author} {\bibfnamefont {S.}~\bibnamefont {Corde}}, \bibinfo {author} {\bibfnamefont {S.~A.}\ \bibnamefont {Edstrom}}, \bibinfo {author} {\bibfnamefont {A.~S.}\ \bibnamefont {Fisher}}, \emph {et~al.},\ }\bibfield  {title} {\bibinfo {title} {Laser ionized preformed plasma at facet},\ }\href {https://doi.org/10.1088/0741-3335/56/8/084011} {\bibfield  {journal} {\bibinfo  {journal} {Plasma Phys. Control. Fusion}\ }\textbf {\bibinfo {volume} {56}},\ \bibinfo {pages} {084011} (\bibinfo {year} {2014})}\BibitemShut {NoStop}%
\bibitem [{\citenamefont {O'Connell}\ \emph {et~al.}(2006)\citenamefont {O'Connell}, \citenamefont {Barnes}, \citenamefont {Decker}, \citenamefont {Hogan}, \citenamefont {Iverson}, \citenamefont {Krejcik} \emph {et~al.}}]{OConnell_2006}%
  \BibitemOpen
  \bibfield  {author} {\bibinfo {author} {\bibfnamefont {C.~L.}\ \bibnamefont {O'Connell}}, \bibinfo {author} {\bibfnamefont {C.~D.}\ \bibnamefont {Barnes}}, \bibinfo {author} {\bibfnamefont {F.-J.}\ \bibnamefont {Decker}}, \bibinfo {author} {\bibfnamefont {M.~J.}\ \bibnamefont {Hogan}}, \bibinfo {author} {\bibfnamefont {R.}~\bibnamefont {Iverson}}, \bibinfo {author} {\bibfnamefont {P.}~\bibnamefont {Krejcik}}, \emph {et~al.},\ }\bibfield  {title} {\bibinfo {title} {Plasma production via field ionization},\ }\href {https://doi.org/10.1103/PhysRevSTAB.9.101301} {\bibfield  {journal} {\bibinfo  {journal} {Phys. Rev. ST Accel. Beams}\ }\textbf {\bibinfo {volume} {9}},\ \bibinfo {pages} {101301} (\bibinfo {year} {2006})}\BibitemShut {NoStop}%
\bibitem [{\citenamefont {FACET-II}(2022)}]{slac_2022}%
  \BibitemOpen
  \bibfield  {author} {\bibinfo {author} {\bibnamefont {FACET-II}},\ }\href@noop {} {\bibinfo {title} {Accepted proposals}},\ \bibinfo {howpublished} {\url{https://facet-ii.slac.stanford.edu/proposals/accepted-proposals}} (\bibinfo {year} {2022})\BibitemShut {NoStop}%
\bibitem [{\citenamefont {{Reichwein}}\ \emph {et~al.}(2022)\citenamefont {{Reichwein}}, \citenamefont {{Pukhov}}, \citenamefont {{Golovanov}},\ and\ \citenamefont {{Kostyukov}}}]{reichwein_2022}%
  \BibitemOpen
  \bibfield  {author} {\bibinfo {author} {\bibfnamefont {L.}~\bibnamefont {{Reichwein}}}, \bibinfo {author} {\bibfnamefont {A.}~\bibnamefont {{Pukhov}}}, \bibinfo {author} {\bibfnamefont {A.}~\bibnamefont {{Golovanov}}},\ and\ \bibinfo {author} {\bibfnamefont {I.~Y.}\ \bibnamefont {{Kostyukov}}},\ }\bibfield  {title} {\bibinfo {title} {{Positron acceleration via laser-augmented blowouts in two-column plasma structures}},\ }\href {https://doi.org/10.1103/PhysRevE.105.055207} {\bibfield  {journal} {\bibinfo  {journal} {\pre}\ }\textbf {\bibinfo {volume} {105}},\ \bibinfo {eid} {055207} (\bibinfo {year} {2022})}\BibitemShut {NoStop}%
\bibitem [{\citenamefont {{Silva}}\ \emph {et~al.}(2021)\citenamefont {{Silva}}, \citenamefont {{Amorim}}, \citenamefont {{Downer}}, \citenamefont {{Hogan}}, \citenamefont {{Yakimenko}}, \citenamefont {{Zgadzaj}},\ and\ \citenamefont {{Vieira}}}]{silva_2021}%
  \BibitemOpen
  \bibfield  {author} {\bibinfo {author} {\bibfnamefont {T.}~\bibnamefont {{Silva}}}, \bibinfo {author} {\bibfnamefont {L.~D.}\ \bibnamefont {{Amorim}}}, \bibinfo {author} {\bibfnamefont {M.~C.}\ \bibnamefont {{Downer}}}, \bibinfo {author} {\bibfnamefont {M.~J.}\ \bibnamefont {{Hogan}}}, \bibinfo {author} {\bibfnamefont {V.}~\bibnamefont {{Yakimenko}}}, \bibinfo {author} {\bibfnamefont {R.}~\bibnamefont {{Zgadzaj}}},\ and\ \bibinfo {author} {\bibfnamefont {J.}~\bibnamefont {{Vieira}}},\ }\bibfield  {title} {\bibinfo {title} {{Stable Positron Acceleration in Thin, Warm, Hollow Plasma Channels}},\ }\href {https://doi.org/10.1103/PhysRevLett.127.104801} {\bibfield  {journal} {\bibinfo  {journal} {\prl}\ }\textbf {\bibinfo {volume} {127}},\ \bibinfo {eid} {104801} (\bibinfo {year} {2021})}\BibitemShut {NoStop}%
\bibitem [{\citenamefont {{Zhou}}\ \emph {et~al.}(2021)\citenamefont {{Zhou}}, \citenamefont {{Hua}}, \citenamefont {{An}}, \citenamefont {{Mori}}, \citenamefont {{Joshi}}, \citenamefont {{Gao}},\ and\ \citenamefont {{Lu}}}]{zhou_2021}%
  \BibitemOpen
  \bibfield  {author} {\bibinfo {author} {\bibfnamefont {S.}~\bibnamefont {{Zhou}}}, \bibinfo {author} {\bibfnamefont {J.}~\bibnamefont {{Hua}}}, \bibinfo {author} {\bibfnamefont {W.}~\bibnamefont {{An}}}, \bibinfo {author} {\bibfnamefont {W.~B.}\ \bibnamefont {{Mori}}}, \bibinfo {author} {\bibfnamefont {C.}~\bibnamefont {{Joshi}}}, \bibinfo {author} {\bibfnamefont {J.}~\bibnamefont {{Gao}}},\ and\ \bibinfo {author} {\bibfnamefont {W.}~\bibnamefont {{Lu}}},\ }\bibfield  {title} {\bibinfo {title} {{High Efficiency Uniform Wakefield Acceleration of a Positron Beam Using Stable Asymmetric Mode in a Hollow Channel Plasma}},\ }\href {https://doi.org/10.1103/PhysRevLett.127.174801} {\bibfield  {journal} {\bibinfo  {journal} {Phys. Rev. Lett.}\ }\textbf {\bibinfo {volume} {127}},\ \bibinfo {pages} {174801} (\bibinfo {year} {2021})}\BibitemShut {NoStop}%
\bibitem [{\citenamefont {Zhou}\ \emph {et~al.}(2022{\natexlab{b}})\citenamefont {Zhou}, \citenamefont {Hua}, \citenamefont {Lu}, \citenamefont {An}, \citenamefont {Su}, \citenamefont {Mori},\ and\ \citenamefont {Joshi}}]{zhou_2022_hcob}%
  \BibitemOpen
  \bibfield  {author} {\bibinfo {author} {\bibfnamefont {S.}~\bibnamefont {Zhou}}, \bibinfo {author} {\bibfnamefont {J.}~\bibnamefont {Hua}}, \bibinfo {author} {\bibfnamefont {W.}~\bibnamefont {Lu}}, \bibinfo {author} {\bibfnamefont {W.}~\bibnamefont {An}}, \bibinfo {author} {\bibfnamefont {Q.}~\bibnamefont {Su}}, \bibinfo {author} {\bibfnamefont {W.~B.}\ \bibnamefont {Mori}},\ and\ \bibinfo {author} {\bibfnamefont {C.}~\bibnamefont {Joshi}},\ }\bibfield  {title} {\bibinfo {title} {High efficiency uniform positron beam loading in a hollow channel plasma wakefield accelerator},\ }\href {https://doi.org/10.1103/PhysRevAccelBeams.25.091303} {\bibfield  {journal} {\bibinfo  {journal} {Phys. Rev. Accel. Beams}\ }\textbf {\bibinfo {volume} {25}},\ \bibinfo {pages} {091303} (\bibinfo {year} {2022}{\natexlab{b}})}\BibitemShut {NoStop}%
\bibitem [{\citenamefont {{Chen}}\ \emph {et~al.}(2020)\citenamefont {{Chen}}, \citenamefont {{Schulte}},\ and\ \citenamefont {{Adli}}}]{Chen_2020}%
  \BibitemOpen
  \bibfield  {author} {\bibinfo {author} {\bibfnamefont {J.~B.~B.}\ \bibnamefont {{Chen}}}, \bibinfo {author} {\bibfnamefont {D.}~\bibnamefont {{Schulte}}},\ and\ \bibinfo {author} {\bibfnamefont {E.}~\bibnamefont {{Adli}}},\ }\bibfield  {title} {\bibinfo {title} {{$\mathrm{e}^+$$\mathrm{e}^-$ Beam-beam Parameter Study for a TeV-scale PWFA Linear Collider}},\ }\href@noop {} {\bibfield  {journal} {\bibinfo  {journal} {arXiv e-prints}\ } (\bibinfo {year} {2020})},\ \Eprint {https://arxiv.org/abs/arXiv:2009.13672} {arXiv:2009.13672} \BibitemShut {NoStop}%
\bibitem [{\citenamefont {Lindstrøm}\ \emph {et~al.}(2022)\citenamefont {Lindstrøm}, \citenamefont {Beinortaite}, \citenamefont {Svensson} \emph {et~al.}}]{lindstrom_2023}%
  \BibitemOpen
  \bibfield  {author} {\bibinfo {author} {\bibfnamefont {C.~A.}\ \bibnamefont {Lindstrøm}}, \bibinfo {author} {\bibfnamefont {J.}~\bibnamefont {Beinortaite}}, \bibinfo {author} {\bibfnamefont {J.~B.}\ \bibnamefont {Svensson}}, \emph {et~al.},\ }\bibfield  {title} {\bibinfo {title} {Preservation of beam quality in a plasma-wakefield accelerator}} (\bibinfo {year} {2022}),\ \bibinfo {note} {{Preprint at \url{http://doi.org/10.21203/rs.3.rs-2300900/v1}}}\BibitemShut {NoStop}%
\bibitem [{\citenamefont {Lu}\ \emph {et~al.}(2006{\natexlab{a}})\citenamefont {Lu}, \citenamefont {Huang}, \citenamefont {Zhou}, \citenamefont {Tzoufras}, \citenamefont {Tsung}, \citenamefont {Mori},\ and\ \citenamefont {Katsouleas}}]{Lu_2006_pop}%
  \BibitemOpen
  \bibfield  {author} {\bibinfo {author} {\bibfnamefont {W.}~\bibnamefont {Lu}}, \bibinfo {author} {\bibfnamefont {C.}~\bibnamefont {Huang}}, \bibinfo {author} {\bibfnamefont {M.}~\bibnamefont {Zhou}}, \bibinfo {author} {\bibfnamefont {M.}~\bibnamefont {Tzoufras}}, \bibinfo {author} {\bibfnamefont {F.~S.}\ \bibnamefont {Tsung}}, \bibinfo {author} {\bibfnamefont {W.~B.}\ \bibnamefont {Mori}},\ and\ \bibinfo {author} {\bibfnamefont {T.}~\bibnamefont {Katsouleas}},\ }\bibfield  {title} {\bibinfo {title} {A nonlinear theory for multidimensional relativistic plasma wave wakefields},\ }\href {https://doi.org/10.1063/1.2203364} {\bibfield  {journal} {\bibinfo  {journal} {Phys. Plasmas}\ }\textbf {\bibinfo {volume} {13}},\ \bibinfo {pages} {056709} (\bibinfo {year} {2006}{\natexlab{a}})}\BibitemShut {NoStop}%
\bibitem [{\citenamefont {Lu}\ \emph {et~al.}(2006{\natexlab{b}})\citenamefont {Lu}, \citenamefont {Huang}, \citenamefont {Zhou}, \citenamefont {Mori},\ and\ \citenamefont {Katsouleas}}]{Lu_2006}%
  \BibitemOpen
  \bibfield  {author} {\bibinfo {author} {\bibfnamefont {W.}~\bibnamefont {Lu}}, \bibinfo {author} {\bibfnamefont {C.}~\bibnamefont {Huang}}, \bibinfo {author} {\bibfnamefont {M.}~\bibnamefont {Zhou}}, \bibinfo {author} {\bibfnamefont {W.~B.}\ \bibnamefont {Mori}},\ and\ \bibinfo {author} {\bibfnamefont {T.}~\bibnamefont {Katsouleas}},\ }\bibfield  {title} {\bibinfo {title} {Nonlinear theory for relativistic plasma wakefields in the blowout regime},\ }\href {https://doi.org/10.1103/PhysRevLett.96.165002} {\bibfield  {journal} {\bibinfo  {journal} {Phys. Rev. Lett.}\ }\textbf {\bibinfo {volume} {96}},\ \bibinfo {pages} {165002} (\bibinfo {year} {2006}{\natexlab{b}})}\BibitemShut {NoStop}%
\bibitem [{\citenamefont {Chen}(1987)}]{Chen:1987}%
  \BibitemOpen
  \bibfield  {author} {\bibinfo {author} {\bibfnamefont {P.}~\bibnamefont {Chen}},\ }\bibfield  {title} {\bibinfo {title} {{A Possible Final Focusing Mechanism for Linear Colliders}},\ }\href {https://cds.cern.ch/record/166083/files/p171.pdf} {\bibfield  {journal} {\bibinfo  {journal} {Part. Accel.}\ }\textbf {\bibinfo {volume} {20}},\ \bibinfo {pages} {171} (\bibinfo {year} {1987})}\BibitemShut {NoStop}%
\bibitem [{\citenamefont {van Tilborg}\ \emph {et~al.}(2015)\citenamefont {van Tilborg}, \citenamefont {Steinke}, \citenamefont {Geddes}, \citenamefont {Matlis}, \citenamefont {Shaw}, \citenamefont {Gonsalves} \emph {et~al.}}]{vanTilborg_2015}%
  \BibitemOpen
  \bibfield  {author} {\bibinfo {author} {\bibfnamefont {J.}~\bibnamefont {van Tilborg}}, \bibinfo {author} {\bibfnamefont {S.}~\bibnamefont {Steinke}}, \bibinfo {author} {\bibfnamefont {C.~G.~R.}\ \bibnamefont {Geddes}}, \bibinfo {author} {\bibfnamefont {N.~H.}\ \bibnamefont {Matlis}}, \bibinfo {author} {\bibfnamefont {B.~H.}\ \bibnamefont {Shaw}}, \bibinfo {author} {\bibfnamefont {A.~J.}\ \bibnamefont {Gonsalves}}, \emph {et~al.},\ }\bibfield  {title} {\bibinfo {title} {Active plasma lensing for relativistic laser-plasma-accelerated electron beams},\ }\href {https://doi.org/10.1103/PhysRevLett.115.184802} {\bibfield  {journal} {\bibinfo  {journal} {Phys. Rev. Lett.}\ }\textbf {\bibinfo {volume} {115}},\ \bibinfo {pages} {184802} (\bibinfo {year} {2015})}\BibitemShut {NoStop}%
\bibitem [{\citenamefont {Benedetti}\ \emph {et~al.}(2017)\citenamefont {Benedetti}, \citenamefont {Schroeder}, \citenamefont {Esarey},\ and\ \citenamefont {Leemans}}]{Benedetti_2017}%
  \BibitemOpen
  \bibfield  {author} {\bibinfo {author} {\bibfnamefont {C.}~\bibnamefont {Benedetti}}, \bibinfo {author} {\bibfnamefont {C.~B.}\ \bibnamefont {Schroeder}}, \bibinfo {author} {\bibfnamefont {E.}~\bibnamefont {Esarey}},\ and\ \bibinfo {author} {\bibfnamefont {W.~P.}\ \bibnamefont {Leemans}},\ }\bibfield  {title} {\bibinfo {title} {Emittance preservation in plasma-based accelerators with ion motion},\ }\href {https://doi.org/10.1103/PhysRevAccelBeams.20.111301} {\bibfield  {journal} {\bibinfo  {journal} {Phys. Rev. Accel. Beams}\ }\textbf {\bibinfo {volume} {20}},\ \bibinfo {pages} {111301} (\bibinfo {year} {2017})}\BibitemShut {NoStop}%
\bibitem [{\citenamefont {Panofsky}\ and\ \citenamefont {Wenzel}(1956)}]{panofsky_wenzel}%
  \BibitemOpen
  \bibfield  {author} {\bibinfo {author} {\bibfnamefont {W.~K.~H.}\ \bibnamefont {Panofsky}}\ and\ \bibinfo {author} {\bibfnamefont {W.~A.}\ \bibnamefont {Wenzel}},\ }\bibfield  {title} {\bibinfo {title} {Some considerations concerning the transverse deflection of charged particles in radio‐frequency fields},\ }\href {https://doi.org/10.1063/1.1715427} {\bibfield  {journal} {\bibinfo  {journal} {Rev. Sci. Instrum.}\ }\textbf {\bibinfo {volume} {27}},\ \bibinfo {pages} {967} (\bibinfo {year} {1956})}\BibitemShut {NoStop}%
\bibitem [{\citenamefont {Diederichs}\ \emph {et~al.}(2023)\citenamefont {Diederichs}, \citenamefont {Benedetti}, \citenamefont {Esarey}, \citenamefont {Th\'evenet}, \citenamefont {Sinn}, \citenamefont {Osterhoff},\ and\ \citenamefont {Schroeder}}]{diederichs_temperature_2023}%
  \BibitemOpen
  \bibfield  {author} {\bibinfo {author} {\bibfnamefont {S.}~\bibnamefont {Diederichs}}, \bibinfo {author} {\bibfnamefont {C.}~\bibnamefont {Benedetti}}, \bibinfo {author} {\bibfnamefont {E.}~\bibnamefont {Esarey}}, \bibinfo {author} {\bibfnamefont {M.}~\bibnamefont {Th\'evenet}}, \bibinfo {author} {\bibfnamefont {A.}~\bibnamefont {Sinn}}, \bibinfo {author} {\bibfnamefont {J.}~\bibnamefont {Osterhoff}},\ and\ \bibinfo {author} {\bibfnamefont {C.~B.}\ \bibnamefont {Schroeder}},\ }\bibfield  {title} {\bibinfo {title} {Temperature effects in plasma-based positron acceleration schemes using electron filaments},\ }\href {https://doi.org/10.1063/5.0155489} {\bibfield  {journal} {\bibinfo  {journal} {Phys. Plasmas}\ }\textbf {\bibinfo {volume} {30}},\ \bibinfo {pages} {073104} (\bibinfo {year} {2023})}\BibitemShut {NoStop}%
\bibitem [{\citenamefont {Barklow}\ \emph {et~al.}(2023)\citenamefont {Barklow}, \citenamefont {Gessner}, \citenamefont {Hogan}, \citenamefont {Ng}, \citenamefont {Peskin}, \citenamefont {Raubenheimer} \emph {et~al.}}]{Barklow2023}%
  \BibitemOpen
  \bibfield  {author} {\bibinfo {author} {\bibfnamefont {T.}~\bibnamefont {Barklow}}, \bibinfo {author} {\bibfnamefont {S.}~\bibnamefont {Gessner}}, \bibinfo {author} {\bibfnamefont {M.}~\bibnamefont {Hogan}}, \bibinfo {author} {\bibfnamefont {C.-K.}\ \bibnamefont {Ng}}, \bibinfo {author} {\bibfnamefont {M.}~\bibnamefont {Peskin}}, \bibinfo {author} {\bibfnamefont {T.}~\bibnamefont {Raubenheimer}}, \emph {et~al.},\ }\bibfield  {title} {\bibinfo {title} {Beam delivery and beamstrahlung considerations for ultra-high energy linear colliders},\ }\href {https://doi.org/10.1088/1748-0221/18/09/p09022} {\bibfield  {journal} {\bibinfo  {journal} {Journal of Instrumentation}\ }\textbf {\bibinfo {volume} {18}}\bibinfo  {number} { (09)},\ \bibinfo {pages} {P09022}}\BibitemShut {NoStop}%
\bibitem [{\citenamefont {Schroeder}\ \emph {et~al.}(2016)\citenamefont {Schroeder}, \citenamefont {Esarey}, \citenamefont {Benedetti},\ and\ \citenamefont {Leemans}}]{Schroeder:2016phw}%
  \BibitemOpen
\bibfield  {number} {  }\bibfield  {author} {\bibinfo {author} {\bibfnamefont {C.~B.}\ \bibnamefont {Schroeder}}, \bibinfo {author} {\bibfnamefont {E.}~\bibnamefont {Esarey}}, \bibinfo {author} {\bibfnamefont {C.}~\bibnamefont {Benedetti}},\ and\ \bibinfo {author} {\bibfnamefont {W.~P.}\ \bibnamefont {Leemans}},\ }\bibfield  {title} {\bibinfo {title} {{Efficiency considerations for high-energy physics applications of laser-plasma accelerators}},\ }\href {https://doi.org/10.1063/1.4965590} {\bibfield  {journal} {\bibinfo  {journal} {AIP Conf. Proc.}\ }\textbf {\bibinfo {volume} {1777}},\ \bibinfo {pages} {020001} (\bibinfo {year} {2016})}\BibitemShut {NoStop}%
\bibitem [{\citenamefont {Lindstr{\o}m}(2019)}]{Lindstrom_phd}%
  \BibitemOpen
  \bibfield  {author} {\bibinfo {author} {\bibfnamefont {C.~A.}\ \bibnamefont {Lindstr{\o}m}},\ }\emph {\bibinfo {title} {{Emittance growth and preservation in a plasma-based linear collider}}},\ \href {https://www.duo.uio.no/handle/10852/66134?locale-attribute=no} {Ph.D. thesis},\ \bibinfo  {school} {University of Oslo} (\bibinfo {year} {2019})\BibitemShut {NoStop}%
\bibitem [{\citenamefont {Shiltsev}\ \emph {et~al.}(1999)\citenamefont {Shiltsev}, \citenamefont {Danilov}, \citenamefont {Finley},\ and\ \citenamefont {Sery}}]{Shiltsev_1999}%
  \BibitemOpen
  \bibfield  {author} {\bibinfo {author} {\bibfnamefont {V.}~\bibnamefont {Shiltsev}}, \bibinfo {author} {\bibfnamefont {V.}~\bibnamefont {Danilov}}, \bibinfo {author} {\bibfnamefont {D.}~\bibnamefont {Finley}},\ and\ \bibinfo {author} {\bibfnamefont {A.}~\bibnamefont {Sery}},\ }\bibfield  {title} {\bibinfo {title} {Considerations on compensation of beam-beam effects in the tevatron with electron beams},\ }\href {https://doi.org/10.1103/PhysRevSTAB.2.071001} {\bibfield  {journal} {\bibinfo  {journal} {Phys. Rev. ST Accel. Beams}\ }\textbf {\bibinfo {volume} {2}},\ \bibinfo {pages} {071001} (\bibinfo {year} {1999})}\BibitemShut {NoStop}%
\bibitem [{\citenamefont {Foster}\ \emph {et~al.}(2023)\citenamefont {Foster}, \citenamefont {D'Arcy},\ and\ \citenamefont {Lindstr{\o}m}}]{foster_2023}%
  \BibitemOpen
  \bibfield  {author} {\bibinfo {author} {\bibfnamefont {B.}~\bibnamefont {Foster}}, \bibinfo {author} {\bibfnamefont {R.}~\bibnamefont {D'Arcy}},\ and\ \bibinfo {author} {\bibfnamefont {C.~A.}\ \bibnamefont {Lindstr{\o}m}},\ }\bibfield  {title} {\bibinfo {title} {A hybrid, asymmetric, linear higgs factory based on plasma-wakefield and radio-frequency acceleration},\ }\href {http://iopscience.iop.org/article/10.1088/1367-2630/acf395} {\bibfield  {journal} {\bibinfo  {journal} {New J. Phys.}\ } (\bibinfo {year} {2023})}\BibitemShut {NoStop}%
\bibitem [{\citenamefont {Telnov}(1998)}]{Telnov:1998vs}%
  \BibitemOpen
  \bibfield  {author} {\bibinfo {author} {\bibfnamefont {V.~I.}\ \bibnamefont {Telnov}},\ }\bibfield  {title} {\bibinfo {title} {{Gamma gamma, gamma-electron colliders}},\ }in\ \href@noop {} {\emph {\bibinfo {booktitle} {{17th International Conference on High-Energy Accelerators}}}}\ (\bibinfo {year} {1998})\ p.~\bibinfo {pages} {88},\ \Eprint {https://arxiv.org/abs/hep-ex/9810019} {arXiv:hep-ex/9810019} \BibitemShut {NoStop}%
\bibitem [{\citenamefont {Rosenzweig}\ \emph {et~al.}(1996)\citenamefont {Rosenzweig} \emph {et~al.}}]{rosenzweig:1996}%
  \BibitemOpen
  \bibfield  {author} {\bibinfo {author} {\bibfnamefont {J.}~\bibnamefont {Rosenzweig}} \emph {et~al.},\ }\bibfield  {title} {\bibinfo {title} {{A linear collider based on nonlinear plasma wake-field acceleration}},\ }in\ \href {https://www.slac.stanford.edu/pubs/snowmass96/PDF/ACC067.PDF} {\emph {\bibinfo {booktitle} {{Proceedings of Snowmass 1996}}}}\ (\bibinfo {year} {1996})\BibitemShut {NoStop}%
\bibitem [{\citenamefont {Adli}(2018)}]{adli_2019}%
  \BibitemOpen
  \bibfield  {author} {\bibinfo {author} {\bibfnamefont {E.}~\bibnamefont {Adli}},\ }\bibfield  {title} {\bibinfo {title} {{Plasma Wakefield Linear Colliders---Opportunities and Challenges}},\ }\href {https://doi.org/10.1098/rsta.2018.0419} {\bibfield  {journal} {\bibinfo  {journal} {Phil. Trans. Roy. Soc. Lond. A}\ }\textbf {\bibinfo {volume} {377}},\ \bibinfo {pages} {0419} (\bibinfo {year} {2018})}\BibitemShut {NoStop}%
\end{thebibliography}%

\end{document}